\documentclass{aa} 

\usepackage[T1]{fontenc}
\usepackage{graphicx}
\usepackage{subfigure}
\usepackage{txfonts}	
\usepackage{amsmath}	
\usepackage{tabularx}
\usepackage{booktabs}
\usepackage{multirow}
\usepackage{newtxtext,newtxmath}
\usepackage{comment}
\usepackage{lscape}
\usepackage{xcolor}
\usepackage{txfonts}
\usepackage{hyperref}
 
\usepackage[flushleft]{threeparttable}
\def\4u{4U 1820-303}
\def\fa{f$_a$}
\def\cstat{{\sc C-stat}}

\def\nh{{$N_{\rm H}$}} 
\def\be{\begin{equation}} 
\def\ee{\end{equation}} 
\def\dcstat{$\Delta$\cstat}

\def\msol{M$_\odot$}
\def\ergscm2{erg s$^{-1}$ cm$^{-2}$}
\def\ergss{erg s$^{-1}$}

\begin{document}

   \title{Line detections in photospheric radius expansion bursts from \\ 4U 1820-303}

   \subtitle{Confirmation of previous detections and their temporal evolution}

   \author{F. Barra\inst{1,2,3},
   D. Barret\inst{4}, C. Pinto\inst{3}, T. Di Salvo\inst{1}, N. Weinberg\inst{5}, S. Guichandut\inst{6}
          }
   \institute{Universit\`a degli Studi di Palermo, Dipartimento di Fisica e Chimica, via Archirafi 36, I-90123 Palermo, Italy \\ \email{francesco.barra@unipa.it}
\and
Center for Astrophysics | Harvard  \& Smithsonian, 60 Garden Street, Cambridge, MA 02138, USA
\and
INAF/IASF Palermo, via Ugo La Malfa 153, I-90146 Palermo, Italy 
\and
Institut de Recherche en Astrophysique et Planétologie, 9 avenue du Colonel Roche, Toulouse, 31028, France \\
\email{dbarret@irap.omp.eu} 
\and
Department of Physics, University of Texas at Arlington, Arlington, TX 76019, USA
\and
Department of Physics and Trottier Space Institute, McGill University, 3600 rue University, Montreal, QC, H3A 2T8, Canada
\\
              }
   \date{XXXX; accepted YYYY}

 
  \abstract
   { NICER (Neutron star Interior Composition ExploreR) 
    is the instrument of choice for the spectral analysis of type I X-ray bursts, as it provides high throughput at X-ray CCD resolution, down to 0.3 keV. }
   {Triggered by the detection of absorption and emission lines in the first four photospheric radius expansion (PRE) bursts detected by NICER, 
   we wish to test the dependency of the absorption line energies with the inferred blackbody radius, as it was reported that the absorption line energies were positively correlated with the inferred blackbody radius, tentatively explained by a combination of a weaker gravitational redshift and higher blue shifts in burst showing larger blackbody radius.}
   {We thus re-analyse those 4 bursts and analyse 8 more bursts from \4u, for which we report evidence for PRE. We first follow the spectral evolution of the burst on the shortest possible timescales (tenth of a second). We adopt two parallel continuum descriptions to characterise the photospheric expansion and line evolution. Using the accretion enhanced  model, in which the burst emission is modelled as the sum of a blackbody and a component describing the persistent emission recorded prior to the burst and multiplied by a constant (\fa), 
   we infer maximum equivalent blackbody radii up  to $\sim 900$ km. 

   The peak bolometric (0.1-20 keV) luminosity reached between $4-7 \times 10^{\rm 38}$ {\ergss} (and even higher if absorption from a putative photo-ionised absorber is accounted for) in our sample of bursts, which is  greater than the Eddington luminosity of an Helium accretor. 

   In individual bursts, we detect absorption lines and assess their significance through extensive Monte Carlo simulations. For the characterisation of the spectral lines we use state-of-art plasma codes available within {\sc{spex}} with a phenomenological continuum. A deep search throughout the temperature-velocity parameter space is run to explore Doppler shifts and minimise the chance of getting stuck into local minima.}
   {We detect several significant (> 99.9 \% significance) absorption lines, including the 2.97 keV line previously reported. 
   On the one hand, we do not confirm the correlation between the line energies and the inferred blackbody radius, while on the other in some bursts showing the larger radii, up to four lines are reported and the line strength is higher. From the modelling of the feature lines, a photo-/collisionally ionised slightly redshifted (almost rest-frame) gas in emission is suggested in the most cases, although, in particular for the burst presenting the greatest PRE, a combination of photo-ionisation plasma in emission and absorption is preferred.}
  {}
\authorrunning{F. Barra et al. }
\titlerunning{Line detections in photospheric radius expansion bursts from \4u}

   \maketitle
%

\section{Introduction}\label{sec1}

\4u\ is a well studied type I X-ray burster \citep{Grindlay1976ApJ...205L.127G,Vacca1986MNRAS.220..339V,Paradijs1987A&A...172L..20V,Haberl1987ApJ...314..266H,Kuulkers2002A&A...382..503K,Strohmayer2002ApJ...566.1045S,Cumming2003ApJ...595.1077C,Ballantyne2004ApJ...602L.105B,Guver2010ApJ...719.1807G,Boutloukos2010ApJ...720L..15B,Kusmierek2011MNRAS.415.3344K,Zand2012A&A...547A..47I, Ozel2016ApJ...820...28O, Suleimanov2017MNRAS.472.3905S,Keek2018ApJ...856L..37K,Strohmayer2019ApJ...878L..27S,Galloway2021ASSL..461..209G}. The X-ray binary is located in the globular cluster NGC 6624 at a distance of 8.0 kpc, according to the latest measurements combining Gaia EDR3, HST, and older data \citep{Baumgardt2021MNRAS.505.5957B}. \4u~has an orbital period of only 11.4 min implying a low mass Helium white dwarf companion for the neutron star \citep{Stella1987ApJ...312L..17S}. Thus matter accreting onto the neutron star is predominantly Helium \citep[e.g.][]{Cumming2003ApJ...595.1077C}. 

X-ray bursts are very energetic events, resulting from unstable thermonuclear burning of accreted fuel on the surface of neutron stars \cite[][for a recent review]{Galloway2021ASSL..461..209G}. \4u\ shows a wide variety of bursts, from superbursts \citep{Strohmayer2002ApJ...566.1045S,Ballantyne2004ApJ...602L.105B} to strong photospheric radius expansion (PRE) burst \citep[e.g.][]{Keek2018ApJ...856L..37K,Galloway2017PASA...34...19G,Suleimanov2017MNRAS.472.3905S} and regular bursts \cite[][and reference therein]{Galloway2020ApJS..249...32G}. From the burst properties, constraints on the mass and radius of the neutron star can be derived \citep{Guver2010ApJ...719.1807G,Kusmierek2011MNRAS.415.3344K,Ozel2016ApJ...820...28O,Suleimanov2017MNRAS.472.3905S}. Overall, these measurements point to a rather small radius for the neutron star, about $10-12$ km, while its mass seems on the other hand on the high side (1.6 \msol) \citep{Ozel2016ApJ...820...28O}.
Interestingly, \cite{Ballantyne2004ApJ...602L.105B}, following a 3 hour long superburst with the Rossi X-ray Timing Explorer \citep[RXTE;][]{Jahoda1996SPIE.2808...59J} detected a reflection component, whose parameters suggested that the inner disc was pushed away during the event, before recovering its initial state about 1000 seconds later. Using the same RXTE data, \cite{zand2010A&A...520A..81I} reported evidence for a variable $\sim 10$ keV  edge, interpreted as the spectral signature of heavy element ashes expected to be present in thermonuclear X-ray bursts with photospheric super expansion \citep{Weinberg2006ApJ...639.1018W}.

The launch of the Neutron star Interior Composition ExploreR \citep[NICER;][]{Gendreau2012SPIE.8443E..13G} offers the unique opportunity to study X-ray bursts, being very bright events, with a spectral resolution, comparable to X-ray CCDs ($\sim 100$ eV), avoiding at the same time, issues such as pile-up. As important, NICER enables exploring the energy band pass below 2.5-3 keV, in which the Rossi X-ray Timing Explorer was not sensitive, and in which the bulk of the blackbody flux is emitted in the PRE phase. It is also an energy range in which the impact of the burst on the accretion disc can be probed \citep[e.g.][]{Speicher2022MNRAS.509.1736S}. Finally thanks to the improved spectral resolution, NICER offers unprecedented sensitivity for detecting narrow emission and absorption features \citep[see][for recent reviews]{Degenaar2018SSRv..214...15D,Galloway2021ASSL..461..209G}. All this clearly makes NICER the instrument of choice for deep spectral investigations of type I X-ray bursts.

\4u being a prototype X-ray burster, it is therefore not surprising that NICER observed it soon after launch, and quite intensively afterwards. The first burst observed with NICER showed evidence for photospheric radius expansion \citep{Keek2018ApJ...856L..37K} (see below). That one, and the subsequent four bursts, 3 of which showed PRE, were searched for X-ray spectral features by \cite{Strohmayer2019ApJ...878L..27S}. Combining pairs of bursts showing PRE, \cite{Strohmayer2019ApJ...878L..27S} reported a prominent  $\sim 1$ keV emission line, and two absorption lines at 1.7 and 3 keV respectively. \cite{Strohmayer2019ApJ...878L..27S} interpreted the lines in the context of burst-driven wind models \citep{Yu2018ApJ...863...53Y}  finding that the line energies in the co-added spectrum of the first burst pair (reaching higher photospheric radii) appeared blueshifted by a factor of $1.046 \pm 0.006$ compared to the line energies of the second pair of bursts. This was interpreted in a scenario in which pair 1 bursts showing larger radii suffered from weaker gravitational redshifts, but being supposed to generate faster outflows experienced higher blueshifts. 

In this paper, we extend the analysis of \cite{Keek2018ApJ...856L..37K} and \cite{Strohmayer2019ApJ...878L..27S} to a larger sample of 12 bursts present in the NICER archival data of \4u, with the goal of searching for more PRE bursts to further investigate the presence of spectral features. We first analyse the data with the so-called enhanced accretion model (hereafter referred to as \fa\ model) following \cite{Worpel2013ApJ...772...94W}, over the shortest possible timescales (a few tenths of a second). We then focus the spectral analysis of the time interval of the burst, during which the inferred blackbody radius is larger than 100 km, as derived from the \fa\ model. We repeat the analysis with the \fa\ model with the {\sc{xspec}} code \citep{Arnaud1996ASPC..101...17A} using its Python wrapper Py{\sc{xspec}} (Sect. \ref{Sect: continuum_modelling_xspec}). Then the residuals are searched for absorption edges that may be best detected, during the PRE phase of the burst. Finally, we characterise the spectral lines with models of optically-thin plasmas with the state-of-art models available in {\sc{spex}} (Sect. \ref{Sect: line_modelling_spex}). In all spectral fits we adopt Cash statistics \citep{Cash1979ApJ...228..939C} and optimal spectra binning \citep{Kaastra2016}.

\section{Spectra extraction and burst continuum modelling}
\label{Sect: continuum_modelling_xspec}

We have retrieved all the archival data of \4u\ from HEASARC up to December 2021, and processed them with standard filtering criteria with the \texttt{nicerl2} script provided as part of  the \texttt{HEASOFTV6.31.1} software suite, as recommended from the NICER data analysis web page (NICER software version : \texttt{{NICER\_2022-12-16\_V010a}}). Similarly, the latest calibration files of the instrument are used throughout this paper (reference from the \texttt{CALDB} database is \texttt{xti20221001}). A systematic error of 1.5\% was added as suggested by the NICER data analysis web page.

The cleaned light curves from each OBSID were produced with \texttt{nicerl3} between the 0.3 and 7\,keV band, with a time resolution of 120\,ms, and searched for X-ray bursts, using a standard scanning technique (searching deviations above a local mean). Over the archival data set, 7 more bursts were detected (see Table \ref{table_bfit_persistent_emission}) in addition to those reported by \cite{Keek2018ApJ...856L..37K,Strohmayer2019ApJ...878L..27S}. 

\subsection{The persistent emission before the burst }

We first extract a spectrum of the persistent emission through \texttt{nicerl3}, with a 200 second exposure, over an interval ending 10 seconds before the onset of the burst (in two cases, the segment is shortened to $\sim 60$ seconds as the burst occurred at the start of the observations). The spectra are grouped with the recommended optimal binning \citep[\texttt{ftgrouppha,}][]{Kaastra2016}\footnote{https://heasarc.gsfc.nasa.gov/lheasoft/help/ftgrouppha.html}. The overall spectral shape of the persistent emission combines a soft and a harder component. The persistent emission spectrum is modelled in {\sc{xspec}} as a sum of an absorbed disc blackbody and a power law. Alternative models for the hard component, such as comptonised component (\texttt{nthcomp} or \texttt{comptt}) provides equally good fits, although the seed photon temperature or the electron temperature cannot be constrained (see Section \ref{Sect: line_modelling_spex} for more details). 

We record the quality of the fit, as the deviation of the \cstat\ against the expected value following  \cite{Kaastra2017A&A...605A..51K}. In order to avoid the fitting to be trapped in local minima, we initialise the fits with 20 random values drawn uniformly within plausible ranges. The best fit results of the persistent emission recorded prior to the burst are listed in Table \ref{table_bfit_persistent_emission}. To compute the flux/luminosity and associated errors, we generate a set of 40 parameters from the best-fit values and the covariance matrix and then within each set of parameters, we set the absorption column to zero and calculate the flux between 0.1 and 20 keV. The fluxes are ordered and the interval containing 90\% of the fluxes. As can be noticed, the parameters of the persistent emission remain close to each other over the burst data set. The 0.1-20 keV unabsorbed luminosity varied between $\sim 2.0$ and $\sim 6.0 \times 10^{37}$ {\ergss} (assuming a distance of 8 kpc). These are typical luminosities for the hard spectral state of the source \citep{bloser2000ApJ...542.1000B}, when type I X-ray bursts occur.

\begin{table*}[!t]
\caption{Results of the best fit continuum modelling of the persistent emission with the \texttt{TBabs*(diskbb+powerlaw)} model in \sc{xspec}.} 
\begin{minipage}{\textwidth}
\footnotesize
    \centering
\begin{tabular}{cccccccccc}
\hline\hline
{ObsID} & {Date} & {Rate} & {Nh} & {kTin} & {Norm} &{$\Gamma$} & {Norm} & {L$_x$} & {\cstat\ (dev)} \\
\hline
1050300108 & 2017-08-29 & 1799.4 & $0.22^{+0.01}_{-0.01}$ & $0.91^{+0.04}_{-0.04}$ & $0.11^{+0.02}_{-0.02}$ & $1.46^{+0.05}_{-0.05}$ & $0.59^{+0.05}_{-0.05}$ & $7.32\pm 0.08$ & 393.5 (2.7) \\ 
1050300108 & 2017-08-29 & 1780.5 & $0.22^{+0.01}_{-0.01}$ & $0.86^{+0.03}_{-0.03}$ & $0.13^{+0.02}_{-0.02}$ & $1.48^{+0.04}_{-0.04}$ & $0.60^{+0.04}_{-0.04}$ & $7.24\pm 0.06$ & 408.7 (3.3) \\ 
1050300109 & 2017-08-30 & 1459.7 & $0.21^{+0.01}_{-0.01}$ & $0.63^{+0.01}_{-0.01}$ & $0.40^{+0.04}_{-0.04}$ & $1.45^{+0.05}_{-0.05}$ & $0.44^{+0.04}_{-0.04}$ & $5.64\pm 0.04$ & 342.4 (0.7) \\ 
1050300109 & 2017-08-30 & 1577.5 & $0.21^{+0.01}_{-0.01}$ & $0.69^{+0.02}_{-0.02}$ & $0.29^{+0.03}_{-0.03}$ & $1.44^{+0.05}_{-0.05}$ & $0.48^{+0.04}_{-0.04}$ & $6.25\pm 0.05$ & 364.4 (1.6) \\ 
1050300109 & 2017-08-30 & 1682.4 & $0.22^{+0.01}_{-0.01}$ & $0.74^{+0.02}_{-0.02}$ & $0.22^{+0.03}_{-0.03}$ & $1.44^{+0.04}_{-0.05}$ & $0.53^{+0.04}_{-0.04}$ & $6.84\pm 0.05$ & 395.6 (2.8) \\ 
2050300104 & 2019-06-07 & 999.2 & $0.21^{+0.01}_{-0.01}$ & $0.44^{+0.02}_{-0.02}$ & $0.92^{+0.11}_{-0.10}$ & $1.50^{+0.05}_{-0.05}$ & $0.34^{+0.03}_{-0.02}$ & $3.86\pm 0.04$ & 303.5 (-0.9) \\ 
2050300108 & 2019-06-12 & 607.7 & $0.22^{+0.01}_{-0.01}$ & $0.22^{+0.02}_{-0.02}$ & $7.5^{+3.5}_{-2.3}$ & $1.54^{+0.03}_{-0.03}$ & $0.26^{+0.01}_{-0.01}$ & $2.67\pm 0.03$ & 325.9 (-0.1) \\ 
2050300110 & 2019-06-14 & 643.0 & $0.20^{+0.01}_{-0.01}$ & $0.28^{+0.02}_{-0.02}$ & $2.8^{+1.0}_{-0.7}$ & $1.50^{+0.03}_{-0.03}$ & $0.26^{+0.01}_{-0.01}$ & $2.76\pm 0.02$ & 376.5 (1.9) \\ 
2050300119 & 2019-06-26 & 968.1 & $0.20^{+0.01}_{-0.01}$ & $0.43^{+0.01}_{-0.02}$ & $0.99^{+0.12}_{-0.10}$ & $1.44^{+0.05}_{-0.05}$ & $0.31^{+0.02}_{-0.02}$ & $3.85\pm 0.04$ & 392.4 (2.6) \\ 
2050300119 & 2019-06-26 & 1175.4 & $0.21^{+0.01}_{-0.01}$ & $0.53^{+0.01}_{-0.01}$ & $0.60^{+0.06}_{-0.05}$ & $1.45^{+0.05}_{-0.05}$ & $0.35^{+0.03}_{-0.03}$ & $4.52\pm 0.04$ & 334.1 (0.3) \\ 
2050300124 & 2019-07-01 & 1095.7 & $0.19^{+0.01}_{-0.01}$ & $0.51^{+0.02}_{-0.02}$ & $0.83^{+0.08}_{-0.07}$ & $1.21^{+0.07}_{-0.07}$ & $0.24^{+0.03}_{-0.03}$ & $4.72\pm 0.08$ & 341.7 (0.6) \\ 
4680010101 & 2021-05-02 & 971.2 & $0.20^{+0.01}_{-0.01}$ & $0.41^{+0.02}_{-0.02}$ & $1.13^{+0.16}_{-0.13}$ & $1.49^{+0.04}_{-0.04}$ & $0.34^{+0.02}_{-0.02}$ & $3.86\pm 0.03$ & 319.0 (-0.3) \\ 
\hline
\end{tabular}

\tablefoot{\label{table_bfit_persistent_emission}	\footnotesize The best fit spectral parameters of the persistent emission as recorded prior to the burst.  
The fit is performed between 0.3 and 10 keV. The integration time of the spectrum is 200 seconds, with the exception of the 2nd and 11th observation, for which only $\sim 120$ seconds were available prior to the burst. The spectra are grouped by 3 consecutive channels. For the \cstat\, the deviation of the best fit, compared to the expected value expressed in $\sigma$ and the number of degrees of freedom are given within the parenthesis. The bolometric luminosity is computed between 0.2 and 20 keV, assuming a distance of 8 kpc \citep{Baumgardt2021MNRAS.505.5957B} and expressed in units of $10^{37}$ {\ergss}. The normalisation parameter of the \texttt{diskbb} component is expressed in $10^3$ unit. Errors  on the best fit parameters are given at the 90\% confidence levels. The error on the luminosity is derived from the standard deviation of fluxes computed from the multivariate Gaussian distribution of the best fit parameters, from which 40 parameter values were randomly drawn (using \texttt{simpar} command in {\sc{xspec}}). }
\bigskip  
\end{minipage}
\end{table*}

\subsection{Burst profile modelling on short timescales with the \fa\ model }

We generate a set of spectra, over the burst profiles with a varying exposure time, ensuring that each spectrum has about 1000 burst counts, on top of the average persistent emission recorded prior to the burst (1000 burst counts is sufficient to constrain the \fa\ model, yet enabling short timescales to be probed). The integration timescales of the spectra range from tenths of a second at the peak to a second in the tail of the bursts. Following \cite{Worpel2013ApJ...772...94W}, the burst emission is modelled as the sum of the persistent emission spectrum multiplied by the so-called $f_a$ factor and a blackbody spectrum ($f_a$ greater than 1, suggests that the persistent emission increases during a burst, likely resulting from the effects of Poynting-Robertson drag on the disc material). In {\sc{xspec}} terminology, the model is then \texttt{TBabs$\times$(bbodyrad+f$_a\times$(diskbb+powerlaw))}.

Given the extreme count rate reached during the burst, the background is negligible. The column density and the other parameters of the persistent emission are frozen to the values derived from the pre-burst emission, only $f_a$ and the blackbody parameters are allowed to vary. The blackbody flux was converted to an equivalent radius assuming a source distance of 8 kpc \citep{Baumgardt2021MNRAS.505.5957B}. Although the bulk of the burst photons are detected below a few keV for the most part of the burst, close to touch down, the blackbody temperature reaches a few keV, hence can only be constrained by including the data up to 10 keV. The fit was performed again considering again a constant binning of 3 channels for the spectra, but we have verified that alternative binning scheme yielded fully consistent results. As to avoid the fit to get trapped in a local minimum, the first burst spectrum fit is initialised with a sample of 50 input values within plausible ranges. The fit of the subsequent burst spectra always starts from the fitted parameters of the previous burst spectrum, but is still shacked around, with 20 input values drawn randomly from plausible ranges.

\begin{figure*}
    \centering
    \includegraphics[width=0.84\textwidth]{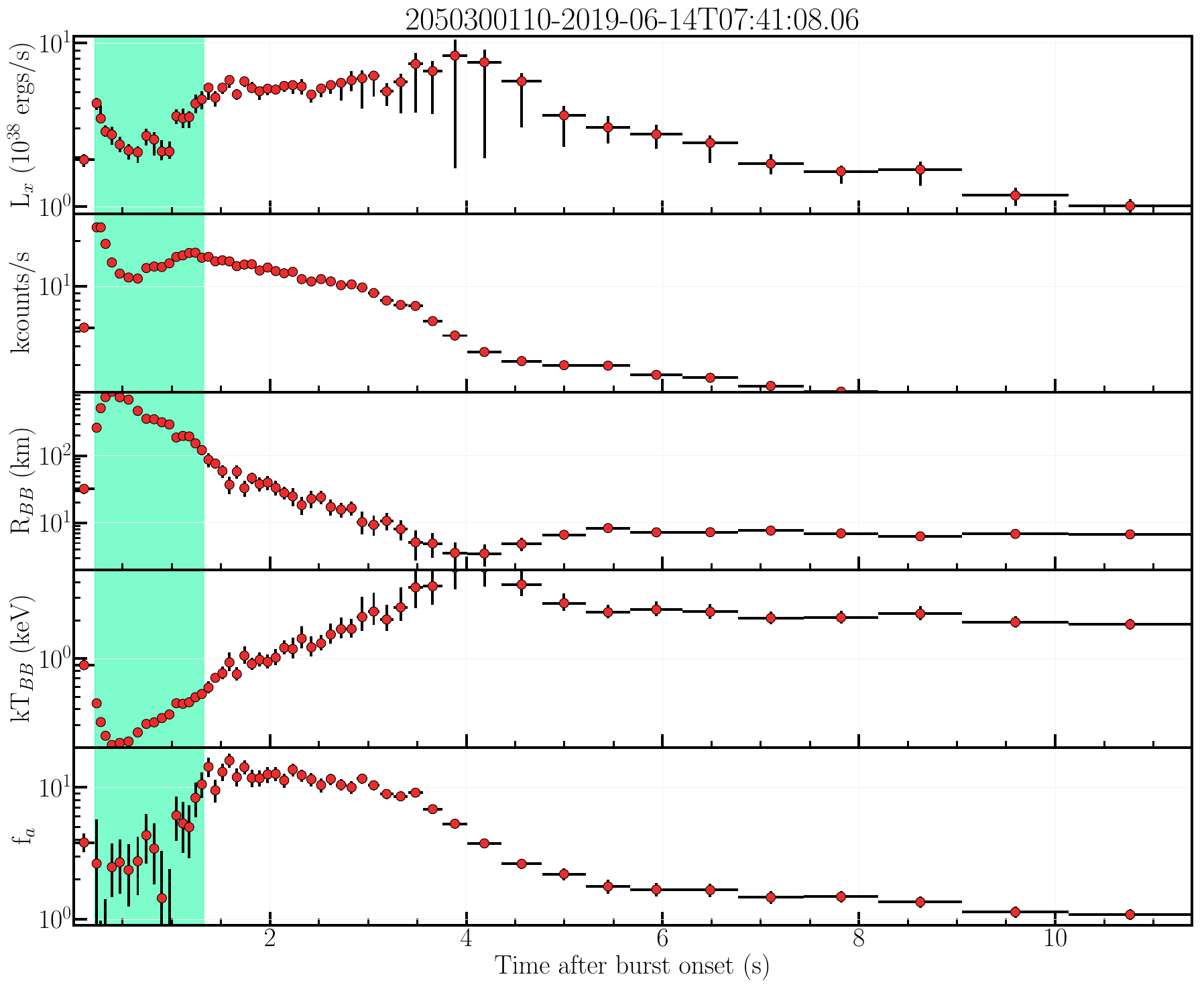}
    \caption{The best fit parameters of the burst recorded in the OBSID 2050300110. From top to bottom: The 0.1-20 keV bolometric X-ray luminosity in units of $10^{38}$ {\ergss} assuming a 8 kpc distance, the 0.3-10 keV count rate (counts/s), the inferred blackbody radius (km), the blackbody temperature (keV), and the \fa\ parameter. The green area defines the time period during which the inferred blackbody radius is larger than 100 km. All errors are given at the 90\% confidence level. }

    \label{fig:burst_parameters}
\end{figure*}

Of the 12 bursts detected, 10 showed cleared signs of PRE, see Figure \ref{fig:burst_parameters} for one representative event of our sample. Our analysis is similar to the one performed by \cite{Keek2018ApJ...856L..37K} on the first burst of the sample. We have also tested a different model for the persistent emission composed of an absorbed blackbody (\texttt{bbody}) and a comptonisation (\texttt{comptt}) component, which is used later on in {\sc{spex}}; our results are fully consistent with each another. In both analysis, the maximum radius reaches a plateau around 200 km, while the blackbody temperature drops to $\sim 0.45$ keV. Our values of \fa\ are also fully consistent, ceiling up around $\sim 8$ for that burst. From the peak of the PRE phase to touch down, we record a total bolometric (0.1-20 keV) flux of $\sim 8 \pm 0.1~\times 10^{-8}$ \ergscm2, slighly larger than the $7.5 \pm 0.1 \times 10^{-8}$ \ergscm2\ measured by \cite{Keek2018ApJ...856L..37K}. For the event presented in Figure \ref{fig:burst_parameters}, the total 0.1-20 keV bolometric X-ray luminosity reached about $7\times 10^{38}$ {\ergss}, before dropping down at the peak of the PRE, and later returning to its plateau value. This is also the period during with the \fa\ parameter quickly drops. All this happens within $\sim 0.5$ seconds. At a distance of 8 kpc, the source reached a super-Eddington luminosity for an Helium accretor (a neutron star with a pure helium photosphere has an isotropic Eddington luminosity of $L_{\rm Edd} \sim 2.5 \times 10^{38}~(M/M_\odot)~\sqrt{1-2.9644(M/M_\odot)/R_{km}}$ {\ergss}, e.g. \cite{Strohmayer2002ApJ...566.1045S}, translating to 3.1$\times 10^{38}$ {\ergss} for an 1.6 \msol\ neutron star of 12 km radius).

\subsection{Spectral analysis of the burst peak (0.7 second from the peak)}
\label{Spectral analysis of the burst peak (0.7 second from the peak)}

Following \cite{Strohmayer2019ApJ...878L..27S}, we extract spectra corresponding to a time interval of 0.7 seconds from the burst peak onwards. We do that by summing the individual $\sim 1000 $ counts spectra falling into this time interval. We first fit the spectra with the \fa\ model and  in all cases the fit is not acceptable from a statistical point of view. We model the burst emission as the sum of  a thermally comptonised continuum \texttt{nthcomp} model \citep{Zdziarski1996MNRAS.283..193Z,Zycky1999MNRAS.309..561Z}, accounting for the persistent emission, and a blackbody. We set the electron temperature to 10 keV and the seed photon temperature to 0.1 keV. The fit is also not acceptable from a statistical point of view. If we leave the column density as a free parameter, this improves the fit, but the best fitted value of \nh\ is significantly lower than the one derived from fitting the pre-burst emission inferred from the $f_a$ model ($\sim 0.17$ versus $\sim 0.21$ $\times 10^{22}$ cm$^{-2}$). This is problematic and likely due to the fact that the fit tries to compensate for the presence of the unmodelled $\sim 0.5$ keV excess feature, identified by \cite{Strohmayer2019ApJ...878L..27S}. The best fit parameters of the burst peak spectra fitted by the $f_a$ model, together with the inferred equivalent blackbody radii are listed in Table \ref{tab:0.7seconds}  (freezing the column density to $0.21 \times 10^{22}$ cm$^{-2}$). As can be seen, the range of blackbody radii probed by our sample extends significantly over the first 5 bursts reported by \cite{Keek2018ApJ...856L..37K,Strohmayer2019ApJ...878L..27S}, reaching out $900$ km.

\cite{Strohmayer2019ApJ...878L..27S} noticed that in addition to the 1 keV lines, excess residuals were also present at 0.5-0.6 keV and between 2.0 - 2.4 keV. Those excesses were attributed to known unmodelled residuals present in the NICER response function, at the time their analysis was performed (although we notice that such residuals are not present in the persistent emission spectra bearing the same number of counts). Unmodelled emission lines may artificially create dips in the spectra, that may be in turn falsely interpreted as absorption lines.  It is therefore worth to investigate first the properties of the emission features, beyond the 1 keV line, as other emission lines from different species may be present. In order to do so, we scan the spectrum, use a sliding Gaussian, over a fixed energy grid, stepping over the normalisation and the with of the Gaussian (in fixed ranges). We do the scanning iteratively, until the improvement of the overall \cstat\  does not exceed a critical threshold (set to 8 to begin with). Once an excess is so detected, a Gaussian function is added to the best fit model, freeing its energy and its normalisation, but retaining the width from the scan. This leads to improving the \cstat\, by a better ajustement of the energy and normalisation of the Gaussian. We will assess through simulations the critical threshold of the \cstat\ corresponding to a significance larger than 99\%. In addition to the 1 keV line, several additional highly significant emission features are detected at $\sim 2.0$ keV and $\sim 2.5-2.6$ keV.

\begin{figure*}[h!]
    \centering
    \includegraphics[width=0.4975\textwidth]{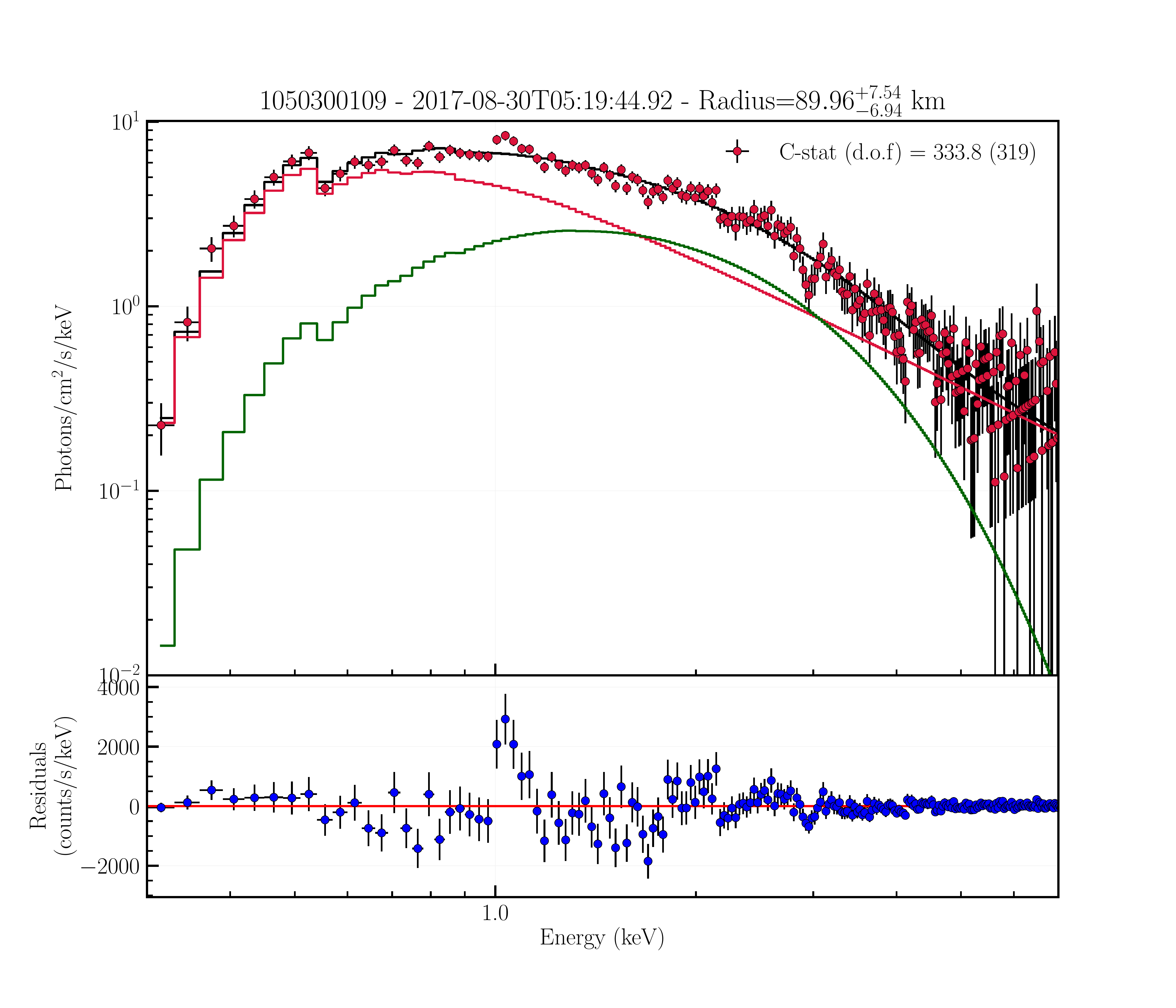}
     \includegraphics[width=0.4975\textwidth]{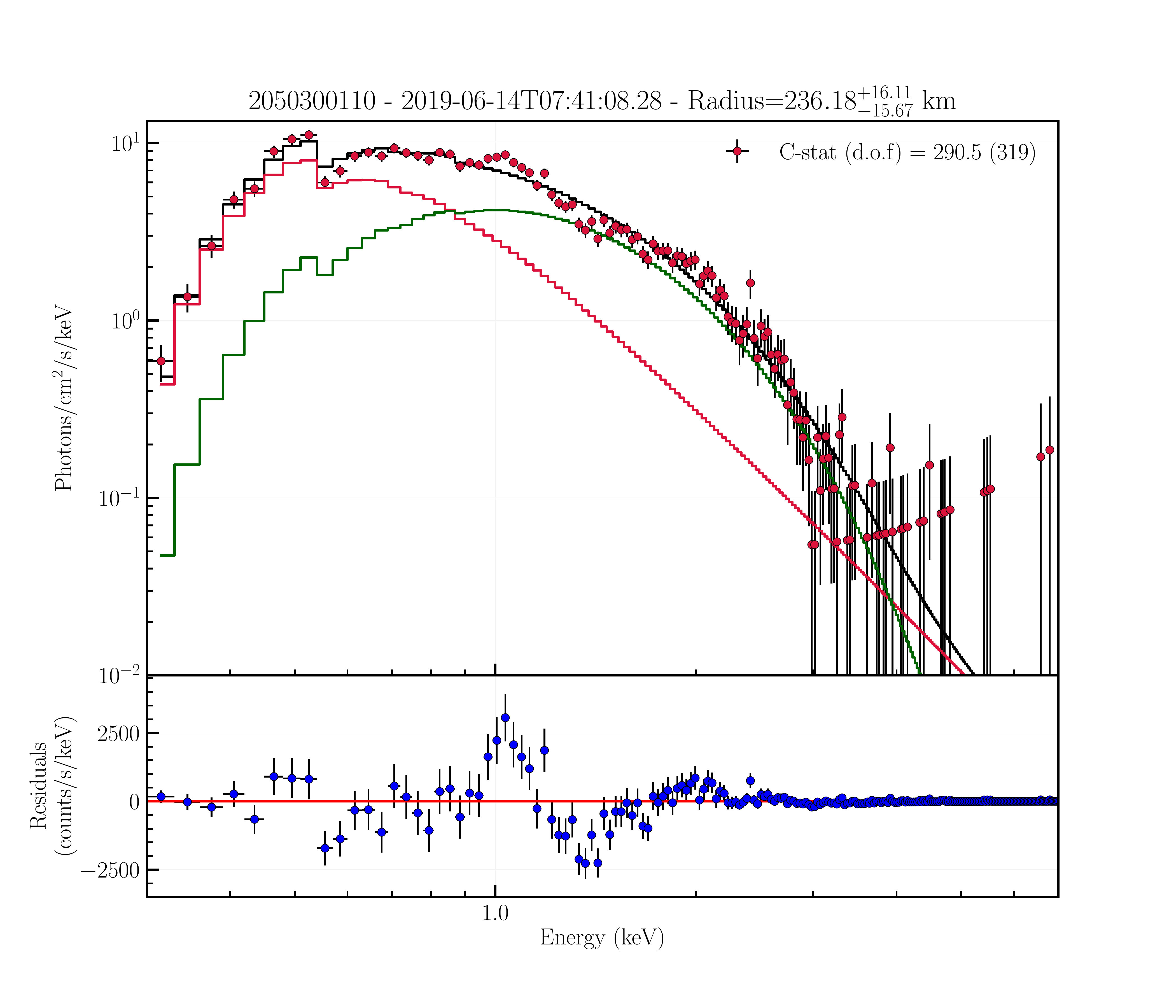}
    \caption{The unfolded spectra of the first burst recorded in OBSID 1050300109 and the one recorded in OBSID 2050300110. The best fit model is the sum of an absorbed \texttt{nthcomp} (in red) and blackbody (in dark green) components, with the column density set to  0.21 $\times 10^{22}$ cm$^{-2}$. The $\sim 1$ keV line as well as the $\sim 1.7$ and $\sim 3$ keV reported by \cite{Strohmayer2019ApJ...878L..27S}  are present also in our analysis.
    }
    \label{fig:0.7seconds}
\end{figure*}

The unfolded spectra and associated residuals of the fit for two representative bursts are shown in Figure \ref{fig:0.7seconds}. 
The lines reported by \cite{Strohmayer2019ApJ...878L..27S} at 1 keV in emission and 3 keV in absorption are clearly visible in the residuals (to improve the statistics, \cite{Strohmayer2019ApJ...878L..27S} grouped burst in pairs, but also noted that the lines may be present in individual bursts. The absorption line at 3 keV is also detected in a later observation, when the equivalent blackbody radius was 120 km.

\begin{table*}[h!]
\caption{Results of the continuum modelling for all bursts with the \fa\ model in {\sc{xspec}}.} 
\centering
	\begin{tabular}{ccccccccc}
    \hline\hline
	\# & OBSID & Date & $\Gamma$ & Norm & kT$_{BB}$ & Radius & L$_x$ & \cstat\\
	\hline
1 & 1050300108 & 2017-08-29T08:53:16.07 & $1.64^{+0.05}_{-0.05}$ & $5.8^{+0.3}_{-0.3}$ & $0.59^{+0.03}_{-0.03}$ & $0.89^{+0.09}_{-0.08}$ & $6.3\pm 0.3$ & 389.7 (2.6) \\ 
2 & 1050300108 & 2017-08-29T11:03:40.41 & $1.49^{+0.05}_{-0.05}$ & $5.6^{+0.3}_{-0.3}$ & $0.75^{+0.09}_{-0.08}$ & $0.46^{+0.09}_{-0.07}$ & $7.6\pm 0.4$ & 342.6 (0.3) \\ 
3 & 1050300109 & 2017-08-30T05:19:44.92 & $1.81^{+0.06}_{-0.05}$ & $6.5^{+0.3}_{-0.3}$ & $0.62^{+0.03}_{-0.03}$ & $0.90^{+0.07}_{-0.07}$ & $5.7\pm 0.2$ & 333.8 (0.6) \\ 
4 & 1050300109 & 2017-08-30T08:06:51.31 & $1.67^{+0.06}_{-0.05}$ & $6.6^{+0.3}_{-0.3}$ & $0.69^{+0.05}_{-0.05}$ & $0.64^{+0.07}_{-0.06}$ & $6.7\pm 0.3$ & 403.4 (3.1) \\ 
5 & 1050300109 & 2017-08-30T14:27:36.30 & $1.68^{+0.06}_{-0.05}$ & $6.4^{+0.3}_{-0.3}$ & $0.67^{+0.05}_{-0.04}$ & $0.68^{+0.08}_{-0.07}$ & $6.3\pm 0.4$ & 353.9 (1.1) \\ 
6 & 2050300104 & 2019-06-07T06:56:50.46 & $2.31^{+0.10}_{-0.09}$ & $5.9^{+0.3}_{-0.3}$ & $0.56^{+0.02}_{-0.02}$ & $1.21^{+0.07}_{-0.06}$ & $4.3\pm 0.1$ & 398.1 (5.3) \\ 
7 & 2050300108 & 2019-06-12T20:00:34.79 & $2.59^{+0.13}_{-0.11}$ & $4.5^{+0.3}_{-0.3}$ & $0.48^{+0.01}_{-0.01}$ & $1.51^{+0.08}_{-0.08}$ & $3.3\pm 0.1$ & 351.5 (4.8) \\ 
8 & 2050300110 & 2019-06-14T07:41:08.28 & $3.62^{+0.26}_{-0.21}$ & $4.1^{+0.4}_{-0.4}$ & $0.36^{+0.01}_{-0.01}$ & $2.36^{+0.16}_{-0.16}$ & $2.8\pm 0.1$ & 290.5 (5.9) \\ 
9 & 2050300119 & 2019-06-26T07:21:30.42 & $2.63^{+0.13}_{-0.11}$ & $4.9^{+0.4}_{-0.3}$ & $0.48^{+0.01}_{-0.01}$ & $1.87^{+0.08}_{-0.08}$ & $4.1\pm 0.1$ & 390.1 (6.7) \\ 
10 & 2050300119 & 2019-06-26T18:21:25.29 & $2.00^{+0.07}_{-0.06}$ & $6.3^{+0.3}_{-0.3}$ & $0.57^{+0.02}_{-0.02}$ & $1.10^{+0.07}_{-0.07}$ & $5.0\pm 0.1$ & 372.4 (2.9) \\ 
11 & 2050300124 & 2019-07-01T14:05:57.58 & $2.07^{+0.07}_{-0.07}$ & $6.0^{+0.3}_{-0.3}$ & $0.54^{+0.02}_{-0.02}$ & $1.26^{+0.08}_{-0.07}$ & $4.7\pm 0.1$ & 398.5 (4.3) \\ 
12 & 4680010101 & 2021-05-02T14:35:35.51 & $2.34^{+0.11}_{-0.10}$ & $5.8^{+0.3}_{-0.3}$ & $0.59^{+0.02}_{-0.02}$ & $1.20^{+0.06}_{-0.05}$ & $4.5\pm 0.1$ & 407.9 (5.8) \\ 
	\hline
	\end{tabular}  
\tablefoot{\label{tab:0.7seconds}{\footnotesize The best fit spectral parameters of the burst peak emission for the \fa\ model with the underlying persistent emission accounted for. Labels are same as in Table \ref{table_bfit_persistent_emission}. The bolometric luminosity is computed between 0.2 and 20 keV, assuming a distance of 8 kpc and expressed in units of $10^{38}$ {\ergss}. The radius parameter is expressed in unit of $10^2$ km. Errors on the best fit parameters are given at the 90\% confidence levels. }}
\end{table*}

\begin{figure*}[h!]
    \centering
    \includegraphics[width=0.9\textwidth]{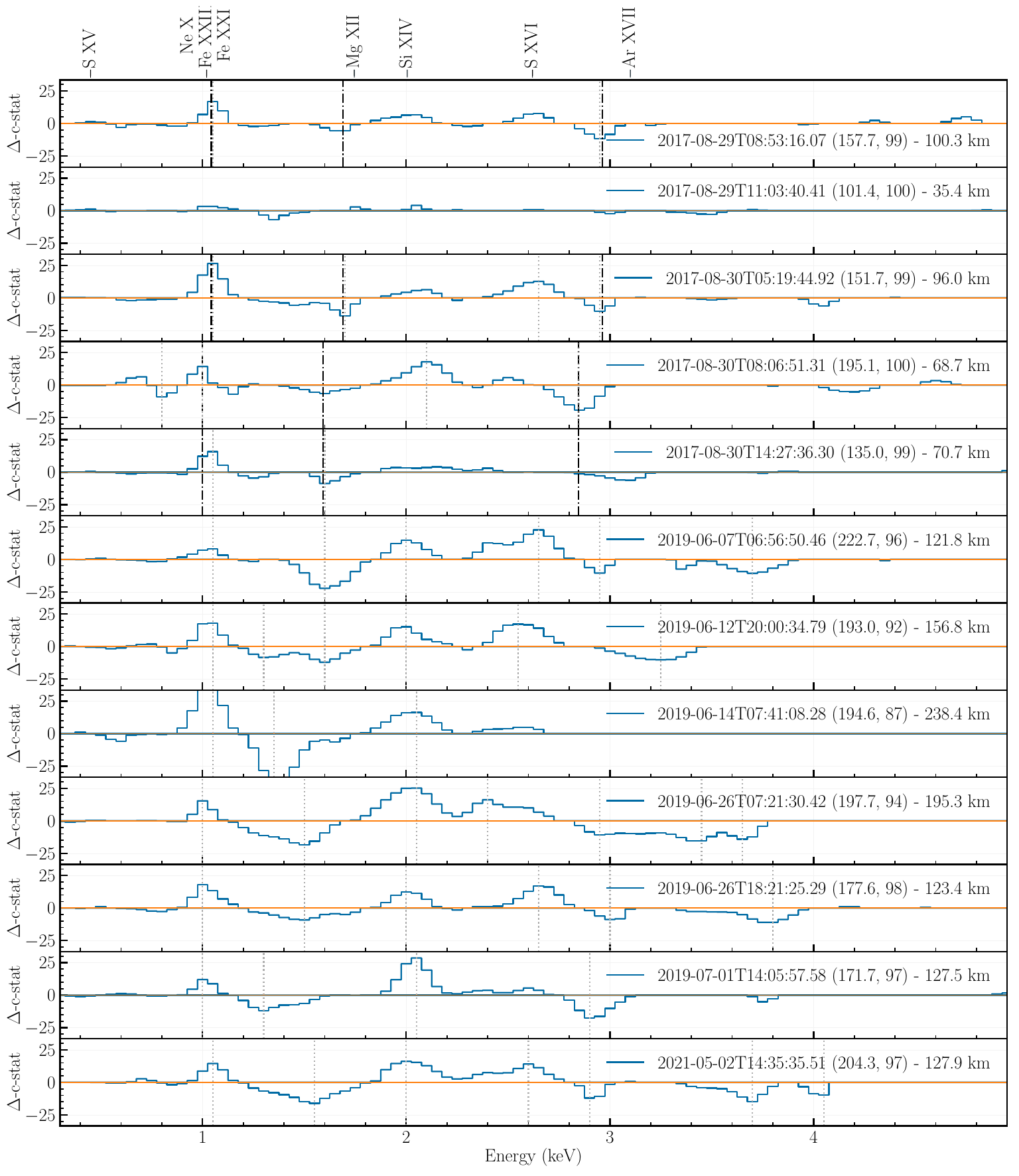}
    \caption{Gaussian line scan for all bursts (0.7s exposure). Bursts are sorted from top to bottom. The radii values provided for each burst represent the average values during periods when the blackbody radius exceeds 100 km.}
    \label{fig:gaus_scan}
\end{figure*}

\subsection{Scanning for edges and lines in individual burst peak spectra }

There is strong evidence that the 1 keV line reported by \cite{Strohmayer2019ApJ...878L..27S} is also present in the subsequent bursts. Before dealing with absorption features, we first add a Gaussian centred around 1 keV with a fixed width ($\sigma=0.05$ keV) to the \texttt{nthcomp} model, and fit the data. The lines is statistically significant in several observations, leading to an improvement of the fit by up to a $\Delta$\cstat\ of 25 (and even larger for Burst 8).

Starting from the best fit, one can now scan the spectra for the presence of additional features in the form of absorption edges or lines. For this purpose, using the \texttt{steppar} command we slide an edge function, with depth of the edge (\texttt{MaxTau} parameter) allowed to vary between 0 and 2, sampled with 100 steps. We do the same for a Gaussian function, allowing the normalisation to vary between -1 and 1 in 100 steps, and assuming a fixed width ($\sigma=0.05$ keV, consistent with the value reported by \cite{Strohmayer2019ApJ...878L..27S}). At each energy, we record the improvement of the fit, through the difference between the \cstat\ without any feature and the \cstat\ with the emission/absorption feature included: the largest the difference, the more significant the feature. We use the center of the energy bins of each spectrum as the energy grid for the scanning. The results from the Gaussian line scan are shown in Fig. \ref{fig:gaus_scan}. Most spectral scans confirm the 1 keV feature among other features including some around 2-3 keV. Interestingly, the significance of the 1 keV line is minimum for burst 2 and maximum in burst 8 which show by far the lowest and higher photospheric expansion, respectively.

We do not expect significantly difference results adopting narrower lines widths due to the limited spectral resolution. Given that the line modelling strongly depends on the assumption of emission or absorption - i.e. most individual features can be either modelled with emission or absorption lines - we prefer to describe more features simultaneously with plasma models. A more detailed attempt with models of optically-thin plasmas is shown later on in Sect. \ref{Sect: line_modelling_spex}.

\subsection{Significance of the absorption features detected}

Assuming that the edge features detected are not instrumental, it remains to assess their significance. For this purpose, we follow \cite{Hurkett2008ApJ...679..587H}. The covariance matrix evaluated at the best-fit point was used to construct the multivariate Gaussian distribution from which 1000 parameter values were randomly drawn (using \texttt{simpars} command in {\sc{xspec}}). For each set of model parameter values, a fake spectrum was simulated with the appropriate response and ancillary files and exposure time, with counts in each channel drawn from a Poisson distribution, and binned in the same manner as the observed data. Each simulated spectrum was then fitted and scanned for the presence of a spectral feature the same way as for the real data. We took burst 1 and 8 as test cases. Considering the edge scan, no simulated spectra showed an improvement of \dcstat\ larger than 10, while the maximum value observed in the data reaches 13 and 33. This means that the probability that the edge is a spurious feature, caused by photon statistics is thus $<0.1$\%; i.e., the edge features are detected at $>99.9$\% confidence. For observation 8, one can scale up the distribution resulting from the 1000 simulations to meet the observed \dcstat\, to determine that it would take about $10^8-10^9$ simulations, which is technically unfeasible. However, given the very high \dcstat\ value for some features (e.g. $\sim30$ for the 1 keV feature in burst 8), the chance that the many of these features are not caused by the noise and look-elsewhere effects still remain very high ($\gg 3 \sigma$) when multiplying the p-value for the number of independent spectral channels.

\section{Spectral analysis with models of optically-thin plasmas}
\label{Sect: line_modelling_spex}

The {\sc{spex}} fitting package (v3.07.03, \citealt{Kaastra_1996,kaastra2023_spex}) has state-of-art optically-thin plasma models which may provide further insights on the nature of the lines. As default in {\sc{spex}} we assumed all uncertainties at the 68\% level and group the spectra with the recommended optimal binning \citep{Kaastra2016}. For the purpose of studying the behaviour of emission and absorption lines coming from the bursts, we initially proceed to describe the continuum model of the spectra. All the models take into account the absorption from the circumstellar and interstellar medium by using the \textit{hot} model (with a gas temperature of $10^{-6}$ keV, which yields a neutral gas phase in {\sc{spex}} and is equivalent to \texttt{TBabs} in {\sc{xspec}}). We adopted the state-of-art Solar abundances \citep[][default in {\sc{spex}}]{Lodders2009} due to the limited spectral resolution and the confusion between the Fe L, Ne K and Mg K lines with the exception of the oxygen, fixed at 1.2 of the Solar value as recommended by the results from high-resolution X-ray spectroscopy \citep{Costantini2012}.

\subsection{Testing different spectral model for the Obs. 2050300110 }
First of all, we tested several double-component continuum model to have a reasonable description of the spectra before accounting for the narrow features. In order to do this, we focused on the spectrum of the burst 8 (Obs. ID 2050300110) which presents better statistics with respect to the others. In order to avoid degeneracies due to the model applied, the total column density is set to the value of 1.63 $\times$ $10^{21} \rm cm^{-2}$ estimated with high-resolution X-ray grating spectrometers \citep{Costantini2012}. The models tested are combinations of blackbody (\textit{bb}),  multi-temperature disc blackbody (\textit{dbb}) and inverse comptonisation of soft photons in a hot plasma  (\textit{comt}) model. The analysis of the burst 8 shown that the spectrum is reasonably described with a blackbody \textit{bb}+\textit{comt} model (see {\sc{spex}} manual) with a \cstat / d.o.f $\sim$ 159/57. The temperature of the blackbody is 0.161 keV to which is coupled the temperature of the seed photons of the \textit{comt} component. 

The optical depth of the \textit{comt} model is not constrained pegging to the limit of 100 that we imposed, perhaps because the photosphere is optically thick after the expansion and we are photon-starved above a few keV.  The results are shown in Table \ref{table: Results of the continuum modeling for the 12 bursts. }. The best-fit model confirms strong residuals around 1 keV, 2 keV and 2.6 keV in emission while 3 and 3.4 keV in absorption with a shape that agrees with the {\sc{xspec}} modelling. These features will be analysed in Sect. \ref{Grids multidimensional scan}.

\subsection{Continuum modelling for the 12 bursts}

We proceed to the fitting of the 12 burst spectra in order to describe the emission and absorption features. The temperature of the blackbody is ranging between 0.13 and 0.18 keV (of course with same range obtained for the temperature of the \textit{comt} seed photons). A constraint on the optical depth is difficult to achieve, but in some cases it is found to be around 13-18 (e.g. in burst 2, where there is no photospheric expansion). This is probably due to the fact that the portion of the photosphere that we are observing is optically thick. The best-fit \textit{bb+comt} model provides a good representation of the data with the exception of the residuals at 1 keV, 2 keV, 2.6 keV and the absorption features between 3 and 4 keV in full agreement with the results obtained through the \texttt{nthcomp} + \texttt{bbody} in {\sc{xspec}}. We will attempt at describing these features through plasma models that adopt either collisional-ionisation or photo-ionisation equilibrium in Sect. \ref{section:spex_grids}. We show the results of the continuum modelling for all the bursts spectra in Table \ref{table: Results of the continuum modeling for the 12 bursts. }. The X-ray and bolometric luminosities are slightly lower than those measured with the powerlaw models because the latter require stronger curvature in the soft X-ray band, higher column density and, thereby, higher luminosities. The remarkably higher {\cstat} in the spectra which exhibit strong photospheric expansion (and their absence in the persistent spectra) indicates that the features likely have an astrophysical origin and are somehow related to this phenomenology.

\begin{center}
\begin{table*}[h!]
\caption{Results of the continuum modelling for all bursts with a \textit{bb+comt} model in {\sc{spex}}.}  
 \renewcommand{\arraystretch}{1.}
 \small\addtolength{\tabcolsep}{-3pt}
 \vspace{-0.1cm}
	\centering
	\scalebox{.76}{%
	\begin{tabular}{cccccccccccccc}
    \hline\hline
    \toprule

Fit parameter              & Burst 1             & Burst 2          & Burst 3            & Burst 4                  & Burst 5             & Burst 6                    & Burst 7                       & Burst 8                  & Burst 9                           & Burst 10            & Burst 11                      & Burst 12               \\
\hline
   $\rm n_{\rm H}$                  & 1.63              & 1.63              & 1.63               & 1.63                      & 1.63              & 1.63                     & 1.63                          & 1.63                       & 1.63                         & 1.63                   & 1.63                             & 1.63                   \\
  $\rm L_{X,bb} $                & 0.09 $\pm$ 0.03    & 0.05 $\pm$ $_{0.05}^{0.08}$   & 0.06    $\pm$ $_{0.06}^{0.08}$  & 0.25  $\pm$ 0.19        &   0.18   $\pm$ $_{0.08}^{0.28}$    &  0.91 $\pm$ 0.06          &  0.76  $\pm$ 0.02              & 0.93 $\pm$ 0.04         &  0.86 $\pm$ 0.05                 &  0.04 $\pm$  0.11   &  0.68 $\pm$ 0.14             &  0.83  $\pm$ 0.09          \\
  $\rm L_{X,comt}$              & 3.70  $\pm$ 0.17    & 4.05  $\pm$ 0.26 & 3.98   $\pm$ 0.25  & 3.78  $\pm$ 0.34             &   3.73  $\pm$ 0.22  &  2.63  $\pm$ 0.06          &  2.09  $\pm$ 0.05             & 1.27  $\pm$ 0.06         &  2.80 $\pm$  0.06                 &  3.78  $\pm$ 0.14   &  2.92  $\pm$ 0.46             &  3.04  $\pm$ 0.06          \\
   $\rm L_{Bol,tot}$              & 3.82  $\pm$ 0.17    & 4.24  $\pm$ 0.26 & 4.16  $\pm$ 0.25  & 4.16  $\pm$ 0.34             &   4.00  $\pm$ 0.22  &  3.66  $\pm$ 0.06          &  2.97  $\pm$ 0.05             & 2.39 $\pm$ 0.06         &  3.82 $\pm$  0.06                 &  3.99  $\pm$ 0.14   &  3.72  $\pm$ 0.46             &  3.99  $\pm$ 0.06          \\
  $kT_{\rm bb}$                    & 0.13 $\pm$ 0.01    & 0.16  $\pm$ 0.01 & 0.13  $\pm$ 0.01  & 0.15 $\pm$ 0.02        &   0.13 $\pm$ 0.01  &  0.182 $\pm$ 0.004         &  0.175 $\pm$ 0.003            & 0.161 $\pm$ 0.004        &  0.161 $\pm$ 0.001                &  0.10  $\pm$ 0.01   &  0.16 $\pm$ 0.02             &  0.177 $\pm$ 0.002          \\
  $\rm kT_{seed}$ (coupled)        & 0.13               & 0.16             & 0.13              & 0.15               &   0.13             &  0.182                     &  0.175                        & 0.161                    &  0.161                            &  0.10               &  0.16                        &  0.177                       \\
 $\rm kT_{e}$                     & 0.88  $\pm$ 0.03    & 1.27  $\pm$ 0.08 & 0.83  $\pm$ 0.02  & 0.91 $\pm$ 0.05      &   0.97 $\pm$ 0.04  &  0.562 $\pm$ 0.005         &  0.505 $\pm$ 0.006            & 0.377 $\pm $ 0.004       &  0.480 $\pm$ 0.003                 &  0.73 $\pm$ 0.02    &  0.64 $\pm$ 0.03             &  0.58 $\pm$ 0.01            \\
  $\tau$                       & 16.4 $\pm$ 0.07    & 13.33 $\pm$ 0.76      & 16.54  $\pm$ 0.74  & 16.5  $\pm$ 1.5  &   15.09 $\pm$ 0.07  &  100   $\pm$ $_{48.6}^{0}$ &  100  $\pm$ $_{51.35}^{0}$    & 100 $\pm$ $_{55.11}^{0}$ &  100 $\pm$ $_{30.69}^{0}$           &  18.15 $\pm$ 0.06   &  23.56  $\pm$ $_{3.52}^{7.80}$  &  100   $\pm$  $_{55}^{0}$         \\
  $C_{\rm stat}/\rm d.o.f$             & 133/64              & 64/65         & 125/65             & 170/66                  &   104/65            &  148/64                    &  140/63                       & 159/57                   &  155/63                           &  146/65             &  122/64                       &  164/65                         \\
  \hline
               
        \label{table: Results of the continuum modeling for the 12 bursts. }
    \end{tabular}}


\tablefoot{The column density of the cold interstellar gas, $n_{\rm H}$, is fixed at $1.63 \times 10^{21}$\,cm$^{-2}$ for all bursts \citep{Costantini2012}. The luminosities $\rm L_{X}$ and $\rm L_{Bol}$  are calculated in the 0.3-10 keV and 0.1-20 keV band, respectively, and expressed in $10^{38}$ {\ergss} unit. The temperatures kT are expressed in keV unit. $\tau$ is the optical depth of the comptonisation component. For more information see the {\sc{spex}} manual.}
     \vspace{-0.3cm}
\end{table*}
\end{center}

\subsection{Grids of photo- / collisionally-ionised gas} \label{section:spex_grids}
The significance of detecting the plasma component can be enhanced by simultaneously considering multiple spectral lines. To achieve this, we conducted an extensive search across a multidimensional parameter space for plasma in photoionisation equilibrium (PIE) and collisional ionisation equilibrium (CIE). One major advantage of this method is that it helps avoid the problem of the fitting process becoming trapped in local minima, a common issue when fitting individual lines separately or starting from arbitrary initial parameters. Moreover, this approach facilitates the identification of multiple components or phases within the plasma, which is crucial for accurately determining its properties. While this technique can be computationally intensive, we have implemented several strategies to enhance its efficiency, such as varying plasma model parameters in incremental steps and leveraging pre-calculated model grids to reduce computational overhead.

{\subsubsection{Collisionally-ionised emitting gas model}} \label{Sect: cie_grids}

As previous done by \cite{Kosec_2018b} and \cite{Pinto_2020b}, we performed a multidimensional scan with an emission model of a plasma in collisional ionisation equilibrium (\textit{cie} model in {\sc{spex}}). CIE plasmas are often observed in stellar coronae and outflows of XRBs (see, e.g., \citealt{Marshall2002,Pinto2014,VDEijnden_2018}). We adopted a logarithmic grid of temperatures between 0.1 and 4 keV for a total of 50 bins, and line-of-sight velocities, $v_{\rm LOS}$, between -0.3c and +0.3c (from blueshifted to redshifted plasma, respectively, with a step of 5000 km/s, which is smaller than the spectral resolution in the soft X-ray band). The $v_{\rm LOS}$ is provided by the model \textit{red} with {\sc{FLAG=1}}, i.e. Doppler rather than Cosmological redshift, that multiplies the \textit{cie} component.
The turbulence of the plasma, $v_{\rm RMS}$, was fixed to 1000 km/s since in \cite{Strohmayer2019ApJ...878L..27S} was reported to be around 5000 km/s or lower and to avoid intrinsic merging of lines from different ions before convolving with the instrument response.
We adopted Solar abundances for the \textit{cie} to avoid unnecessary model degeneracy and to significantly speed the computing time. The emission measure defined as EM=$n_{e}$ $n_{\rm H}$V, was the only free parameter of the \textit{cie} model during the spectral fits. We applied this automated routine to scan the \textit{cie} model grids across the spectra of the 0.7s bursts. The best-fit results showed a significant improvement, particularly for burst 8, compared to the continuum-only model ($\Delta${\cstat}$_{\rm cie}$ = 58). The fit suggests a slight blueshift (\rm $v_{\rm LOS} \sim -0.05$c) and a plasma temperature of approximately 1 keV (see Fig. \ref{fig: CIE grids}).

\begin{figure*}[h!]
    \centering
    \includegraphics[width=0.45\textwidth]{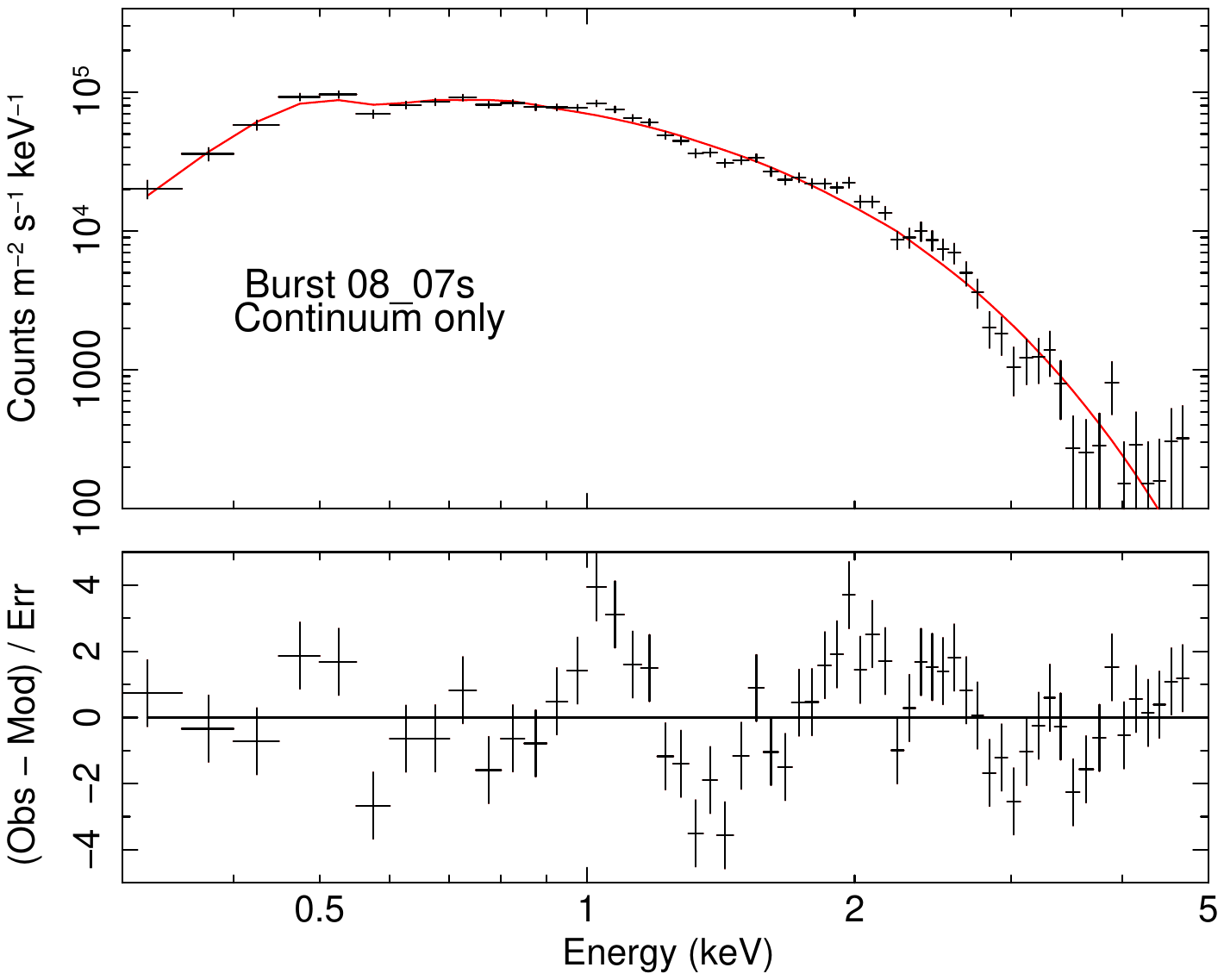}
    \includegraphics[width=0.45\textwidth]{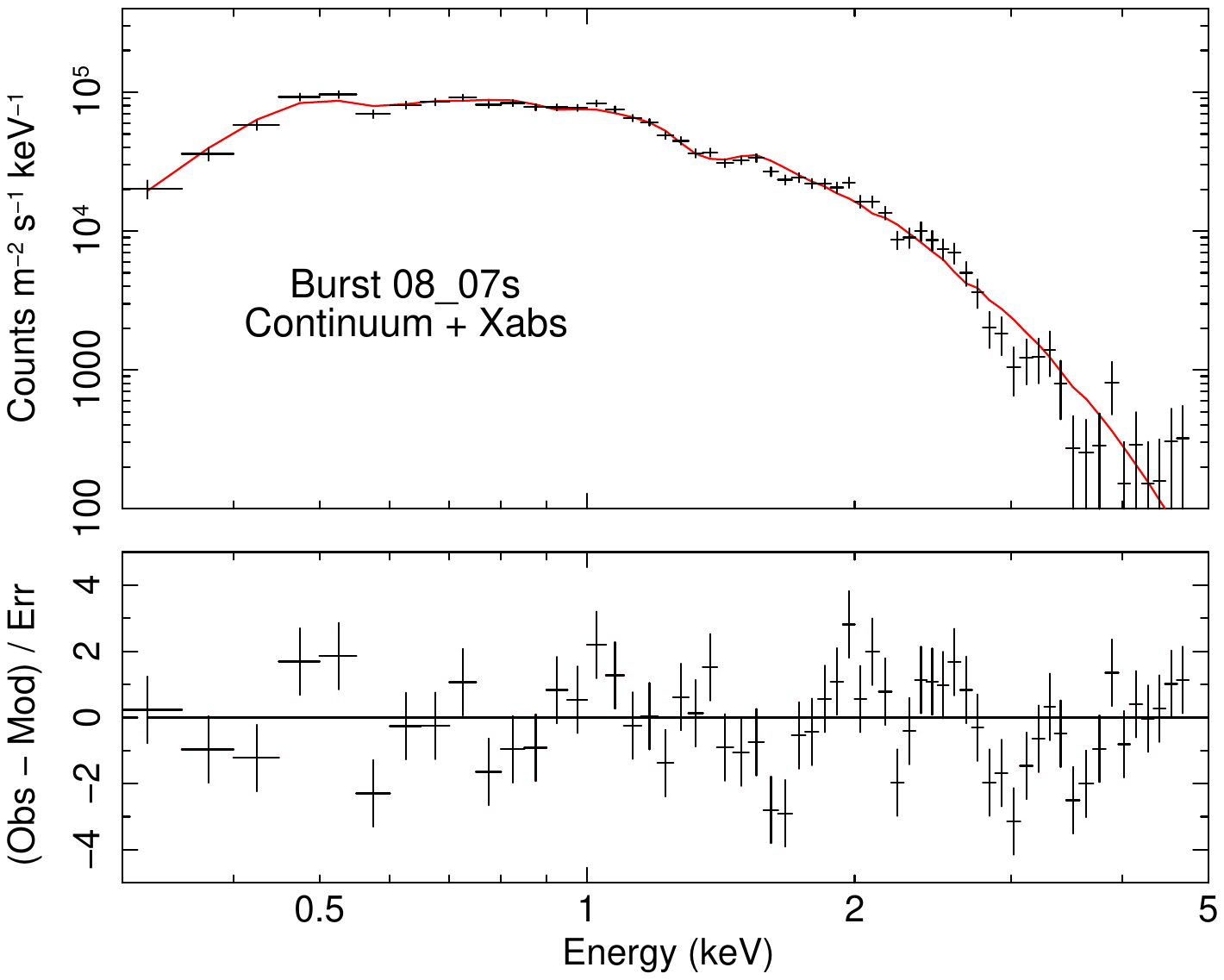}
    \includegraphics[width=0.45\textwidth]{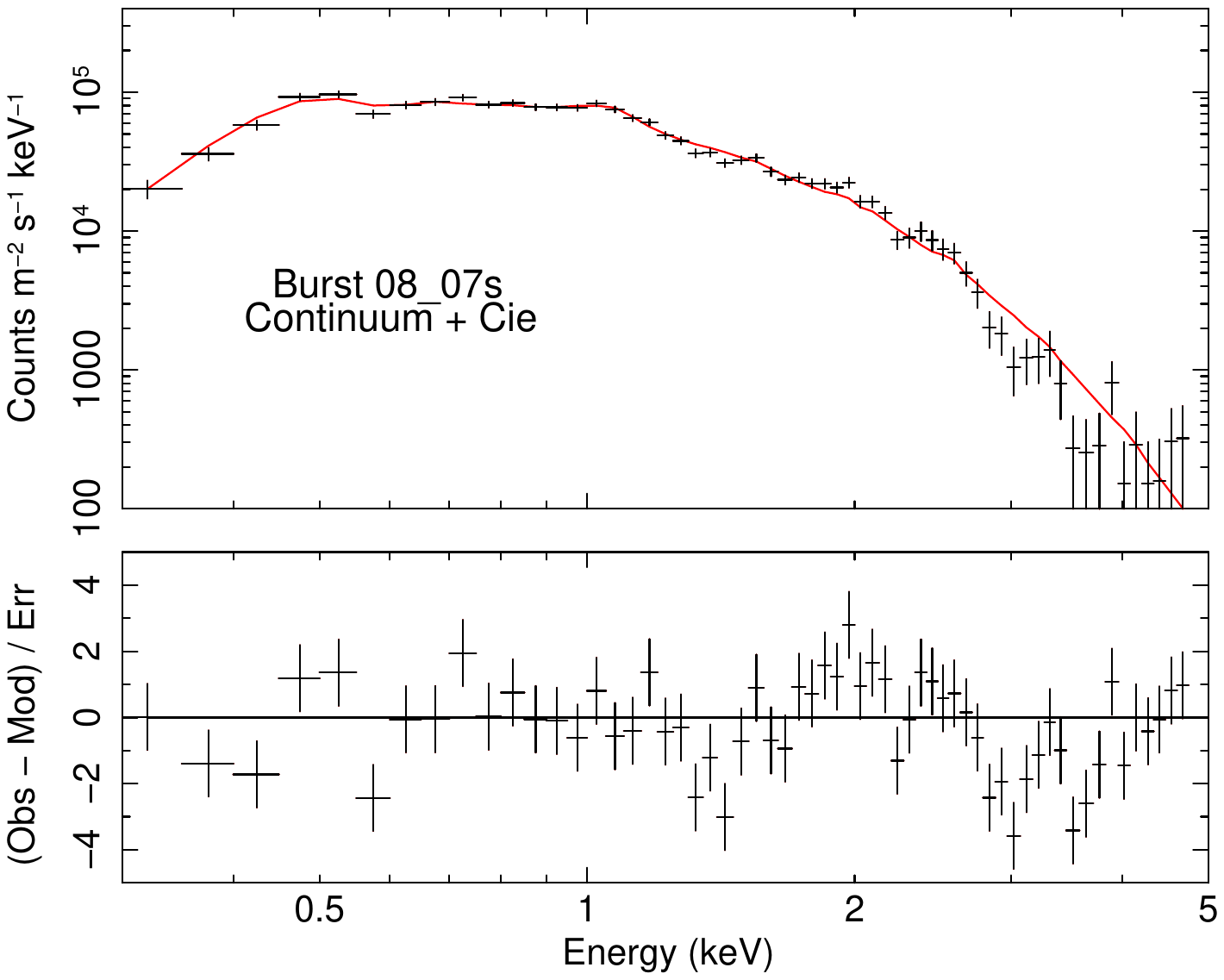}
    \includegraphics[width=0.45\textwidth]{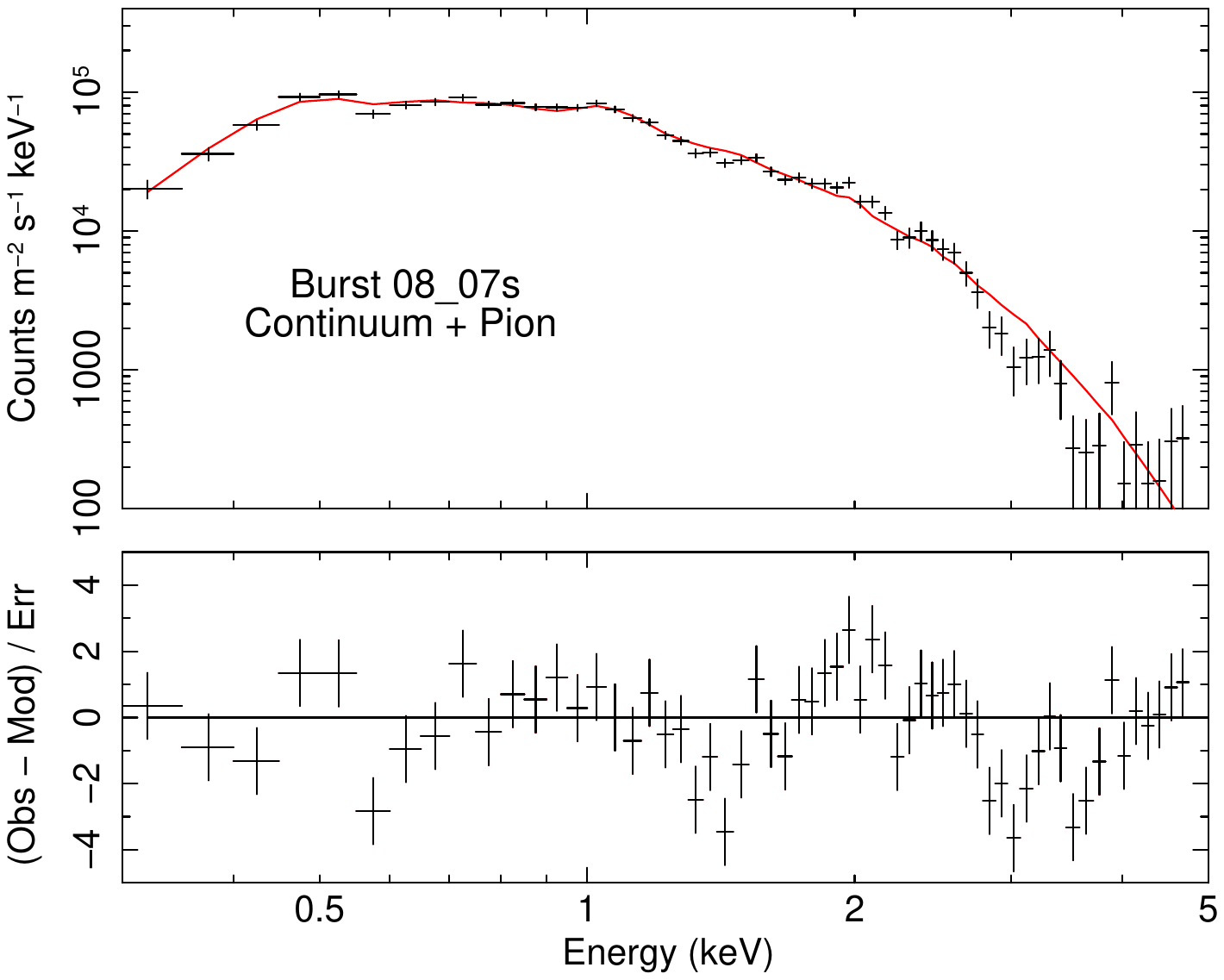}
      \caption{Best-fit spectral modelling of the burst 8 spectrum with different model adopted. From the upper left panel, the model adopted are : {\tt{bb + comt}}, {\tt{(bb + comt)* xabs}}, {\tt{bb + comt + cie}} and {\tt{bb + comt + pion}}. The best-fit results are reported in Table \ref{table: Results of the continuum modeling for the 12 bursts. } and \ref{table: Results of the lines modeling for the 12 bursts.}.}
    \label{fig:burst_08s_CIE_PION}
\end{figure*}

{\subsubsection{Photo-ionisation plasma in emission/absorption model}}
\label{Sect: pie_grids}

Emission and absorption lines can be generated by winds resulting from photospheric expansion during bursts. These features are modeled using detailed photo-ionisation calculations, similar to those applied to describe photospheric absorption lines in Novae during their supersoft phase (e.g., \citealt{Pinto2012, Ness2022}), as well as winds proposed to explain the 1 keV features observed in certain Galactic X-ray binaries (XRBs) and ultraluminous X-ray sources (ULXs) \citep{DelSanto2023, Barra2024}. Such models require an accurate understanding of the radiation field, specifically the spectral energy distribution (SED) spanning from optical to hard X-ray wavelengths.
\vspace{0.5cm}

\begin{figure*}
    \centering
     \includegraphics[width=0.5\textwidth]{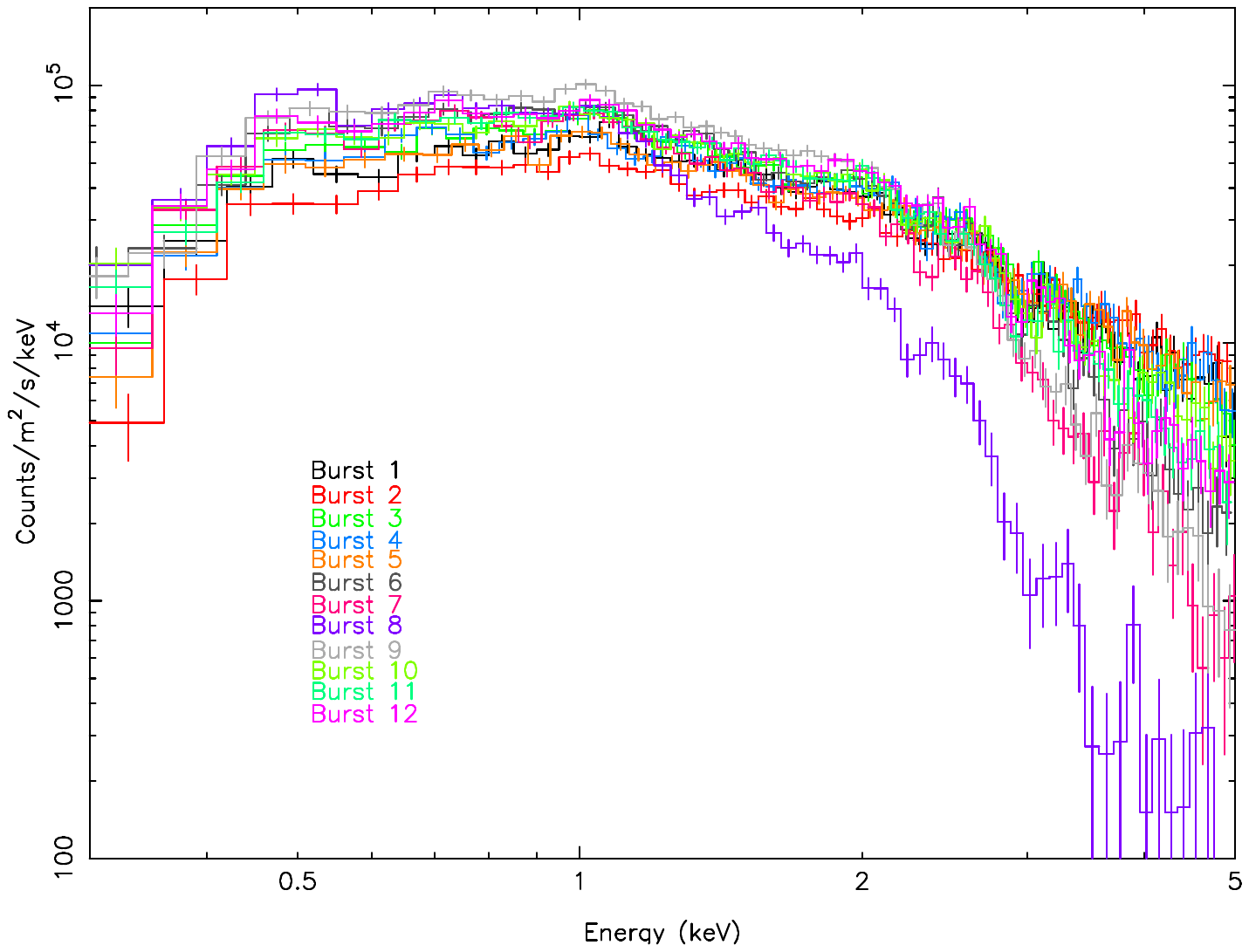}
    \includegraphics[width=0.48\textwidth]{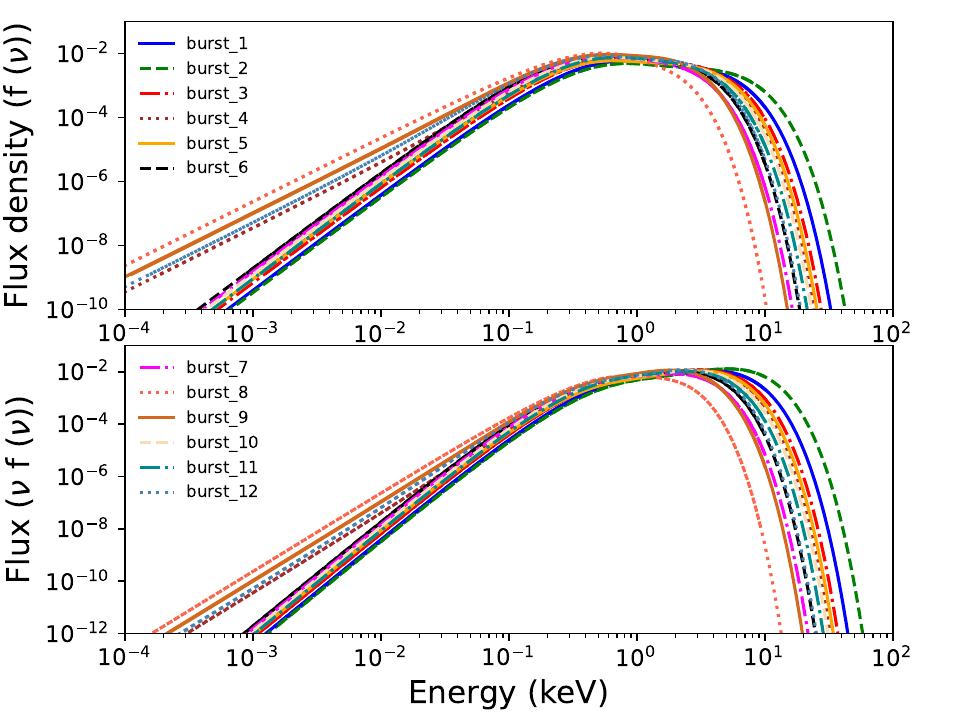}
    \caption{Left panel: NICER spectra of the 12 bursts (0.7s exposure time). Right panel: Spectral energy distribution (SED) of the 12 bursts from the optical to the hard X-rays energy band (0.0001-100 keV). }
        \label{fig:SED_12_bursts}
\end{figure*}

SED and photo-ionisation balance --
As previously done by \cite{Pinto_2020a,Pinto_2020b}, we constructed time-averaged spectra of the bursts from 4U 1820-303 using data from NICER. (see Fig.\,\ref{fig:SED_12_bursts}). The SED strongly peaks in the X-ray band, which is covered by NICER and restrains any systematic uncertainties in the ionisation balance computation due to the limited knowledge of the other energy bands. Therefore, for the X-ray band, i.e. 0.3-10 keV, we use the best fit continuum model \textit{bb+comt}, as reported in Table \ref{table: Results of the continuum modeling for the 12 bursts. }. A very important parameter of the photo-ionisation equilibrium is the ionisation parameter $\xi$ = $L_{\rm ion}$/($n_{\rm H} \ R^2$) (see \citealt{Tarter_1969}),
where $L_{\rm ion}$ is the ionising luminosity (estimated\ in the 13.6 eV - 13.6 keV range), $n_{\rm H}$ the hydrogen volume density and R the distance
from the ionising source. The ionisation balance was calculating with the {\sc{spex}} \textit{pion} model, which calculates the transmission and/or
the emission of a thin slab of photo-ionised gas, self-consistently. We also computed the stability (or
S) curve, which is the relationship between the temperature (or the ionisation parameter) and the ratio between the radiation and the thermal pressure, which can be expressed as $\Xi$ = F/($n_{\rm H}$ckT) = 19222 $\xi/T$ (\citealt{Krolik_1981}). The stability curves are shown in Fig. \ref{fig:burst_08s_pion s curve} for the 12 bursts.
The branches of the S curves with a negative gradient are characterised by thermally unstable gas and, therefore, the plasma is not expected to be found in their correspondence.

Photo-ionised plasma in emission --
We first attempt to describe the emission lines by adding the \textit{pion} model to the continuum model. In a similar way as done for the CIE grids, we created a logarithmic grid with the ionisation parameter log $\xi$ [{\ergss} cm] ranging between 0 and 5 with 0.2 steps and the $v_{\rm LOS}$ from the -0.3c (blueshift) to 0.3c (redshift) with a 5000 km/s step, taking free only the \textit{pion} column density $n_{\rm H}$. Therefore, we set the \textit{pion} covering fraction $f_{cov}$ to zero, such that \textit{pion} model produces exclusively emission line, and $\Omega$, the solid angle, equals to 4$\pi$ (1 in {\sc{spex}} units). This allows to save substantial computational time. For the burst 8 (see Fig. \ref{fig: PION grids}), this model produces a similar improvement to that obtained with the \textit{cie} component ($\Delta${\cstat}$_{\rm pion}$ = 55), corresponding to a plasma redshifted solution with log $\xi \sim$ 3 and $v_{\rm LOS}\sim$ 0.02c. However, the  solutions obtained with the two different emission models are compatible with a rest frame solution within a few sigma.

Photo-ionised plasma in absorption --
To conclude, we also conducted a multidimensional grid scan using the absorption model. While the \textit{pion} model is versatile, capable of modelling both emission and absorption, it recalculates the ionisation balance at each iteration, making it computationally demanding. For the absorbing gas, we opted for the more efficient \textit{xabs} model, which is specifically optimised for absorption. This model leverages a pre-calculated ionisation balance from \textit{pion}, significantly reducing the computational load.
The \textit{xabs} model shares several parameters with \textit{pion} except for the opening angle of the line emission which is set to zero since no emission is present in this model. We adopted a covering fraction equal to unity in order to avoid degeneracy and reduce even further the computing time. We calculated the grid of photo-ionised \textit{xabs} models in the same way as the \textit{pion} models, but assuming line-of-sight velocities, $v_{\rm LOS}$, ranging between -0.3c and zero (i.e. only outflowing gas in absorption or a wind). The best-fit result corresponds to $\Delta${\cstat} $\sim$ 48 achieved for the burst 8 (reported in Fig. \ref{fig: XABS grids}) with a log $ \xi \sim$ 2.5 and a velocity $v_{\rm LOS}$ between - 0.2c and - 0.3c, thereby indicating a blueshifted plasma.

\section{Discussion}
\label{Grids multidimensional scan}

This paper focuses on the detailed analysis of the bursts in the low-mass X-ray binary 4U 1820-303. Initially, we examine the bursts observed between 2017 and 2021, with particular attention to the occurrence of photospheric radius expansion (PRE). Subsequently, we characterize the photospheric expansion and the evolution of spectral lines by employing two parallel continuum descriptions using {\sc{xspec}} and {\sc{spex}} which provided consistent results. Finally, we conduct a thorough analysis of the spectral lines through multidimensional scan grids, using advanced plasma codes in {\sc{spex}}, and compare these results with the radii of the PRE bursts.

\subsection{Broadband spectral properties}
4U 1820-303 exhibits bursts with varying levels of photospheric radius expansion (PRE). We conducted an in-depth X-ray spectral analysis of the 0.7-second spectra following the burst peak, focusing on the period when the PRE is at its maximum. Specifically, we performed a dual analysis using two software tools, {\sc{xspec}} and {\sc{spex}}. The first one was employed for both spectral feature detection and general spectral modelling, while {\sc{spex}} was used mostly for modelling the spectral features with the state of art optically-thin plasma models. Among the various model tested, that took into account a fixed column density of the cold interstellar gas at $1.63 \times 10^{21} \rm cm^{-2}$ \citep{Costantini2012}, our results show that the spectra are reasonable described with a double component model composed of a blackbody (\textit{bb} in {\sc{spex}}) and a comptonisation (\textit{comt} in {\sc{spex}}). These take into account the burst and the persistent emission coming from the source. The temperature of the blackbody is 0.1-0.2 keV and was coupled with the seed photons of the \textit{comt} component. 
The optical depth seems unconstrained ranging between 13, for the burst 2 where there is not a photosperic expansion, and 100 probably due to the fact that the region of photosphere that we are looking at is optically thick and an electronic temperature  between 0.38 to 1.27. The notable \cstat/d.o.f is primarily due to the presence of emission residuals at 1 keV (Ne X, Fe XXI and Fe XXII), 2 keV (Si XIV), 2.6 keV (S XVI), and between 3-4 keV (Ar XVII) in absorption, which will be analysed in the following section. These results are in agreement with those obtained with {\sc{xspec}} and with the Gaussian line scan reported in Fig. \ref{fig:gaus_scan} . These residuals, which are absent in the persistent spectra or very weak PRE bursts (e.g. burst 2), suggest an astrophysical origin related to the burst phenomena that will be discussed in the following section.

\subsection{Bursts properties}
The results of this analysis are consistent with those reported by \citet{Yu_2024} regarding spectral properties of the bursts, including blackbody radii. Therefore, we do not discuss these results in further detail here; however, they are included in our description of the component evolution that produces the feature lines (see Sect. \ref{sect: Lines vs Photospheric radius expansion}).
In order to describe the significance of the detection of the plasma component, after the continuum model fitting, we proceed to perform a deep search throughout a multimensional scan of plasma in photo/collisional ionisation equilibrium (\textit{pion}/\textit{xabs} models for the photo-ionised plasma in emission/absorption and \textit{cie} model, respectively).  The results for the all burst modelling are shown in Table \ref{table: Results of the lines modeling for the 12 bursts.} and plotted in Fig. \ref{fig:tables cie-pion-xabs}. The scan grids for each model and burst, with the corresponding results discussed below, are provided in the Appendix. 

CIE spectral modelling: 
The temperature of the plasma component ranges from 1.2 to 1.7 keV, except for bursts 10 and 11, which have a characteristic temperature of approximately 4.4 keV (\( \rm kT_{\textit{CIE}} \)). However, the improvement in their fit to the continuum modelling is equal to or less than that of the other bursts.
The velocity of the emitting plasma is almost rest-frame within the uncertainties in all the cases. The total X-ray luminosity (0.3 - 10 keV) associated to the source is ranging between ($2-5) \times 10^{38} \rm erg/s$, slightly greater than the Eddington limit forecasted for a neutron star of 1.6 $M_\odot$ of 12 km radius ($\sim 3.1\times 10^{38} \rm erg/s$, by taking into account that the companion star is an helium white dwarf). \\

PION spectral modelling: For the \textit{pion} model, we obtained from the bursts modelling a plasma moving at low-velocities with a temperature of the order of 1 keV, compatible with the complex of Ne X, Fe XXI and Fe XXII. A deviation from the average values for the column density, the velocity along the LOS and the ionisation parameter is present for the burst 2 where is no a photospheric radius expansion although these values are compatible within the uncertainties with the results of others bursts modelling. The luminosites of the source is ranging in the range (2 - 5)  $\times 10^{38}$ erg/s such as the previous case.   \\

XABS spectral modelling: Using the \textit{xabs} model, the improvement in the fit of the burst spectra is approximately $\Delta$\cstat $\sim$ 20 compared to using only the continuum model, with the exception of bursts 6 and 8, where the improvements are $\Delta$\cstat $\sim$ 42 and 48, respectively. In these cases, the radius of the blackbody is high ($\sim$ 122 km and $\sim$ 238 km, respectively).
For the burst 6, a plasma with a log $\xi \sim 3$ moving towards the observer at $\sim$ 0.2 - 0.3 c is suggested, in agreement with the results obtained from the other bursts. The luminosites of the source during the bursts is of the order of $(2-6) \times 10^{38} \rm erg/s$ although for the bursts 1, 3 and 10 a value of $10^{39-40} \rm erg/s$ is reached due to a high value of the hydrogen column density in the spectral modelling most likely affected by systematics due to the weakness of the features.  \\

\begin{figure}[h!]
    \centering
     \includegraphics[width=0.5\textwidth]{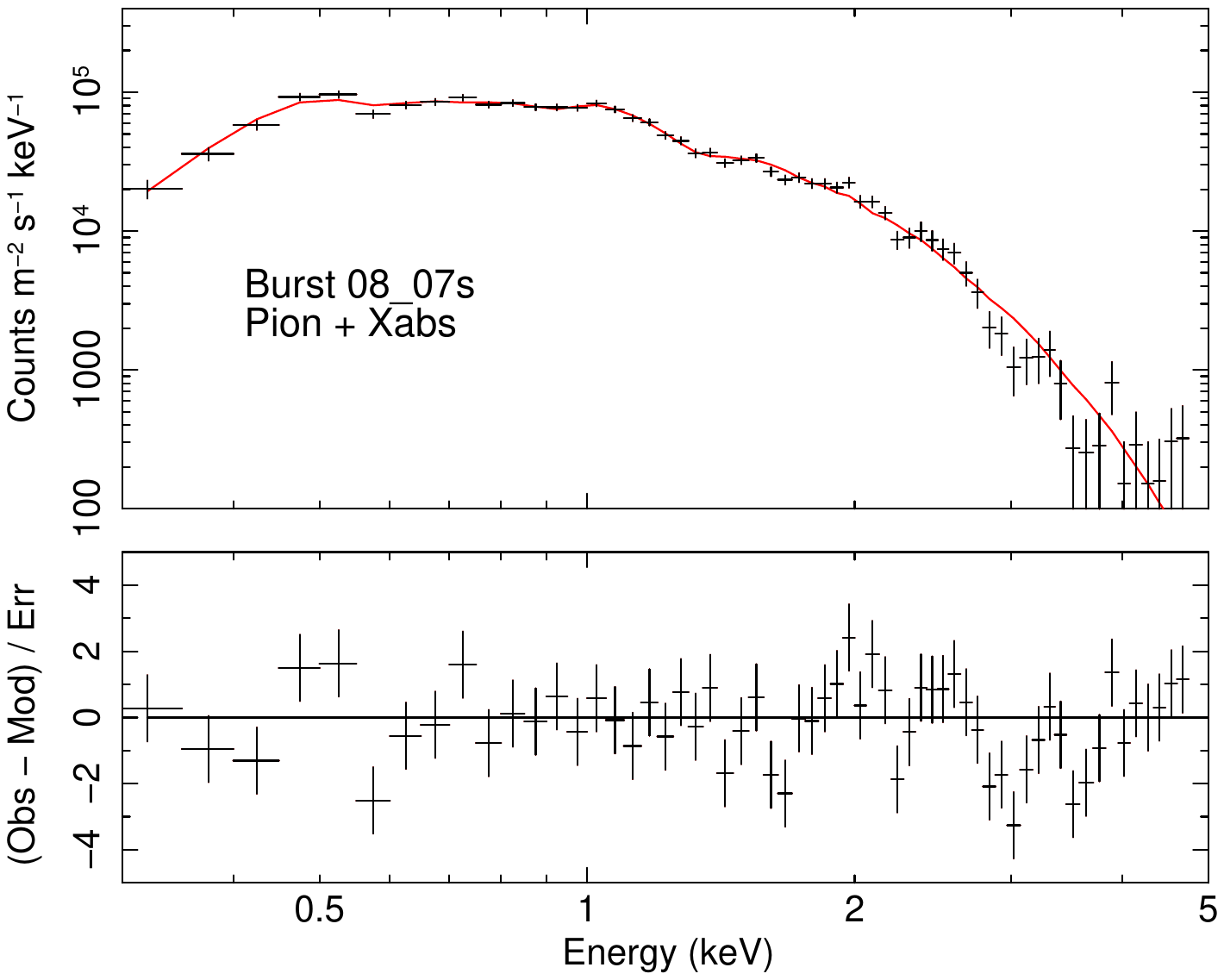}
      \caption{Spectral modelling of the burst 8 spectrum with the {\tt{hot * (pion + xabs * (bb + comt))}} model adopted. The fit results are reported in the main text. }
    \label{fig:burst_08s_PION_XABS}
\end{figure}

\begin{figure*}[h!]
    \centering
    \includegraphics[width=0.45\textwidth]{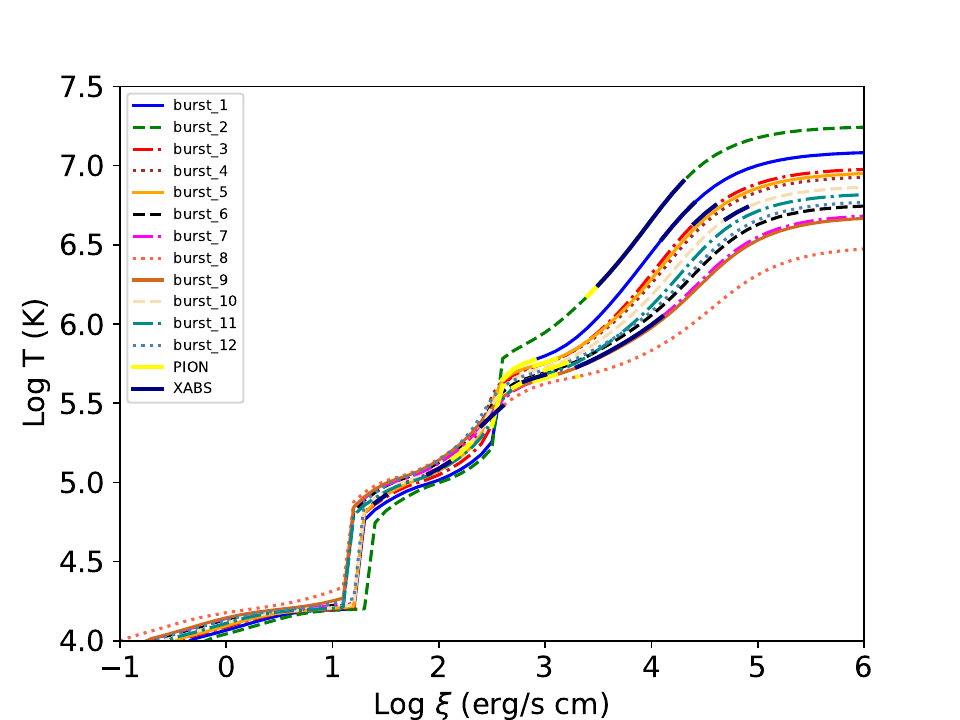}
    \includegraphics[width=0.45\textwidth]{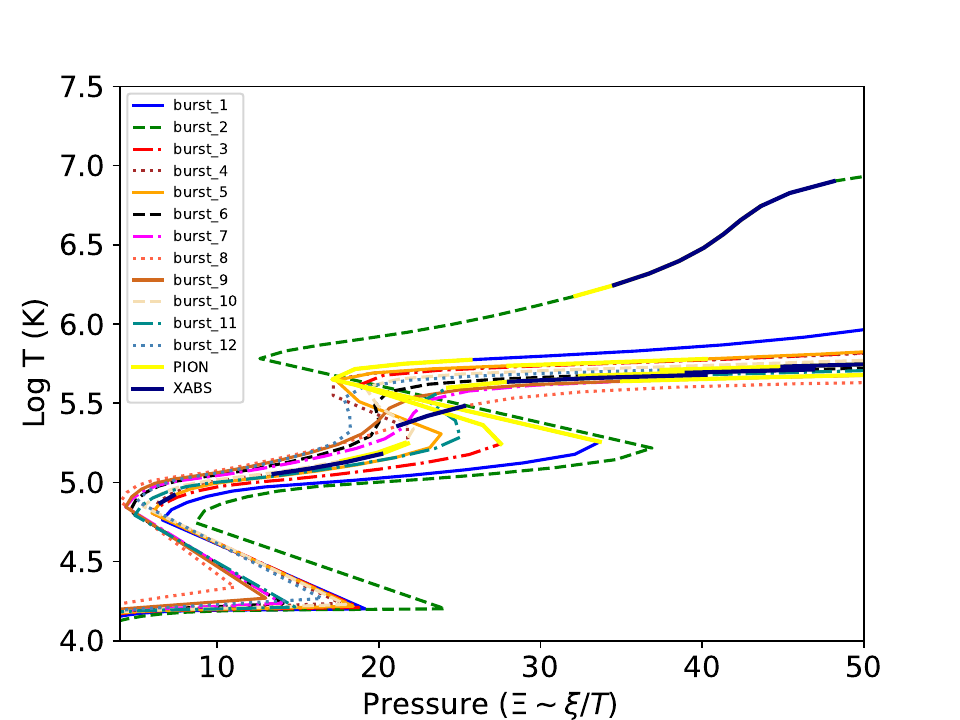}
    \caption{Ionisation balance (left) and thermal-stability curves (right) computed for each burst. The regions in which there are thermal instabilities are identified by the segments with negative slopes (right panel). Thicker segments show the ranges of the best-fitting solutions.}
        \label{fig:burst_08s_pion s curve}
\end{figure*}

From the spectral modelling of the emission lines with both \textit{cie/pion} model, a rest frame plasma solution, within the uncertainties, is forecasted. For the absorption lines, instead, by modelling with the \textit{xabs} component, the results suggested a blueshifted plasma moving along the LOS with velocities ranging between $\sim (0.2-0.35) c$ with the highest $\Delta${\cstat} variation for the burst 6 and 8 (that present the highest photospheric radius expansion with respect to the other bursts). For the latter, which presents more statistics, we also tested a \textit{pion + xabs} component applied to the continuum model resulting in a better quality fit ({\cstat}/d.o.f $\sim$ 83/52) with the respect of the all models previous one tested ($\Delta${\cstat} $\geq$ 20), suggesting the presence of both emitting/absorbing gas with parameters compatible with those obtained in Table \ref{table: Results of the lines modeling for the 12 bursts.} (emitting gas:  $\rm n_{ \rm H,PION} \sim$ 0.05 $\pm$ 0.01 [$10^{24}$/$\rm cm^{2}$],     $\rm v_{ \rm PION} \sim$ 5706 $\pm$ 850 [km/s], log $\xi_{ \rm PION}\sim $  3.39 $\pm$ $_{0.21}^{0.04}$  [erg/s cm]; absorbing gas: $\rm n_{ \rm H,XABS}=$ 0.014 $\pm$ 0.002 [$10^{24}$/$\rm cm^{2}$],$\rm v_{ \rm XABS}\sim$ - 99500 $\pm$ 700 [km/s], log $\xi_{\rm XABS} =$ 2.60 $\pm$ 0.01 [erg/s cm]). The best-fit parameters for the plasma are located within the stability region, as shown in Fig. \ref{fig:burst_08s_pion s curve} (right panel), except for bursts 1 and 3, where the plasma solutions lie in the instability region (characterised by a negative slope). The results obtained with absorption model, in particular for the velocity of the wind, does not agree with the results obtained by \cite{Yu2018ApJ...863...53Y} and \cite{Guichandut2021ApJ...914...49G}, suggesting a wind velocity  always < 0.1c instead of the 0.2 - 0.35c that was estimated from our analysis. This may be attributed to the key limitation that in the latter analysis are strictly within the context of light-element models, meaning a fully ionised, electron-scattering continuum. If a significant amount of metals is present, line-driving could become important and substantially accelerate the gas to higher velocities such as would suggest our analysis. A challenge in achieving more precise line identifications, beyond the need for higher spectral resolution, is the potential presence of both redshifts (in emission, although close to rest in this analysis) and blueshifts (in absorption). These shifts may vary in their contributions across different bursts, as suggested by the results shown in Fig.  \ref{fig:burst_08s_PION_XABS}. \\

Constraints on the line broadening: 
We performed two different tests to constrain the line broadening, using the features at 1 keV (in emission) and 3 keV (in absorption) as probes. In the first test, we performed two simulations of the burst 8 spectrum, both with the same exposure time and based on the best-fit model (continuum + \textit{cie}, see Fig. \ref{fig:burst_08s_CIE_PION}). The simulations assumed broadening velocities of 1000 km/s and 10000 km/s, respectively. From the fits of these simulated spectra, we derived an upper limit for the broadening velocity of 5000 km/s in the 1000 km/s simulation case (as found in the original burst 8 dataset), consistent with the spectral resolution. Similar statistical uncertainties are obtained in the 10000 km/s simulation case. These results are compatible within 1–2$\sigma$. Additionally, we analysed the absorption feature located at 3 keV (where the resolving power is 2-3 times higher than the one at 1 keV and the spectrum is less crowded with lines) by fitting it with a Gaussian component in bursts where the line is stronger (e.g., bursts 4, 8, and 11). The derived broadening velocities span the range 3000-40000 km/s due to their large uncertainties caused by the lower statistics and the still limited resolution at 3 keV. Although strong constraints cannot be placed, it appears that the features may be intrinsically broadened.

\subsection{Lines vs Photospheric radius expansion}
\label{sect: Lines vs Photospheric radius expansion}
From the bursts modelling with the continuum model and the  photo-collisional ionisation models, it is possible to see that the emission/absorption lines are more prominent in those burst that present a larger photospheric radius expansion (i.e. the burst 8, for the residuals at 1 keV). For the burst 2, instead, the residuals are flat and this burst is the which one where there is no photospheric radius expansion (see the $\Delta${\cstat} improvement in the Appendix). Moreover, it seems to be a correlation between the intensity of the residuals at 2.2 keV in emission (Si XIV) with those in absorption at 3 keV (Ar XVII).
These results are in agreement with those obtained by \cite{Weinberg_2006}, and \cite{Strohmayer2019ApJ...878L..27S} for the first 5 burst sorted by ID on Table \ref{tab:0.7seconds}. Simulations indicate that bursts igniting deep within the neutron star envelope have access to more fuel, leading to several significant outcomes (\citealt{Yu2018ApJ...863...53Y}). Firstly, there is an enhance on the metal synthesis and a more intense convection that resulting in a more easily transportation of the nuclear ashes (mainly heavy elements) closer to the photosphere, unveiling their distinct spectral signature  (\cite{zand2010A&A...520A..81I} and references therein). During these powerful burst, there is an increased luminosity at the base, driving stronger winds and resulting in higher photospheric radii. 
In a following paper will examine the correlation between the rapid detection of metals and the intensity of the burst, with a specific focus on the burst's initial rise phase (following the work done by \citealt{Yu2018ApJ...863...53Y,Guichandut2021ApJ...914...49G}).

\subsection{Bursts vs wind properties in XRBs and ULXs}
A useful comparison can be done with the results obtained from photoionisation modelling applied in two different systems such as MAXI J1810-222 \citep{DelSanto2023} and Holmberg II X-1 \citep{Barra2024}. In the first case, the strong feature around 1 keV was interpreted as a blend of blueshift Fe L, Ne X, or O VIII absorption lines. This fact was confirmed by the ionic column densities obtained by the \textit{xabs} component for a log $\xi$ $\sim$ 2. The plasma state solution found varies with the source spectral state. In fact, for the high-flux state, \citet{DelSanto2023} found a highly significant solution of a hot plasma outflowing at > 0.1c, that disagrees with the classical Galactic X-ray binaries (XRB) thermal winds (with velocities of the order of 1000 km/s) and invoking a strong radiation pressure as for winds found in Ultraluminous X-ray sources (ULXs, \cite{pinto2016Natur.533...64P}), although a minimum distance of 100 kpc (outside the Galaxy) is required to identified a source as an Ultraluminous X-ray source. During the soft and intermediate state, the velocity of the plasma are still high for a thermal wind. In the end, for the hard state, the outflow is weaker, due to a low column density, and with wind velocity comparable with those in XRBs. This variability in the properties of the winds with the spectral states might be due to a thermal instability in the plasma or to a different configuration of magnetic field and launching radius. A similar approach is used for the modelling of the features in Holmberg II X-1 \citep{Barra2024} where the features, in emission at 0.5 keV (N VII) and 0.9 keV (Ne IX) and those in absorption between 0.6 and 0.8 keV and above 1 keV were modelled with photoionisation models (\textit{pion} and \textit{xabs} for the emission and absorption lines, respectively). The results of this analysis suggest a multi-phase plasma: the strong N VII emission line (log $ \xi \sim $ 1, v $\sim$ 0.006c) appears to be distinct from the two hotter absorption and emitting components (log $\xi \sim$ 3, v$\sim$ -0.2c). Similar velocity for the absorption lines is required to explain the absorption features in Swift J0243.6+6124 (\cite{VDE_2019}. However, an alternative scenario suggests these features might be linked to iron and magnesium lines, albeit with several caveats. \citet{Paczynski_1986} proposed a relativistic model of the wind driven by X-ray bursts, concluding that all the super-Eddington energy flux gently blows out matter, with the photospheric radius of the outflowing envelope always exceeding 100 km. \cite{Yu2018ApJ...863...53Y} supported this with a hydrodynamic simulation of spherically symmetric super-Eddington winds, demonstrating that the photosphere extends beyond 100 km within the first few seconds, regardless of burst duration and ignition depth, as shown in Fig. \ref{fig:burst_parameters}. 

\section{Conclusion}
In this study, we conducted a comprehensive X-ray spectral analysis of bursts from the low-mass X-ray binary 4U 1820-303 between 2017 and 2021. A model consisting of a blackbody and comptonisation provided a reasonably good fit for the broadband 0.3-10 keV spectrum, although several features, both in emission and absorption, were evident in the residuals. Specifically, emission lines were detected at 1 keV, 2 keV, and 2.4 keV, while absorption lines appeared at 3 keV and 3.4 keV. These features were interpreted using grids of photo-collisionally ionised gas, suggesting a redshifted (near rest-frame) emitting gas in most cases. Notably, the strength of these lines, particularly the 1 keV feature, increased with the blackbody radius, indicating that they are linked to astrophysical processes associated with burst phenomena. However, the need for higher spectral resolution leaves open the possibility that both rest-frame emission and blueshifted absorption (with velocities around 20-30\% the speed of light) could be contributing. Future observations will be crucial in more precisely identifying these lines and the elements involved in these energetic processes with high-resolution (resolving power $\gtrsim$ 1000) detectors such as those onboard XMM-Newton, Chandra and XRISM.

\begin{acknowledgements}
    This work is based on observations obtained with NICER, a NASA science mission funded by the USA. CP acknowledges support for PRIN MUR 2022 SEAWIND 2022Y2T94C funded by European Union - NextGenerationEU and INAF LG 2023 BLOSSOM. T.D.S. acknowledges support from PRIN-INAF 2019 with the project “Probing the geometry of accretion: from theory to observations” (PI: Belloni).
\end{acknowledgements}

\section*{Data Availability}

All the data and software used in this work are publicly available from NASA's HEASARC archive\footnote{https://heasarc.gsfc.nasa.gov/}. Our spectral codes and automated scanning routines are publicly available and can be found on the GitHub\footnote{https://github.com/ciropinto1982}.

\bibliographystyle{aa} 
\bibliography{aanda.bib} 

\begin{thebibliography}{59}
\expandafter\ifx\csname natexlab\endcsname\relax\def\natexlab#1{#1}\fi

\bibitem[{{Arnaud}(1996)}]{Arnaud1996ASPC..101...17A}
{Arnaud}, K.~A. 1996, in Astronomical Society of the Pacific Conference Series,
  Vol. 101, Astronomical Data Analysis Software and Systems V, ed. G.~H.
  {Jacoby} \& J.~{Barnes}, 17

\bibitem[{{Ballantyne} \& {Strohmayer}(2004)}]{Ballantyne2004ApJ...602L.105B}
{Ballantyne}, D.~R. \& {Strohmayer}, T.~E. 2004, \apjl, 602, L105

\bibitem[{{Barra} {et~al.}(2024){Barra}, {Pinto}, {Middleton}, {Di Salvo},
  {Walton}, {G{\'u}rpide}, \& {Roberts}}]{Barra2024}
{Barra}, F., {Pinto}, C., {Middleton}, M., {et~al.} 2024, \aap, 682, A94

\bibitem[{{Baumgardt} \& {Vasiliev}(2021)}]{Baumgardt2021MNRAS.505.5957B}
{Baumgardt}, H. \& {Vasiliev}, E. 2021, \mnras, 505, 5957

\bibitem[{{Bloser} {et~al.}(2000){Bloser}, {Grindlay}, {Kaaret}, {Zhang},
  {Smale}, \& {Barret}}]{bloser2000ApJ...542.1000B}
{Bloser}, P.~F., {Grindlay}, J.~E., {Kaaret}, P., {et~al.} 2000, \apj, 542,
  1000

\bibitem[{{Boutloukos} {et~al.}(2010){Boutloukos}, {Miller}, \&
  {Lamb}}]{Boutloukos2010ApJ...720L..15B}
{Boutloukos}, S., {Miller}, M.~C., \& {Lamb}, F.~K. 2010, \apjl, 720, L15

\bibitem[{{Cash}(1979)}]{Cash1979ApJ...228..939C}
{Cash}, W. 1979, \apj, 228, 939

\bibitem[{{Costantini} {et~al.}(2012){Costantini}, {Pinto}, {Kaastra}, {in't
  Zand}, {Freyberg}, {Kuiper}, {M{\'e}ndez}, {de Vries}, \&
  {Waters}}]{Costantini2012}
{Costantini}, E., {Pinto}, C., {Kaastra}, J.~S., {et~al.} 2012, \aap, 539, A32

\bibitem[{{Cumming}(2003)}]{Cumming2003ApJ...595.1077C}
{Cumming}, A. 2003, \apj, 595, 1077

\bibitem[{{Degenaar} {et~al.}(2018){Degenaar}, {Ballantyne}, {Belloni},
  {Chakraborty}, {Chen}, {Ji}, {Kretschmar}, {Kuulkers}, {Li}, {Maccarone},
  {Malzac}, {Zhang}, \& {Zhang}}]{Degenaar2018SSRv..214...15D}
{Degenaar}, N., {Ballantyne}, D.~R., {Belloni}, T., {et~al.} 2018, \ssr, 214,
  15

\bibitem[{{Del Santo} {et~al.}(2023){Del Santo}, {Pinto}, {Marino},
  {D'A{\`\i}}, {Petrucci}, {Malzac}, {Ferreira}, {Pintore}, {Motta}, {Russell},
  {Segreto}, \& {Sanna}}]{DelSanto2023}
{Del Santo}, M., {Pinto}, C., {Marino}, A., {et~al.} 2023, \mnras, 523, L15

\bibitem[{{Galloway} {et~al.}(2017){Galloway}, {Goodwin}, \&
  {Keek}}]{Galloway2017PASA...34...19G}
{Galloway}, D.~K., {Goodwin}, A.~J., \& {Keek}, L. 2017, \pasa, 34, e019

\bibitem[{{Galloway} {et~al.}(2020){Galloway}, {in't Zand}, {Chenevez},
  {W{\"o}rpel}, {Keek}, {Ootes}, {Watts}, {Gisler}, {Sanchez-Fernandez}, \&
  {Kuulkers}}]{Galloway2020ApJS..249...32G}
{Galloway}, D.~K., {in't Zand}, J., {Chenevez}, J., {et~al.} 2020, \apjs, 249,
  32

\bibitem[{{Galloway} \& {Keek}(2021)}]{Galloway2021ASSL..461..209G}
{Galloway}, D.~K. \& {Keek}, L. 2021, in Astrophysics and Space Science
  Library, Vol. 461, Astrophysics and Space Science Library, ed. T.~M.
  {Belloni}, M.~{M{\'e}ndez}, \& C.~{Zhang}, 209--262

\bibitem[{{Gendreau} {et~al.}(2012){Gendreau}, {Arzoumanian}, \&
  {Okajima}}]{Gendreau2012SPIE.8443E..13G}
{Gendreau}, K.~C., {Arzoumanian}, Z., \& {Okajima}, T. 2012, in Society of
  Photo-Optical Instrumentation Engineers (SPIE) Conference Series, Vol. 8443,
  Space Telescopes and Instrumentation 2012: Ultraviolet to Gamma Ray, ed.
  T.~{Takahashi}, S.~S. {Murray}, \& J.-W.~A. {den Herder}, 844313

\bibitem[{{Grindlay} {et~al.}(1976){Grindlay}, {Gursky}, {Schnopper},
  {Parsignault}, {Heise}, {Brinkman}, \&
  {Schrijver}}]{Grindlay1976ApJ...205L.127G}
{Grindlay}, J., {Gursky}, H., {Schnopper}, H., {et~al.} 1976, \apjl, 205, L127

\bibitem[{{Guichandut} {et~al.}(2021){Guichandut}, {Cumming}, {Falanga}, {Li},
  \& {Zamfir}}]{Guichandut2021ApJ...914...49G}
{Guichandut}, S., {Cumming}, A., {Falanga}, M., {Li}, Z., \& {Zamfir}, M. 2021,
  \apj, 914, 49

\bibitem[{{G{\"u}ver} {et~al.}(2010){G{\"u}ver}, {Wroblewski}, {Camarota}, \&
  {{\"O}zel}}]{Guver2010ApJ...719.1807G}
{G{\"u}ver}, T., {Wroblewski}, P., {Camarota}, L., \& {{\"O}zel}, F. 2010,
  \apj, 719, 1807

\bibitem[{{Haberl} {et~al.}(1987){Haberl}, {Stella}, {White}, {Priedhorsky}, \&
  {Gottwald}}]{Haberl1987ApJ...314..266H}
{Haberl}, F., {Stella}, L., {White}, N.~E., {Priedhorsky}, W.~C., \&
  {Gottwald}, M. 1987, \apj, 314, 266

\bibitem[{{Hurkett} {et~al.}(2008){Hurkett}, {Vaughan}, {Osborne}, {O'Brien},
  {Page}, {Beardmore}, {Godet}, {Burrows}, {Capalbi}, {Evans}, {Gehrels},
  {Goad}, {Hill}, {Kennea}, {Mineo}, {Perri}, \&
  {Starling}}]{Hurkett2008ApJ...679..587H}
{Hurkett}, C.~P., {Vaughan}, S., {Osborne}, J.~P., {et~al.} 2008, \apj, 679,
  587

\bibitem[{{in't Zand} {et~al.}(2012){in't Zand}, {Homan}, {Keek}, \&
  {Palmer}}]{Zand2012A&A...547A..47I}
{in't Zand}, J.~J.~M., {Homan}, J., {Keek}, L., \& {Palmer}, D.~M. 2012, \aap,
  547, A47

\bibitem[{{in't Zand} \& {Weinberg}(2010)}]{zand2010A&A...520A..81I}
{in't Zand}, J.~J.~M. \& {Weinberg}, N.~N. 2010, \aap, 520, A81

\bibitem[{{Jahoda} {et~al.}(1996){Jahoda}, {Swank}, {Giles}, {Stark},
  {Strohmayer}, {Zhang}, \& {Morgan}}]{Jahoda1996SPIE.2808...59J}
{Jahoda}, K., {Swank}, J.~H., {Giles}, A.~B., {et~al.} 1996, in Society of
  Photo-Optical Instrumentation Engineers (SPIE) Conference Series, Vol. 2808,
  EUV, X-Ray, and Gamma-Ray Instrumentation for Astronomy VII, ed. O.~H.
  {Siegmund} \& M.~A. {Gummin}, 59--70

\bibitem[{{Kaastra}(2017)}]{Kaastra2017A&A...605A..51K}
{Kaastra}, J.~S. 2017, \aap, 605, A51

\bibitem[{{Kaastra} \& {Bleeker}(2016)}]{Kaastra2016}
{Kaastra}, J.~S. \& {Bleeker}, J.~A.~M. 2016, \aap, 587, A151

\bibitem[{{Kaastra} {et~al.}(1996){Kaastra}, {Mewe}, \&
  {Nieuwenhuijzen}}]{Kaastra_1996}
{Kaastra}, J.~S., {Mewe}, R., \& {Nieuwenhuijzen}, H. 1996, in UV and X-ray
  Spectroscopy of Astrophysical and Laboratory Plasmas, ed. K.~{Yamashita} \&
  T.~{Watanabe}, 411--414

\bibitem[{Kaastra {et~al.}(2023)Kaastra, Raassen, de~Plaa, \&
  Gu}]{kaastra2023_spex}
Kaastra, J.~S., Raassen, A. J.~J., de~Plaa, J., \& Gu, L. 2023, SPEX X-ray
  spectral fitting package

\bibitem[{{Keek} {et~al.}(2018){Keek}, {Arzoumanian}, {Chakrabarty},
  {Chenevez}, {Gendreau}, {Guillot}, {G{\"u}ver}, {Homan}, {Jaisawal},
  {LaMarr}, {Lamb}, {Mahmoodifar}, {Markwardt}, {Okajima}, {Strohmayer}, \& {in
  't Zand}}]{Keek2018ApJ...856L..37K}
{Keek}, L., {Arzoumanian}, Z., {Chakrabarty}, D., {et~al.} 2018, \apjl, 856,
  L37

\bibitem[{{Kosec} {et~al.}({2018}){Kosec}, {Pinto}, {Walton},
  {et~al.}}]{Kosec_2018b}
{Kosec}, P., {Pinto}, C., {Walton}, D.~J., {et~al.} {2018}, \mnras, 479, 3978

\bibitem[{{Krolik} {et~al.}(1981){Krolik}, {McKee}, \& {Tarter}}]{Krolik_1981}
{Krolik}, J.~H., {McKee}, C.~F., \& {Tarter}, C.~B. 1981, \apj, 249, 422

\bibitem[{{Ku{\'s}mierek} {et~al.}(2011){Ku{\'s}mierek}, {Madej}, \&
  {Kuulkers}}]{Kusmierek2011MNRAS.415.3344K}
{Ku{\'s}mierek}, K., {Madej}, J., \& {Kuulkers}, E. 2011, \mnras, 415, 3344

\bibitem[{{Kuulkers} {et~al.}(2002){Kuulkers}, {in't Zand}, {van Kerkwijk},
  {Cornelisse}, {Smith}, {Heise}, {Bazzano}, {Cocchi}, {Natalucci}, \&
  {Ubertini}}]{Kuulkers2002A&A...382..503K}
{Kuulkers}, E., {in't Zand}, J.~J.~M., {van Kerkwijk}, M.~H., {et~al.} 2002,
  \aap, 382, 503

\bibitem[{{Lodders} \& {Palme}(2009)}]{Lodders2009}
{Lodders}, K. \& {Palme}, H. 2009, Meteoritics and Planetary Science
  Supplement, 72, 5154

\bibitem[{{Marshall} {et~al.}(2002){Marshall}, {Canizares}, \&
  {Schulz}}]{Marshall2002}
{Marshall}, H.~L., {Canizares}, C.~R., \& {Schulz}, N.~S. 2002, \apj, 564, 941

\bibitem[{{Ness} {et~al.}(2022){Ness}, {Beardmore}, {Bezak}, {Dobrotka},
  {Drake}, {Vander Meulen}, {Osborne}, {Orio}, {Page}, {Pinto}, {Singh}, \&
  {Starrfield}}]{Ness2022}
{Ness}, J.~U., {Beardmore}, A.~P., {Bezak}, P., {et~al.} 2022, \aap, 658, A169

\bibitem[{{{\"O}zel} {et~al.}(2016){{\"O}zel}, {Psaltis}, {G{\"u}ver}, {Baym},
  {Heinke}, \& {Guillot}}]{Ozel2016ApJ...820...28O}
{{\"O}zel}, F., {Psaltis}, D., {G{\"u}ver}, T., {et~al.} 2016, \apj, 820, 28

\bibitem[{{Paczynski} \& {Proszynski}(1986)}]{Paczynski_1986}
{Paczynski}, B. \& {Proszynski}, M. 1986, \apj, 302, 519

\bibitem[{{Pinto} {et~al.}(2014){Pinto}, {Costantini}, {Fabian}, {Kaastra}, \&
  {in't Zand}}]{Pinto2014}
{Pinto}, C., {Costantini}, E., {Fabian}, A.~C., {Kaastra}, J.~S., \& {in't
  Zand}, J.~J.~M. 2014, \aap, 563, A115

\bibitem[{{Pinto} {et~al.}(2020{\natexlab{a}}){Pinto}, {Mehdipour}, {Walton},
  {et~al.}}]{Pinto_2020a}
{Pinto}, C., {Mehdipour}, M., {Walton}, D.~J., {et~al.} 2020{\natexlab{a}},
  \mnras, 491, 5702

\bibitem[{{Pinto} {et~al.}(2016){Pinto}, {Middleton}, \&
  {Fabian}}]{pinto2016Natur.533...64P}
{Pinto}, C., {Middleton}, M.~J., \& {Fabian}, A.~C. 2016, \nat, 533, 64

\bibitem[{{Pinto} {et~al.}(2012){Pinto}, {Ness}, {Verbunt}, {Kaastra},
  {Costantini}, \& {Detmers}}]{Pinto2012}
{Pinto}, C., {Ness}, J.~U., {Verbunt}, F., {et~al.} 2012, \aap, 543, A134

\bibitem[{{Pinto} {et~al.}(2020{\natexlab{b}}){Pinto}, {Walton}, {Kara},
  {Parker}, {Soria}, {et~al.}}]{Pinto_2020b}
{Pinto}, C., {Walton}, D.~J., {Kara}, E., {et~al.} 2020{\natexlab{b}}, \mnras,
  492, 4646

\bibitem[{{Speicher} {et~al.}(2022){Speicher}, {Ballantyne}, \&
  {Fragile}}]{Speicher2022MNRAS.509.1736S}
{Speicher}, J., {Ballantyne}, D.~R., \& {Fragile}, P.~C. 2022, \mnras, 509,
  1736

\bibitem[{{Stella} {et~al.}(1987){Stella}, {Priedhorsky}, \&
  {White}}]{Stella1987ApJ...312L..17S}
{Stella}, L., {Priedhorsky}, W., \& {White}, N.~E. 1987, \apjl, 312, L17

\bibitem[{{Strohmayer} {et~al.}(2019){Strohmayer}, {Altamirano}, {Arzoumanian},
  {Bult}, {Chakrabarty}, {Chenevez}, {Fabian}, {Gendreau}, {Guillot}, {in 't
  Zand}, {Jaisawal}, {Keek}, {Kosec}, {Ludlam}, {Mahmoodifar}, {Malacaria}, \&
  {Miller}}]{Strohmayer2019ApJ...878L..27S}
{Strohmayer}, T.~E., {Altamirano}, D., {Arzoumanian}, Z., {et~al.} 2019, \apjl,
  878, L27

\bibitem[{{Strohmayer} \& {Brown}(2002)}]{Strohmayer2002ApJ...566.1045S}
{Strohmayer}, T.~E. \& {Brown}, E.~F. 2002, \apj, 566, 1045

\bibitem[{{Suleimanov} {et~al.}(2017){Suleimanov}, {Kajava}, {Molkov},
  {N{\"a}ttil{\"a}}, {Lutovinov}, {Werner}, \&
  {Poutanen}}]{Suleimanov2017MNRAS.472.3905S}
{Suleimanov}, V.~F., {Kajava}, J. J.~E., {Molkov}, S.~V., {et~al.} 2017,
  \mnras, 472, 3905

\bibitem[{{Tarter} {et~al.}(1969){Tarter}, {Tucker}, \&
  {Salpeter}}]{Tarter_1969}
{Tarter}, C.~B., {Tucker}, W.~H., \& {Salpeter}, E.~E. 1969, \apj, 156, 943

\bibitem[{{Vacca} {et~al.}(1986){Vacca}, {Lewin}, \& {van
  Paradijs}}]{Vacca1986MNRAS.220..339V}
{Vacca}, W.~D., {Lewin}, W.~H.~G., \& {van Paradijs}, J. 1986, \mnras, 220, 339

\bibitem[{{van den Eijnden} {et~al.}(2018){van den Eijnden}, {Degenaar},
  {Pinto}, {Patruno}, {Wette}, {Messenger}, {Hern{\'a}ndez Santisteban},
  {Wijnands}, {Miller}, {Altamirano}, {Paerels}, {Chakrabarty}, \&
  {Fabian}}]{VDEijnden_2018}
{van den Eijnden}, J., {Degenaar}, N., {Pinto}, C., {et~al.} 2018, \mnras, 475,
  2027

\bibitem[{{van Paradijs} \& {Lewin}(1987)}]{Paradijs1987A&A...172L..20V}
{van Paradijs}, J. \& {Lewin}, W.~H.~G. 1987, \aap, 172, L20

\bibitem[{van den Eijnden {et~al.}(2019)van den Eijnden, Degenaar, Schulz,
  Nowak, Wijnands, Russell, Hernández Santisteban, Bahramian, Maccarone,
  Kennea, \& Heinke}]{VDE_2019}
van den Eijnden, J., Degenaar, N., Schulz, N.~S., {et~al.} 2019, Monthly
  Notices of the Royal Astronomical Society, 487, 4355

\bibitem[{{Weinberg} {et~al.}(2006){Weinberg}, {Bildsten}, \&
  {Schatz}}]{Weinberg2006ApJ...639.1018W}
{Weinberg}, N.~N., {Bildsten}, L., \& {Schatz}, H. 2006, \apj, 639, 1018

\bibitem[{Weinberg {et~al.}(2006)Weinberg, Bildsten, \& Schatz}]{Weinberg_2006}
Weinberg, N.~N., Bildsten, L., \& Schatz, H. 2006, The Astrophysical Journal,
  639, 1018

\bibitem[{{Worpel} {et~al.}(2013){Worpel}, {Galloway}, \&
  {Price}}]{Worpel2013ApJ...772...94W}
{Worpel}, H., {Galloway}, D.~K., \& {Price}, D.~J. 2013, \apj, 772, 94

\bibitem[{{Yu} \& {Weinberg}(2018)}]{Yu2018ApJ...863...53Y}
{Yu}, H. \& {Weinberg}, N.~N. 2018, \apj, 863, 53

\bibitem[{{Yu} {et~al.}(2024){Yu}, {Li}, {Lu}, {Pan}, {Yang}, {Chen}, {Zhang},
  \& {Falanga}}]{Yu_2024}
{Yu}, W., {Li}, Z., {Lu}, Y., {et~al.} 2024, \aap, 683, A93

\bibitem[{{Zdziarski} {et~al.}(1996){Zdziarski}, {Johnson}, \&
  {Magdziarz}}]{Zdziarski1996MNRAS.283..193Z}
{Zdziarski}, A.~A., {Johnson}, W.~N., \& {Magdziarz}, P. 1996, \mnras, 283, 193

\bibitem[{{{\.Z}ycki} {et~al.}(1999){{\.Z}ycki}, {Done}, \&
  {Smith}}]{Zycky1999MNRAS.309..561Z}
{{\.Z}ycki}, P.~T., {Done}, C., \& {Smith}, D.~A. 1999, \mnras, 309, 561

\end{thebibliography}

\begin{appendix}
\onecolumn
\section{CIE grids}
\label{appendix-cie}
\begin{figure*}[h!]
    \centering
    \includegraphics[width=0.30\textwidth]{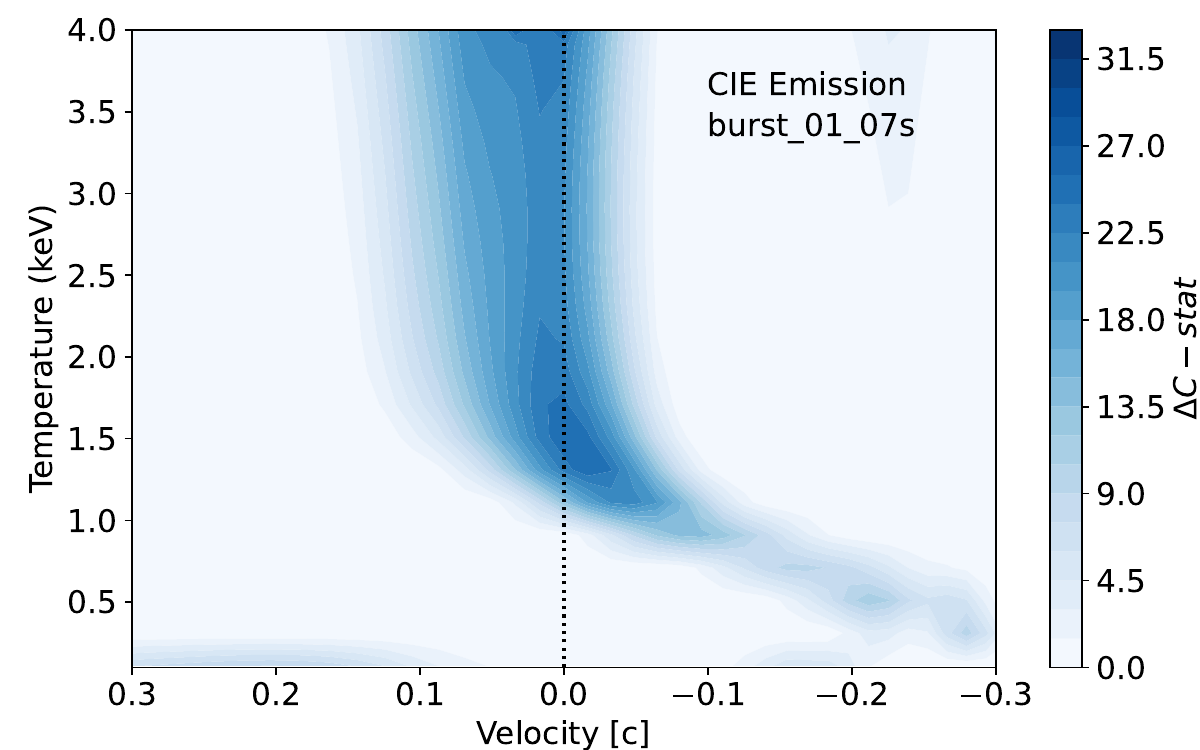}
    \includegraphics[width=0.30\textwidth]{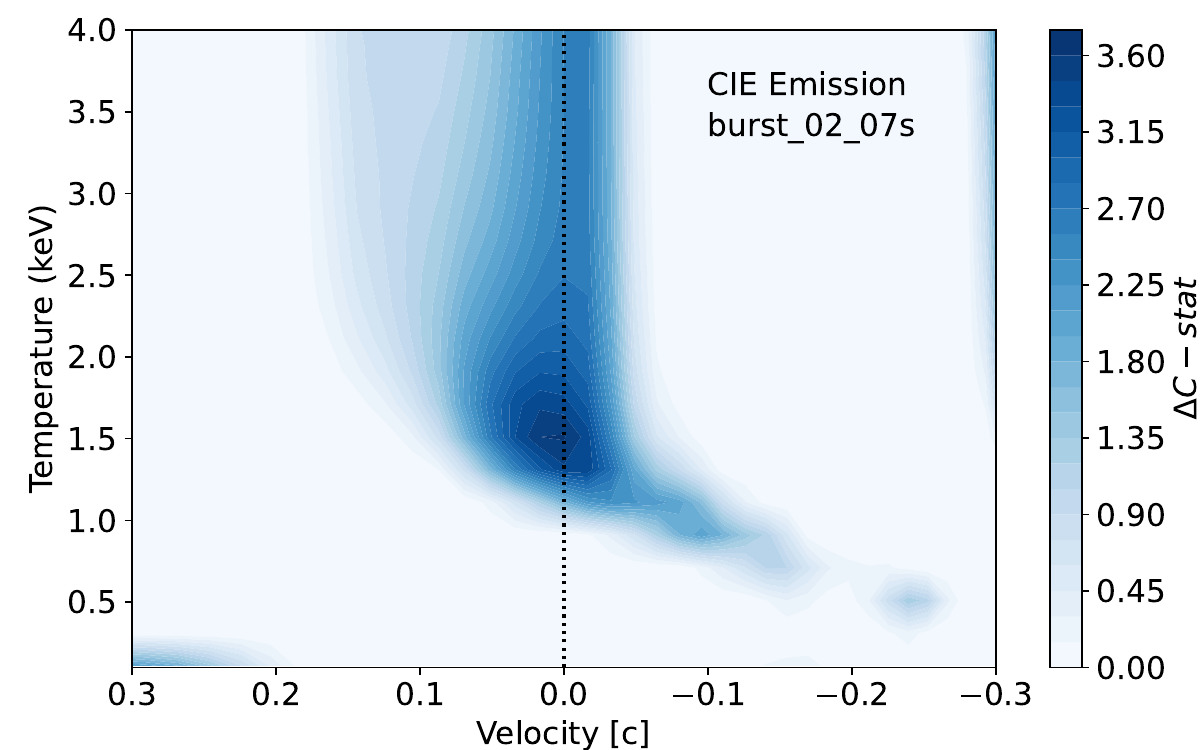}
    \includegraphics[width=0.30\textwidth]{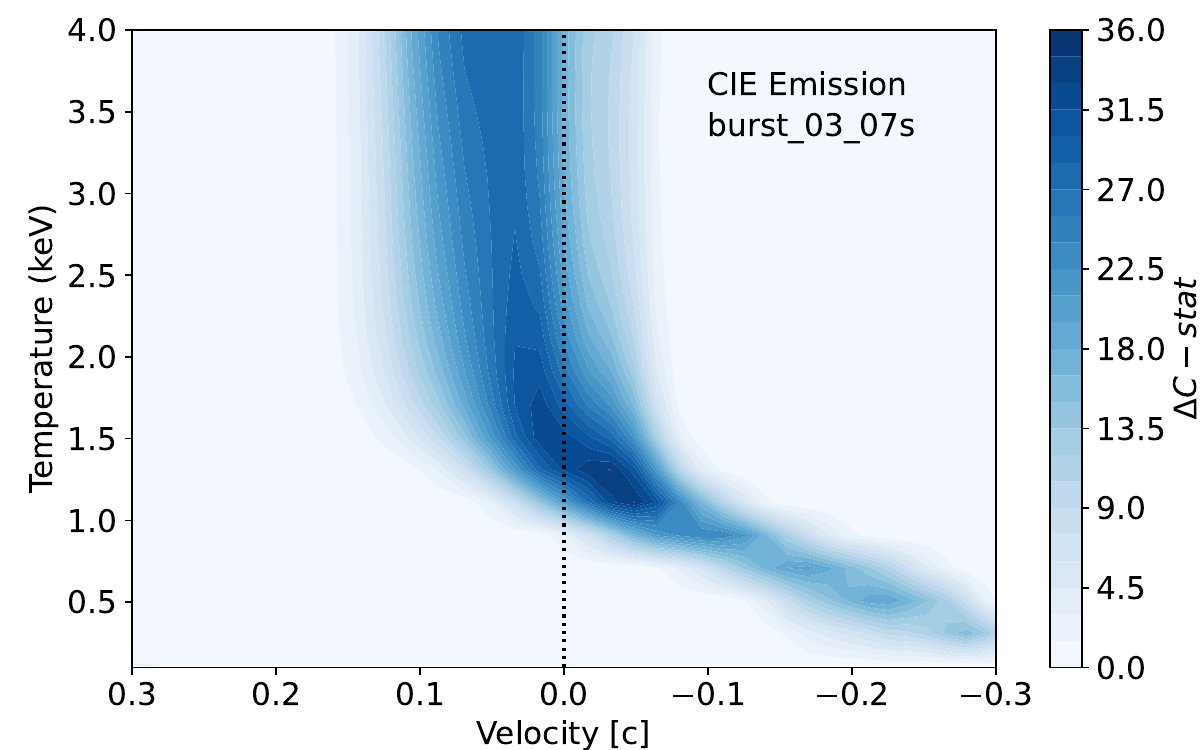}
    \includegraphics[width=0.30\textwidth]{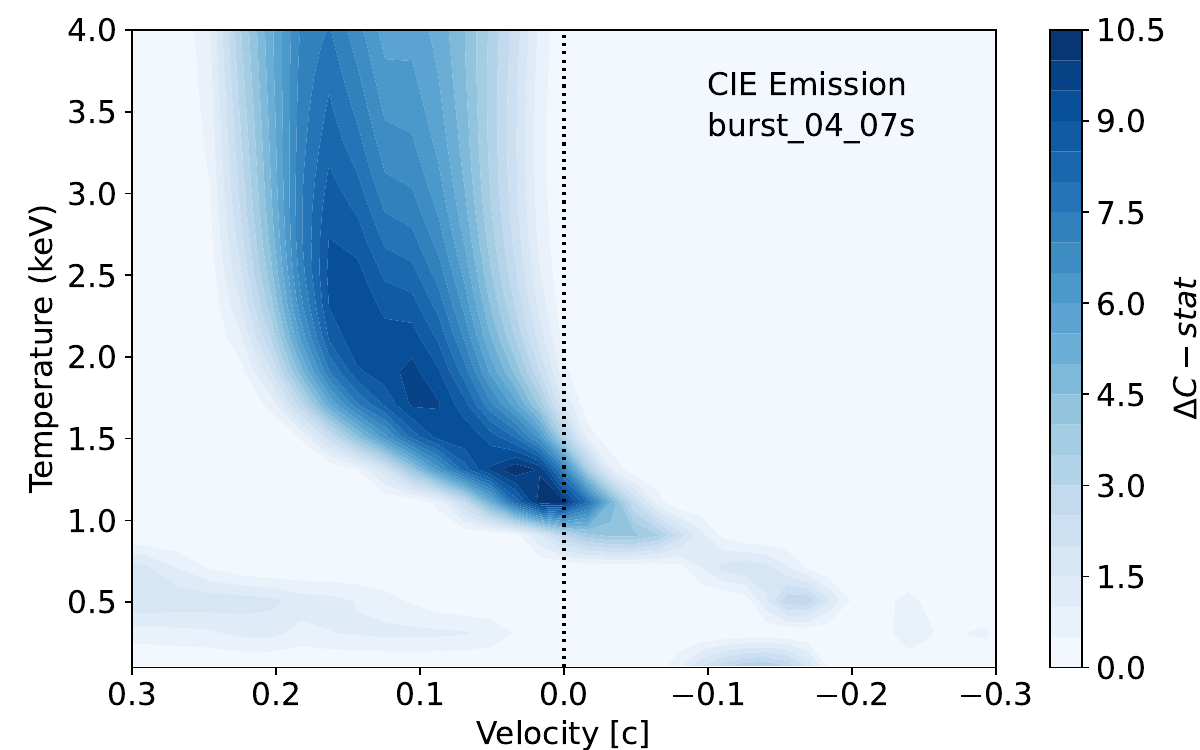}
    \includegraphics[width=0.30\textwidth]{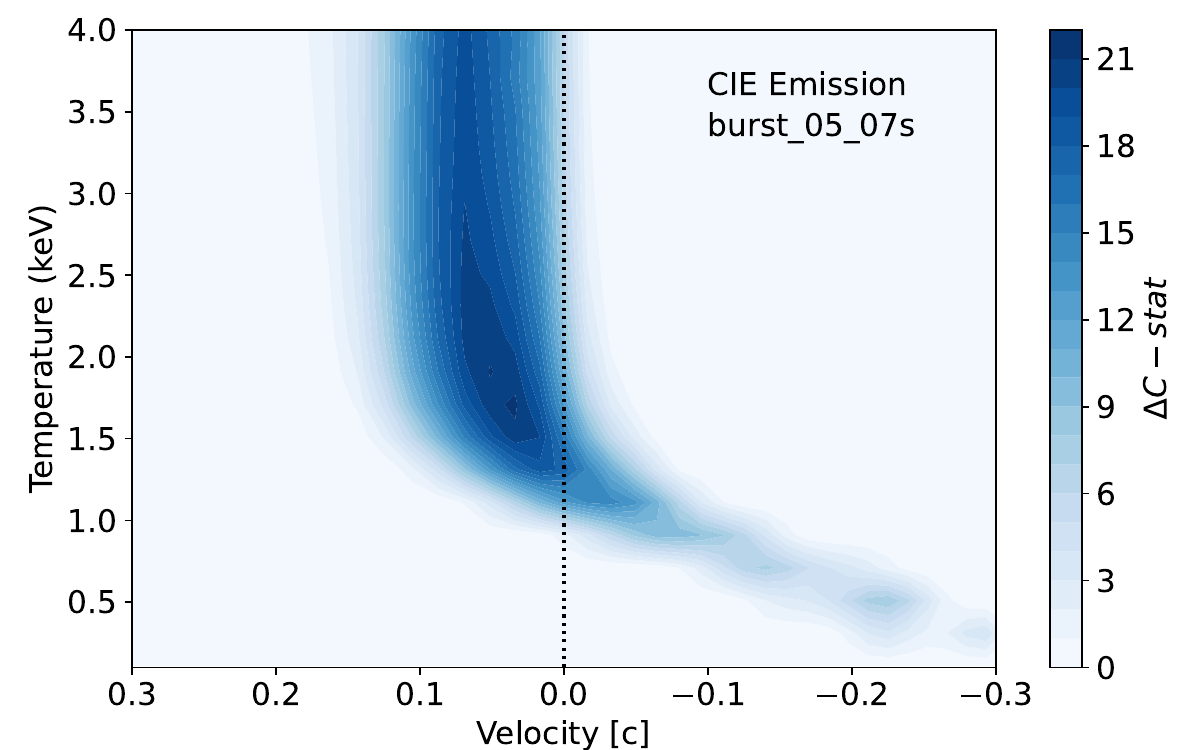}
    \includegraphics[width=0.30\textwidth]{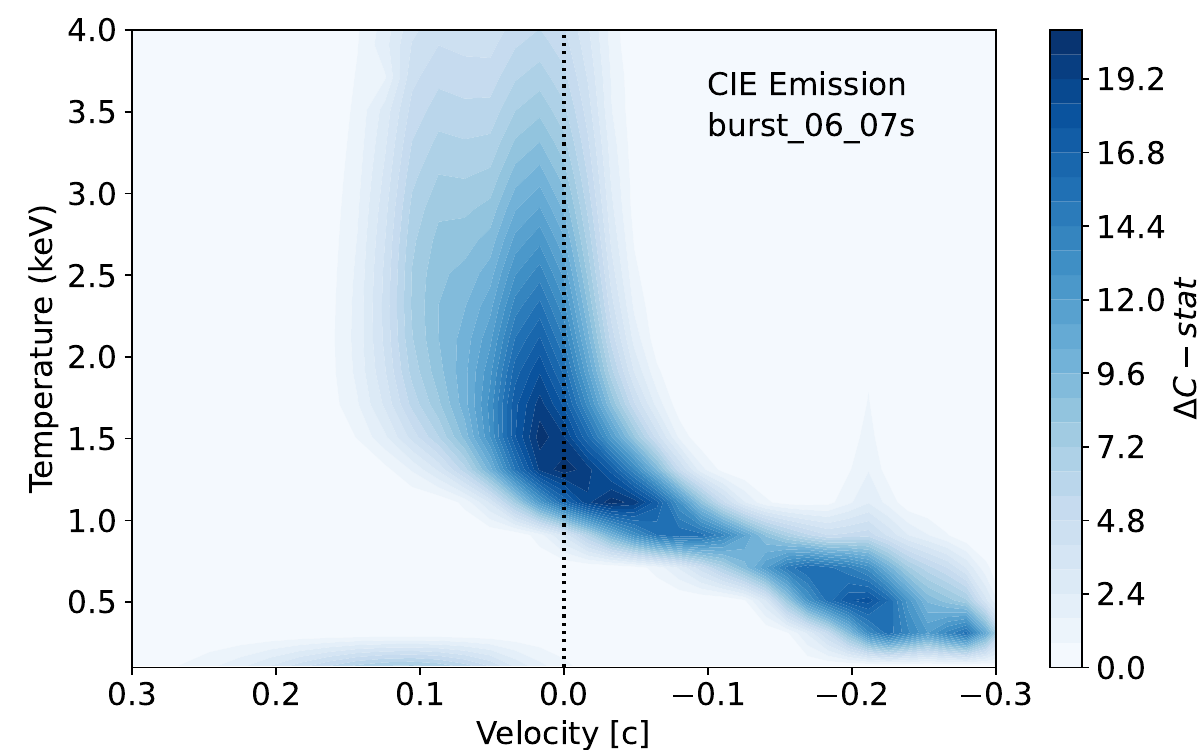}
    \includegraphics[width=0.30\textwidth]{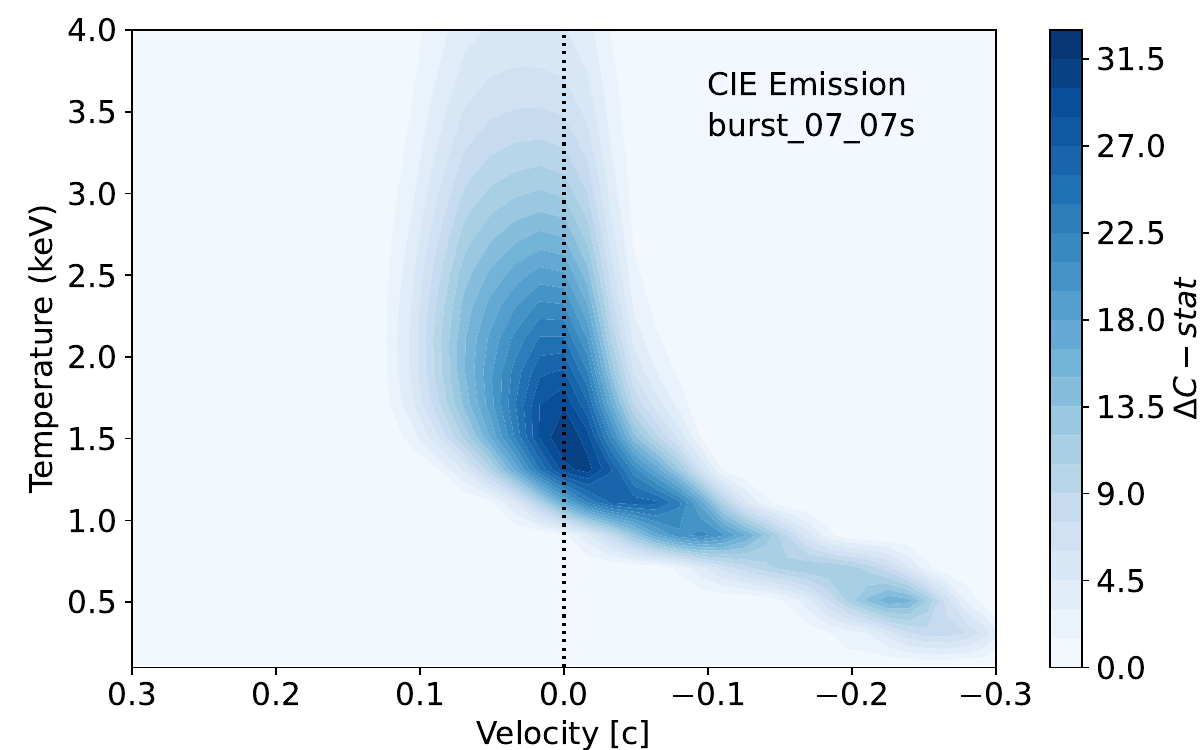}
    \includegraphics[width=0.30\textwidth]{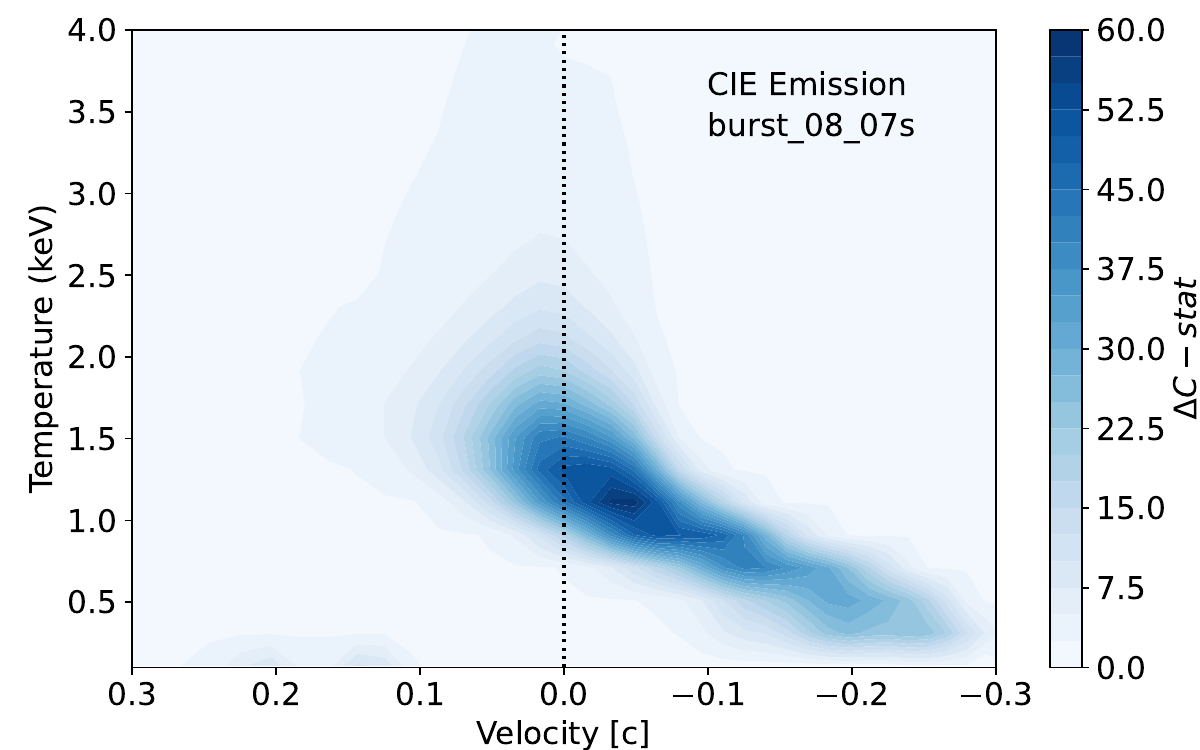}
    \includegraphics[width=0.30\textwidth]{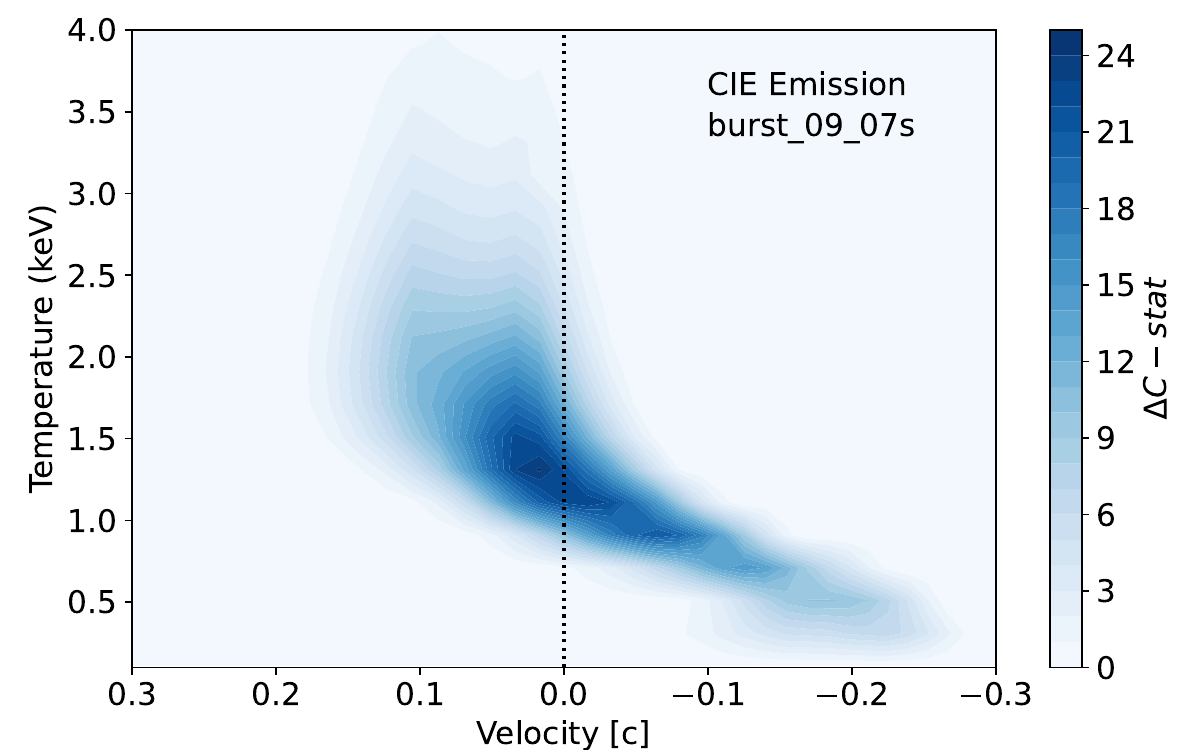}
    \includegraphics[width=0.30\textwidth]{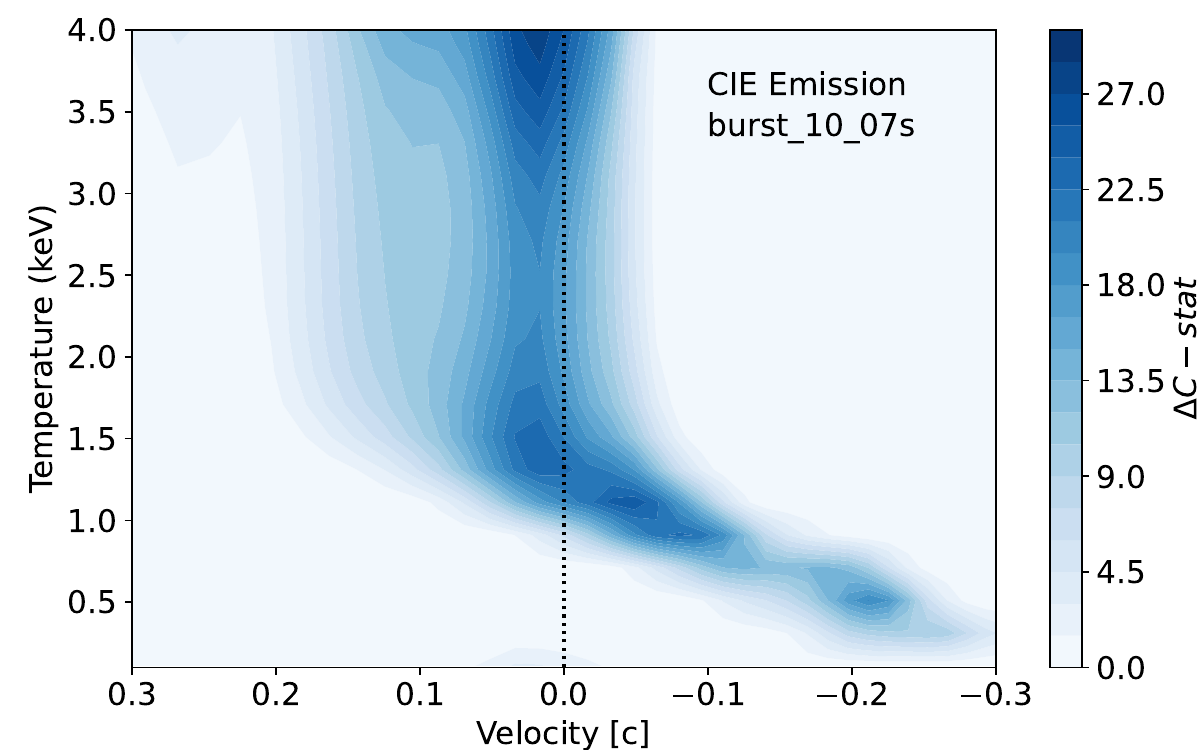}
    \includegraphics[width=0.30\textwidth]{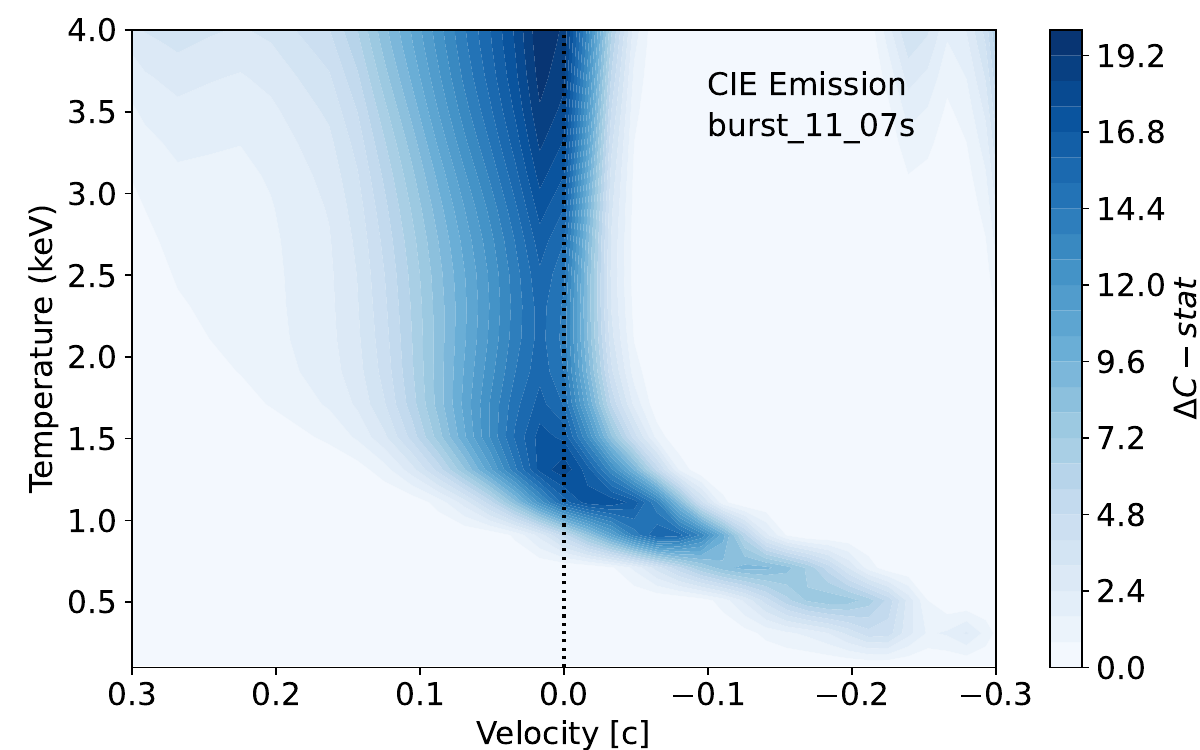}
    \includegraphics[width=0.30\textwidth]{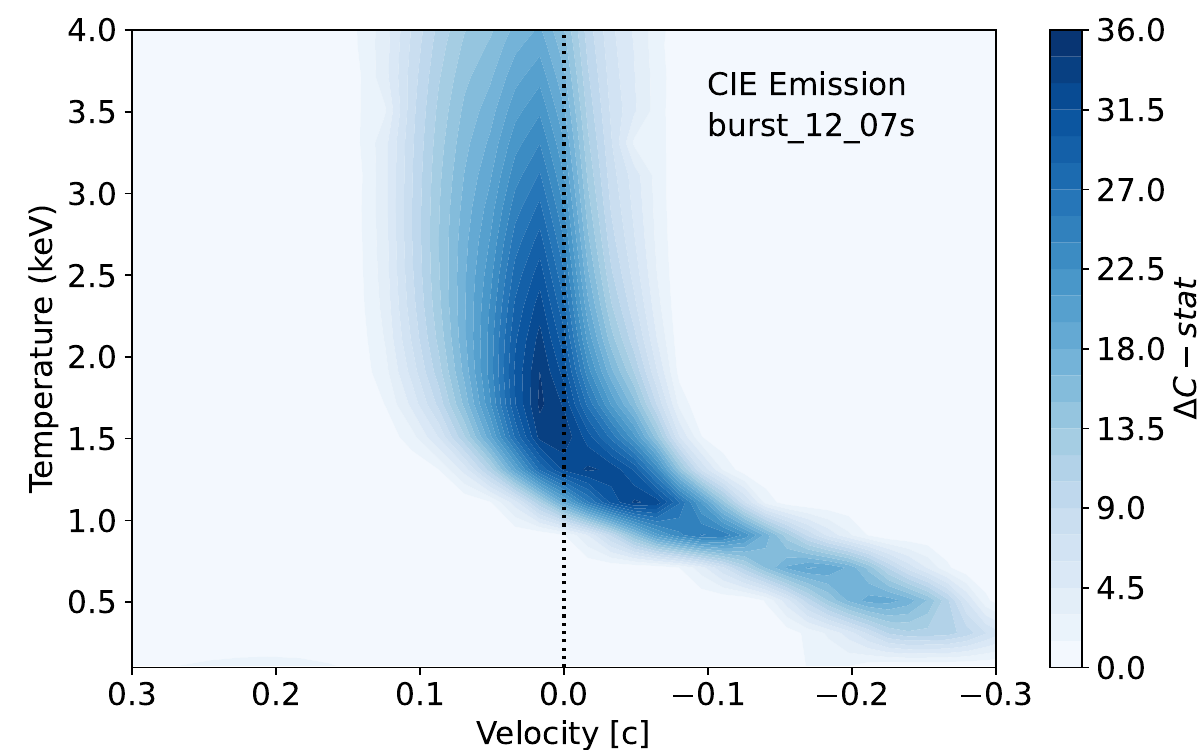}
       \caption{Multidimensional scan grids with the emission models of collisionally-ionised plasma (\textit{cie}) for the 12 bursts. }
       \label{fig: CIE grids}
       \end{figure*}

\newpage
\section{PION grids}
\begin{figure*}[h!]
    \centering
    \includegraphics[width=0.30\textwidth]{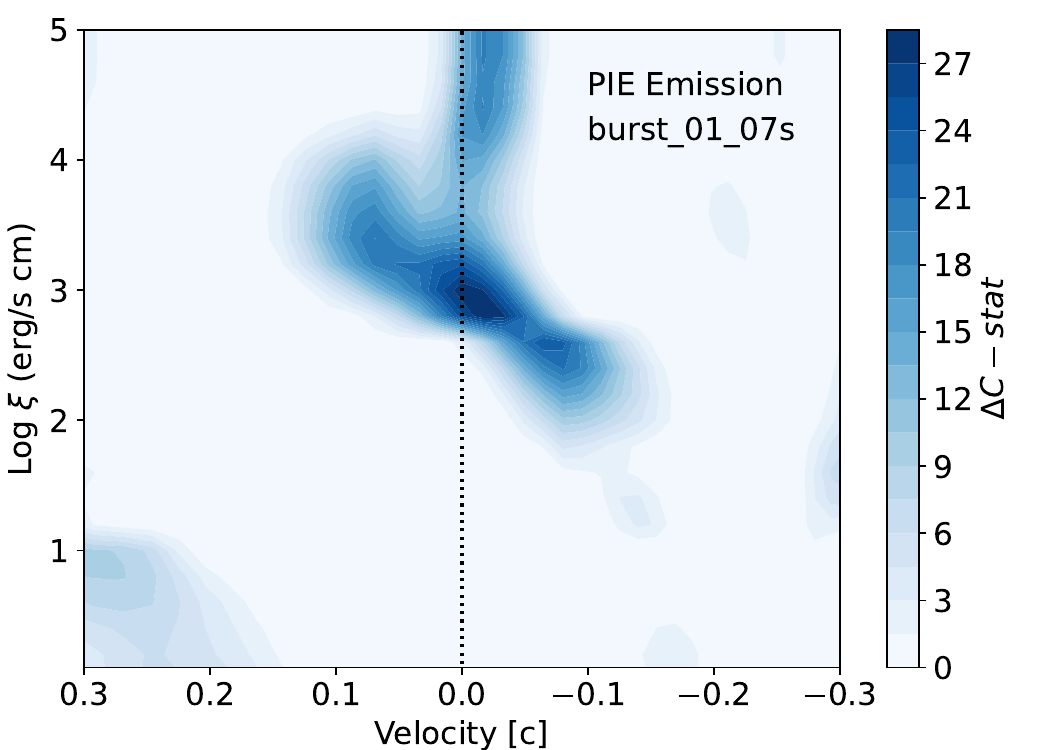}
    \includegraphics[width=0.30\textwidth]{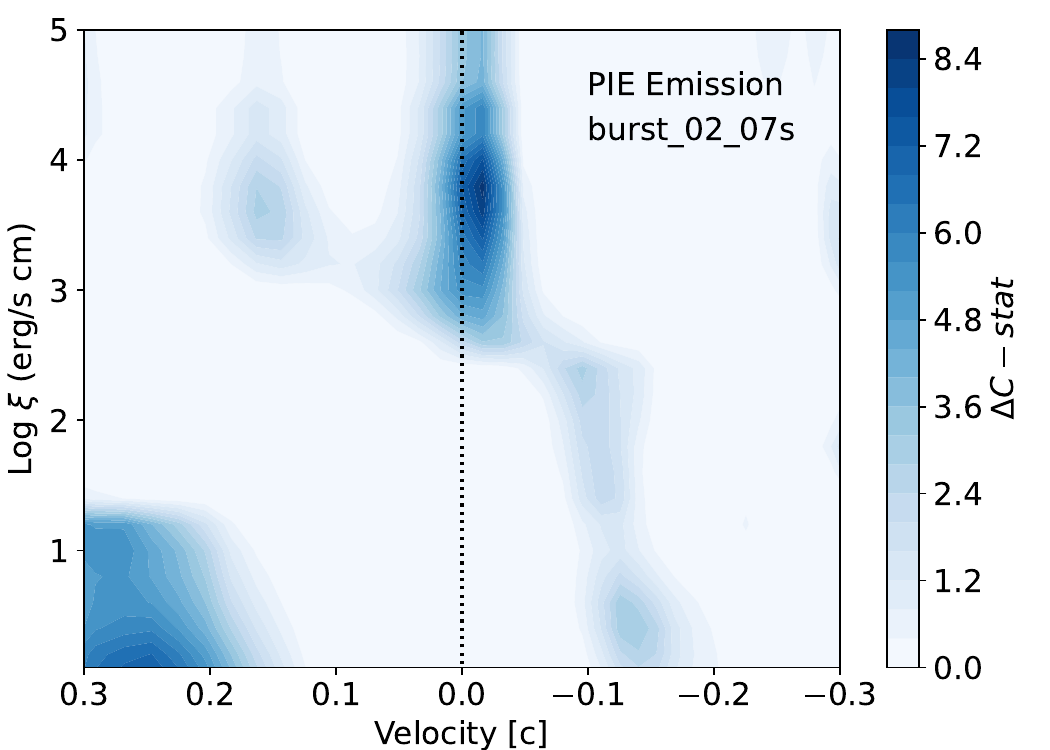}
    \includegraphics[width=0.30\textwidth]{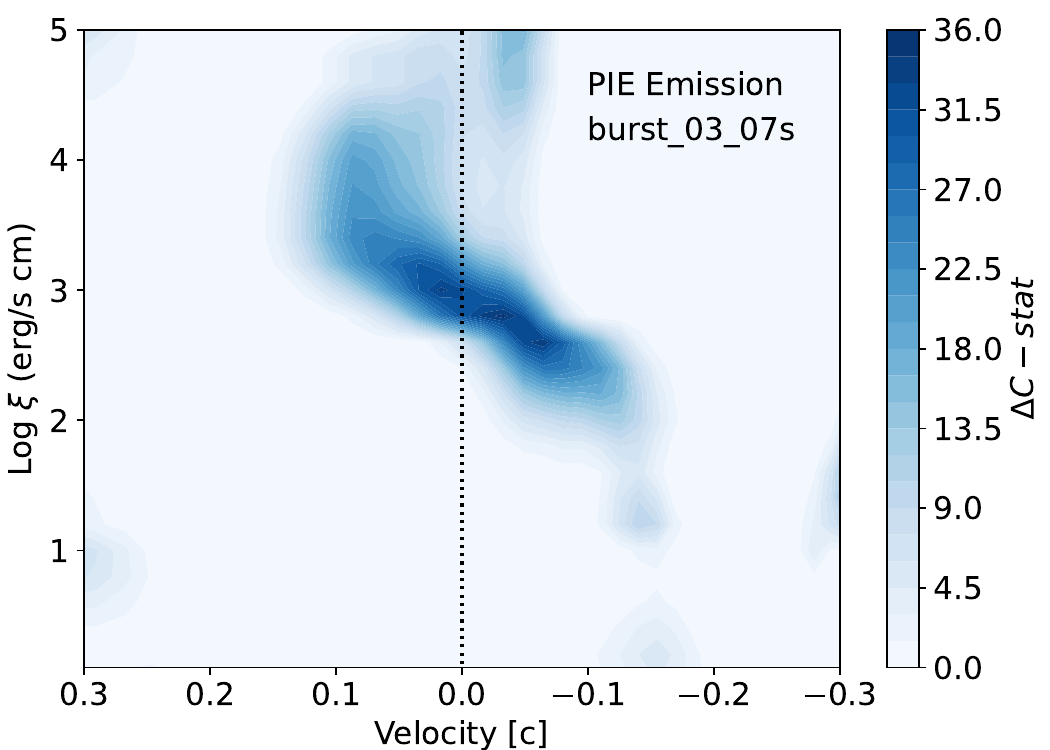}
    \includegraphics[width=0.30\textwidth]{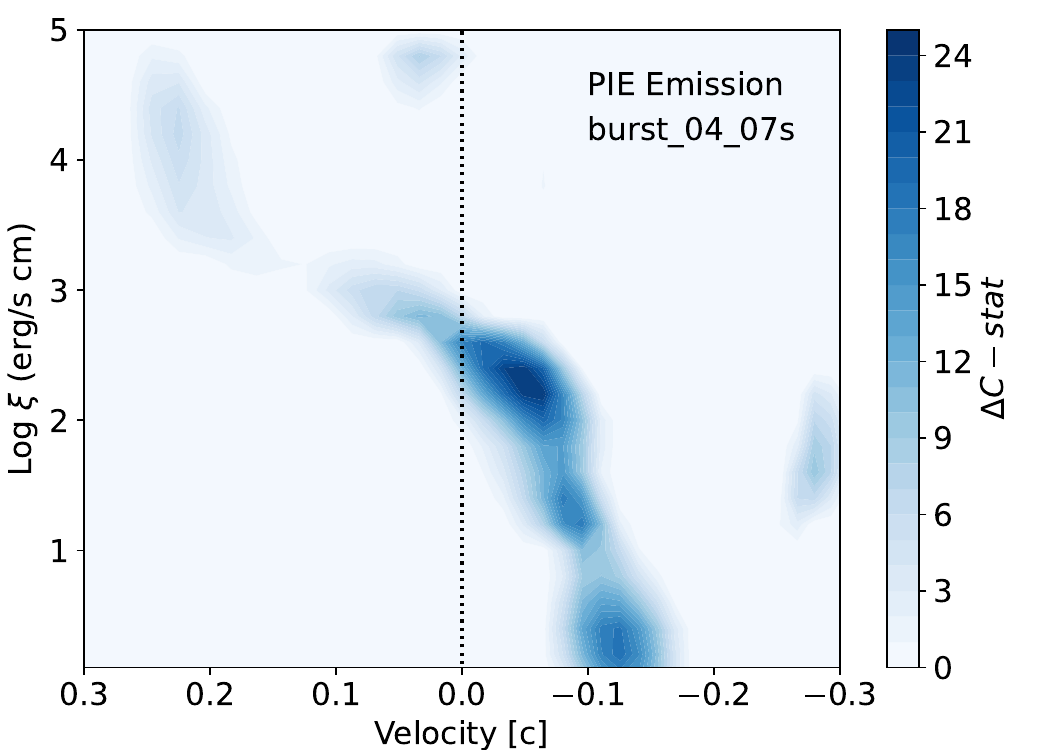}
    \includegraphics[width=0.30\textwidth]{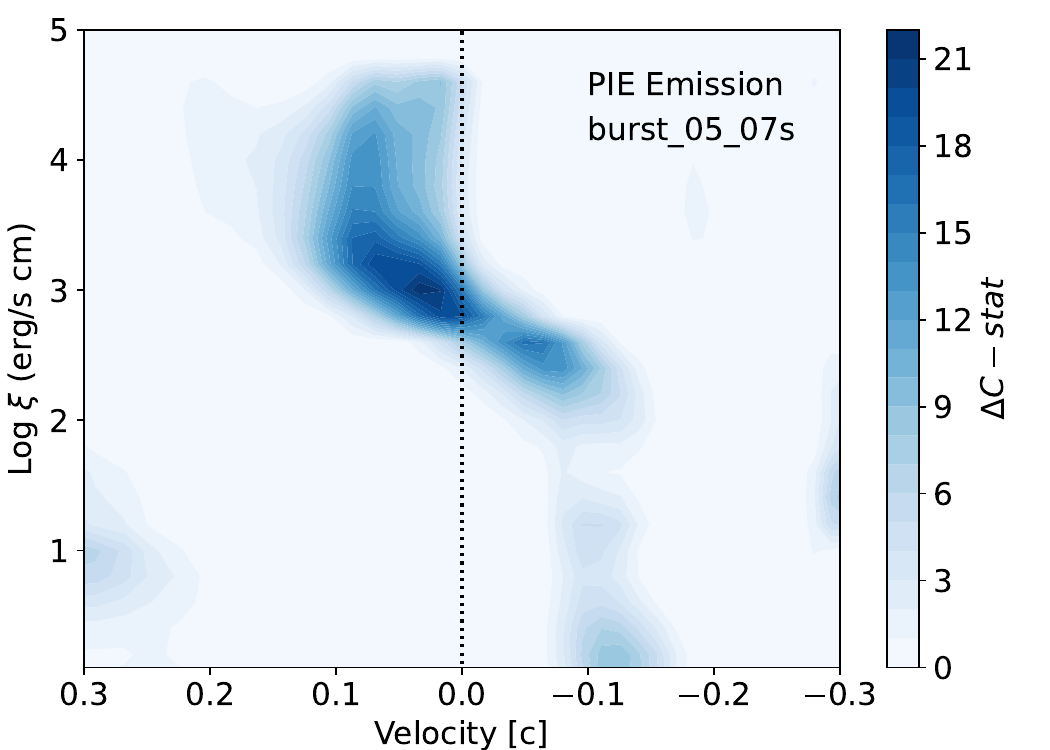}
    \includegraphics[width=0.30\textwidth]{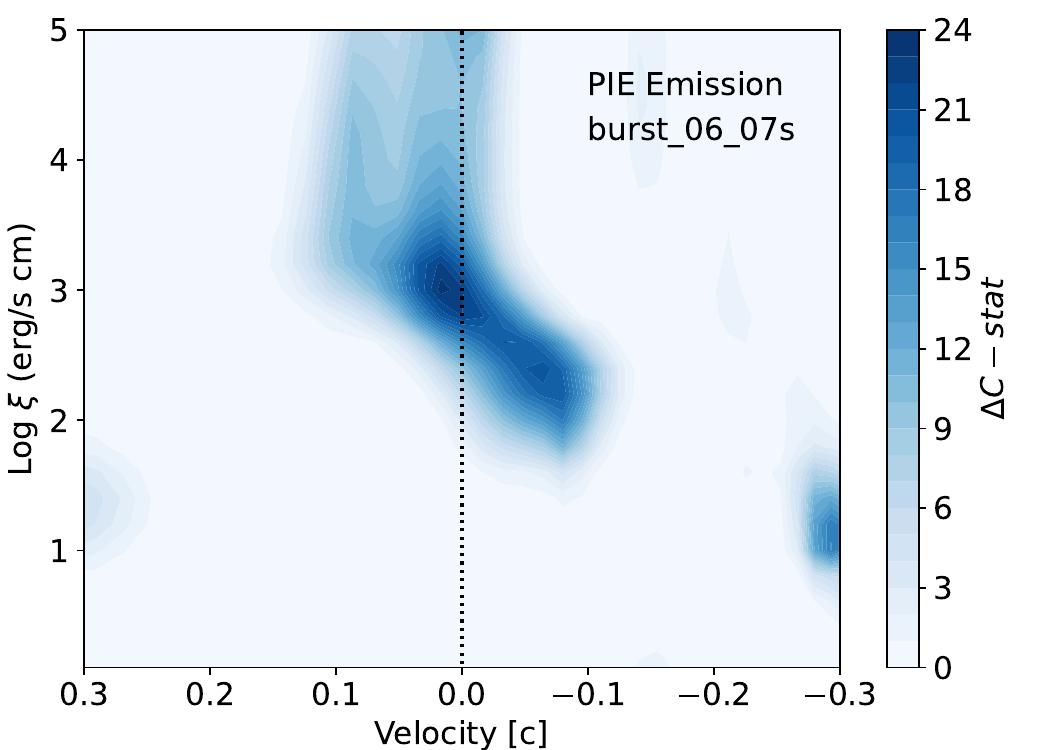}
    \includegraphics[width=0.30\textwidth]{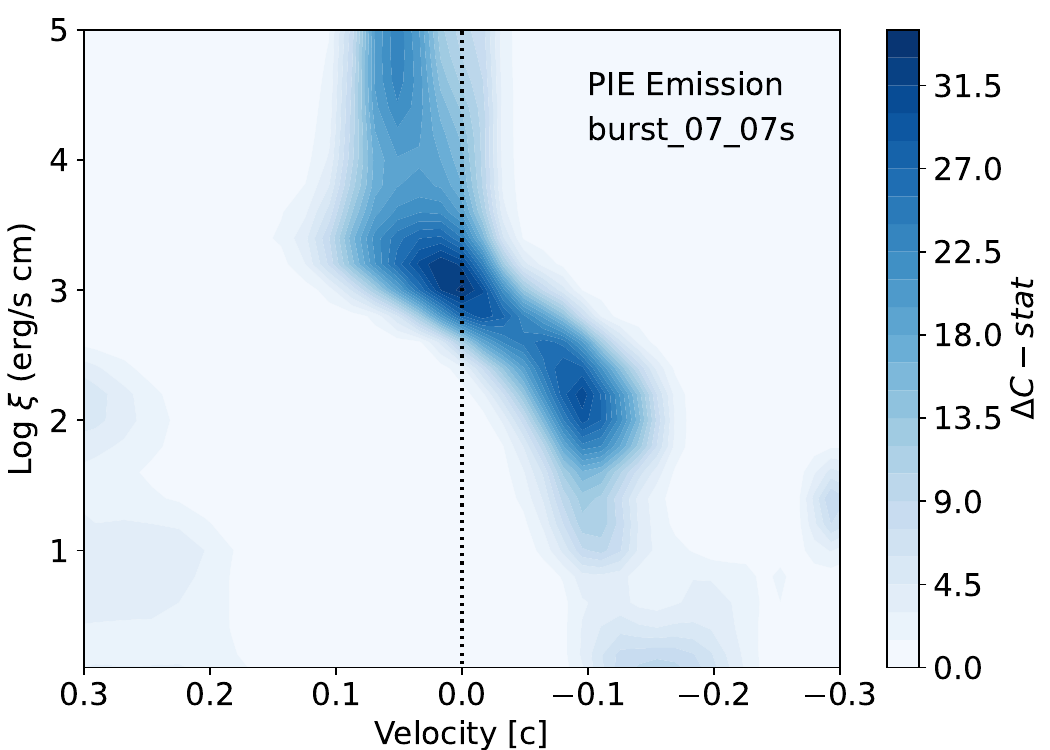}
    \includegraphics[width=0.30\textwidth]{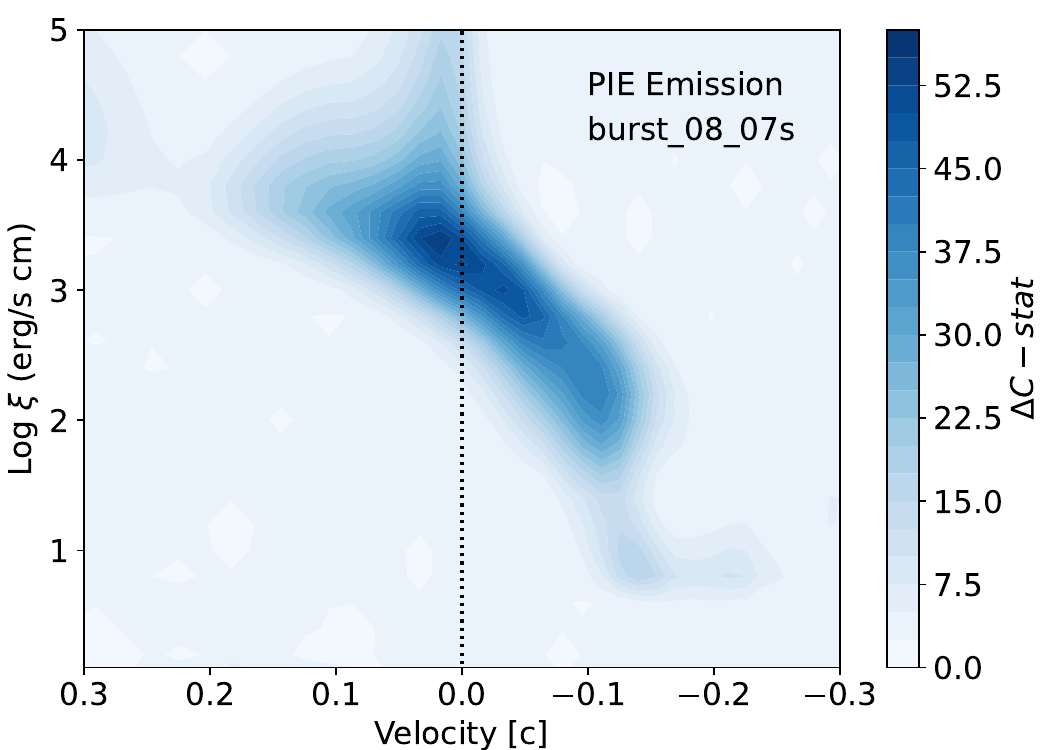}
    \includegraphics[width=0.30\textwidth]{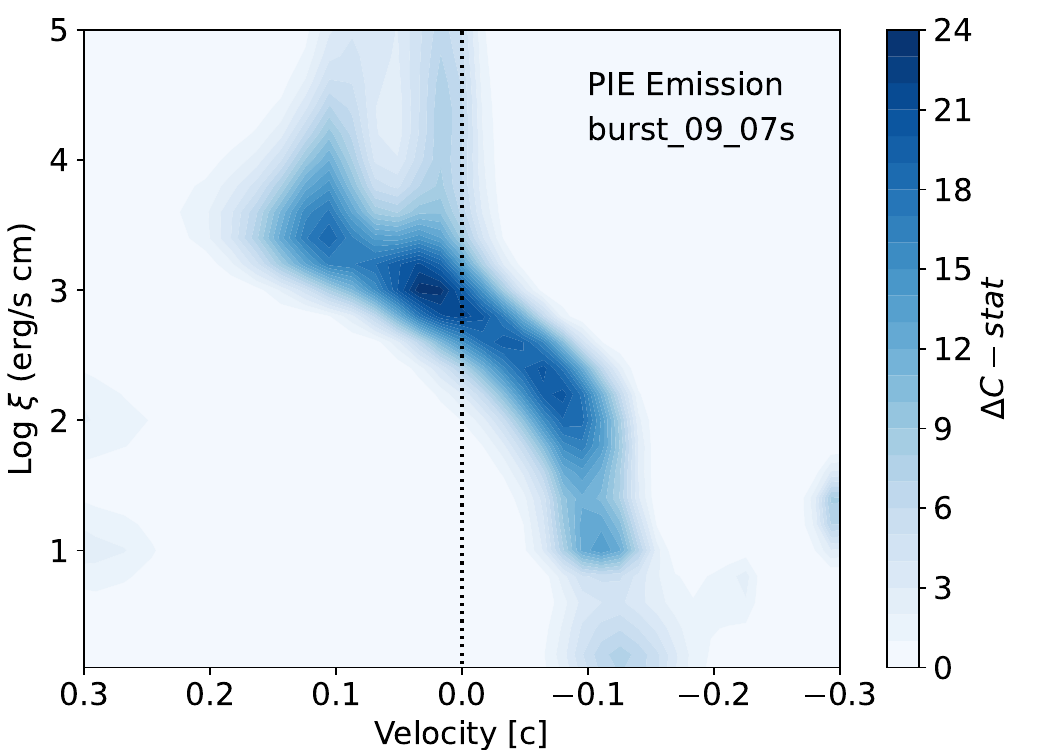}
    \includegraphics[width=0.30\textwidth]{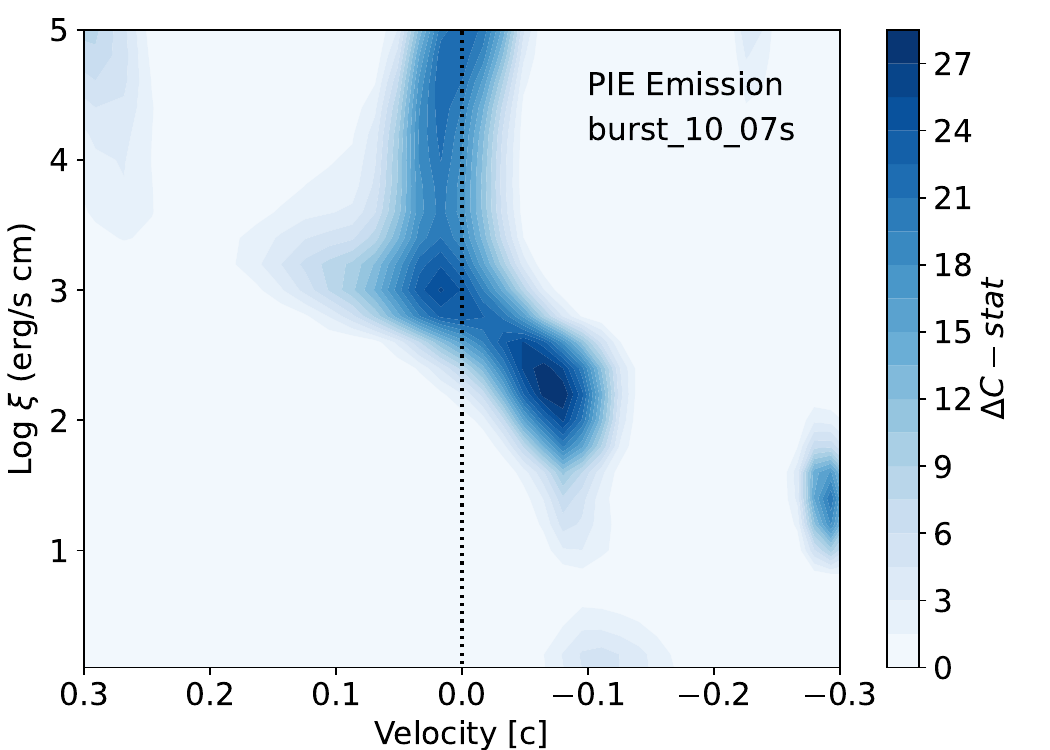}
    \includegraphics[width=0.30\textwidth]{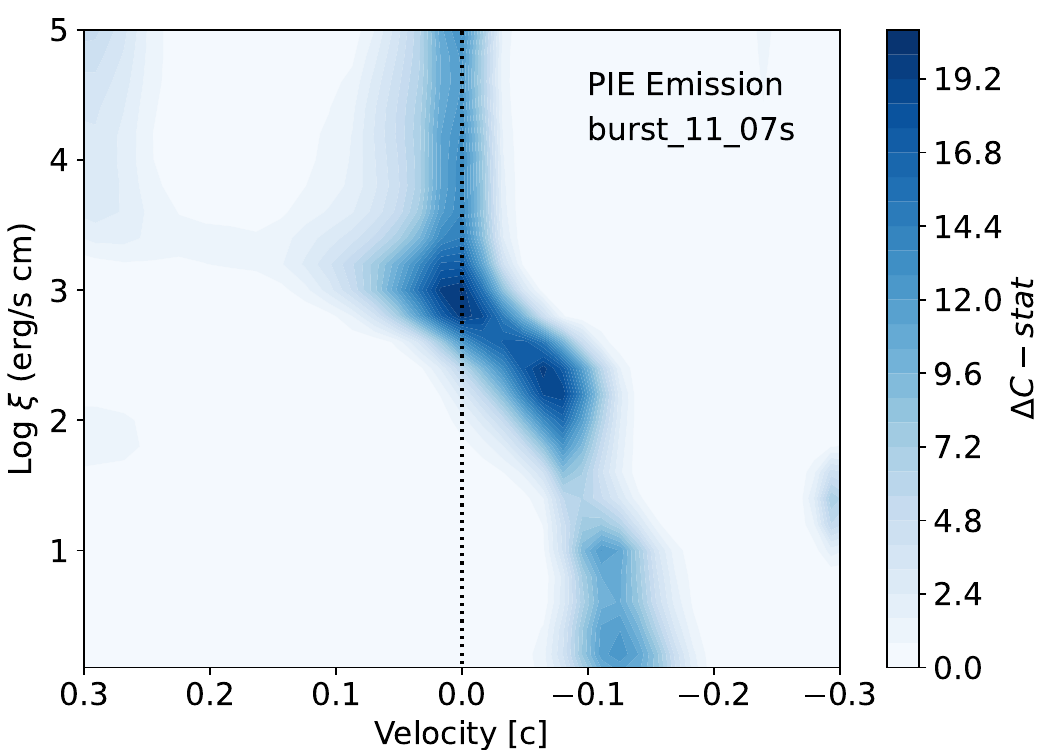}
    \includegraphics[width=0.30\textwidth]{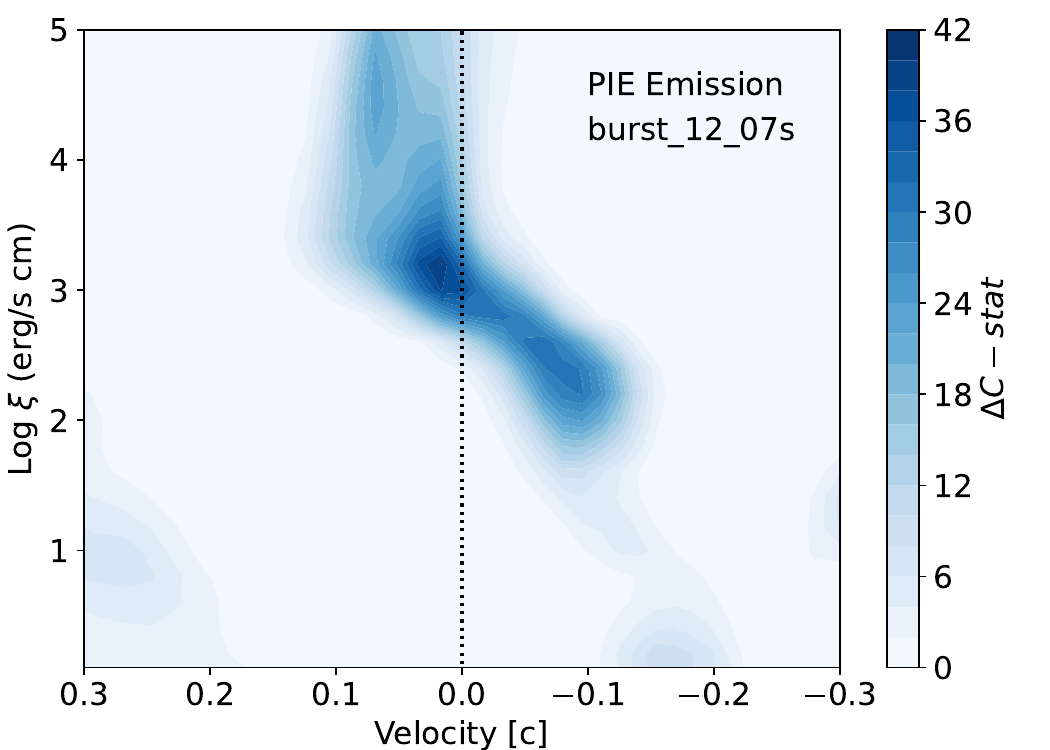}
       \caption{Multidimensional scan grids with the emission models of photo-ionised plasma (\textit{pion}) for the 12 bursts.}
       \label{fig: PION grids}
\end{figure*}

\newpage

\section{XABS grids}
\begin{figure*}[h!]
    \centering
    \includegraphics[width=0.30\textwidth]{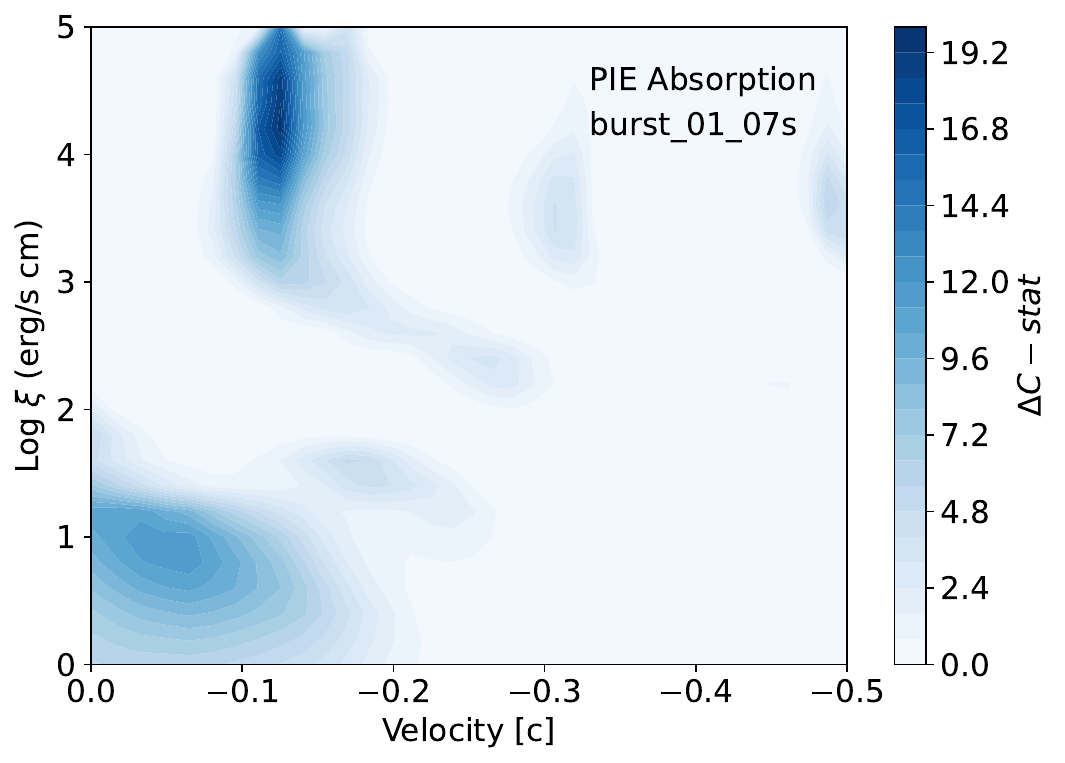}
    \includegraphics[width=0.30\textwidth]{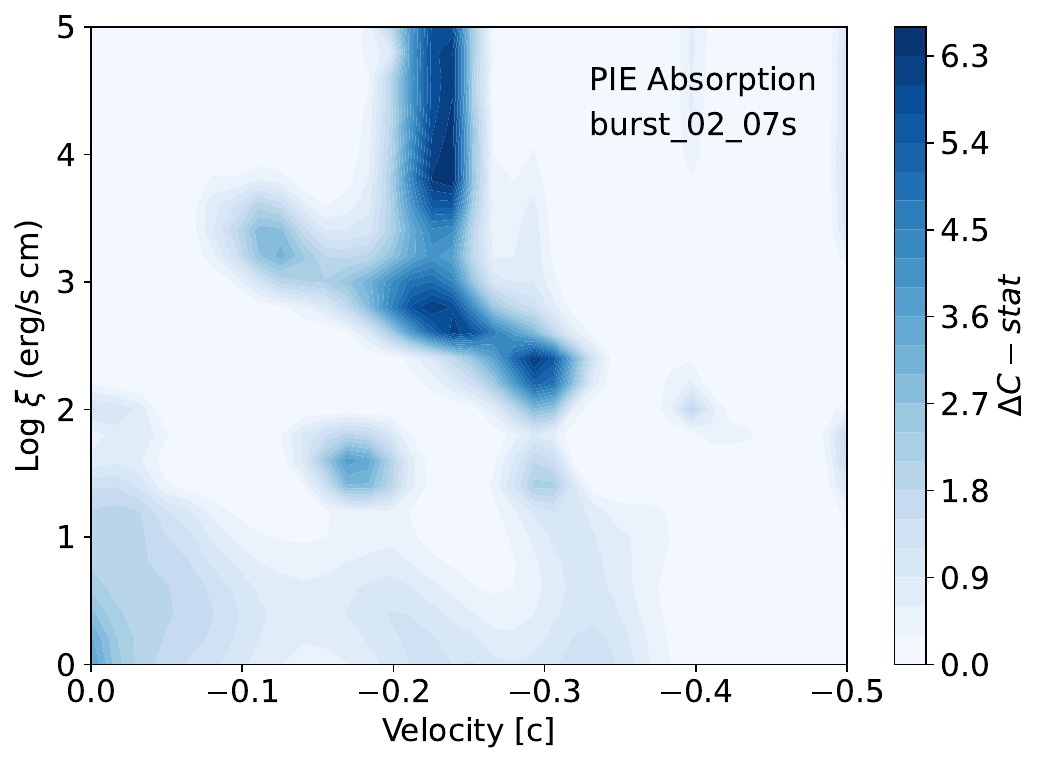}
    \includegraphics[width=0.30\textwidth]{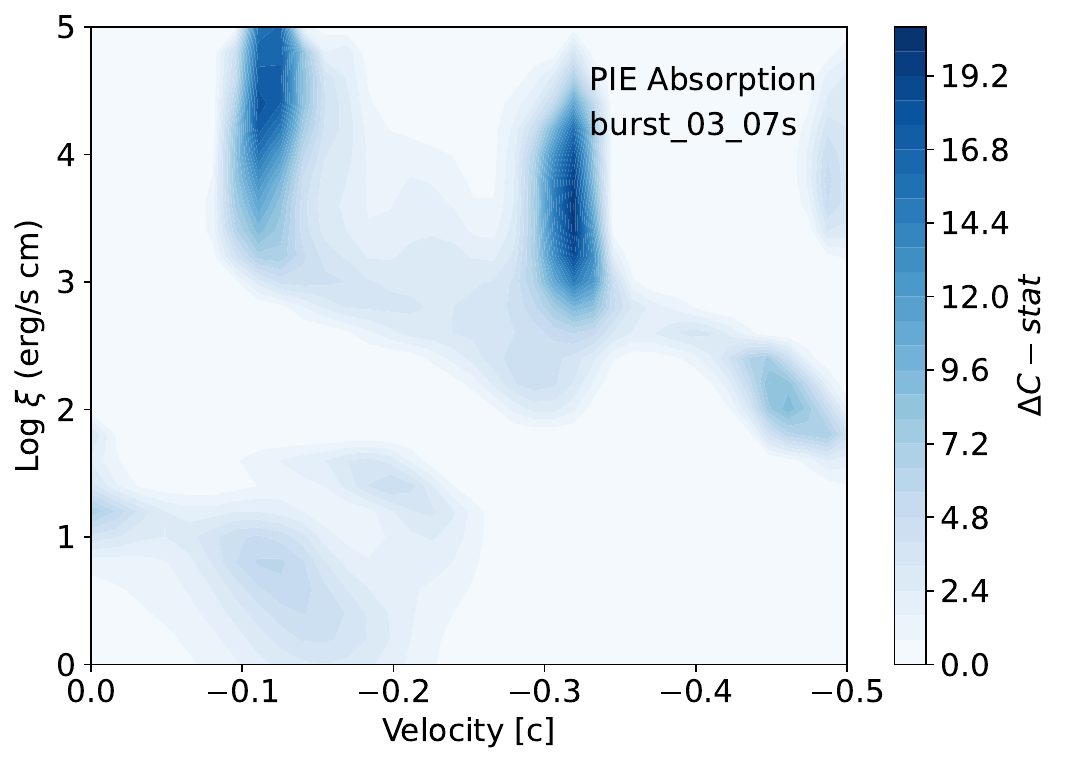}
    \includegraphics[width=0.30\textwidth]{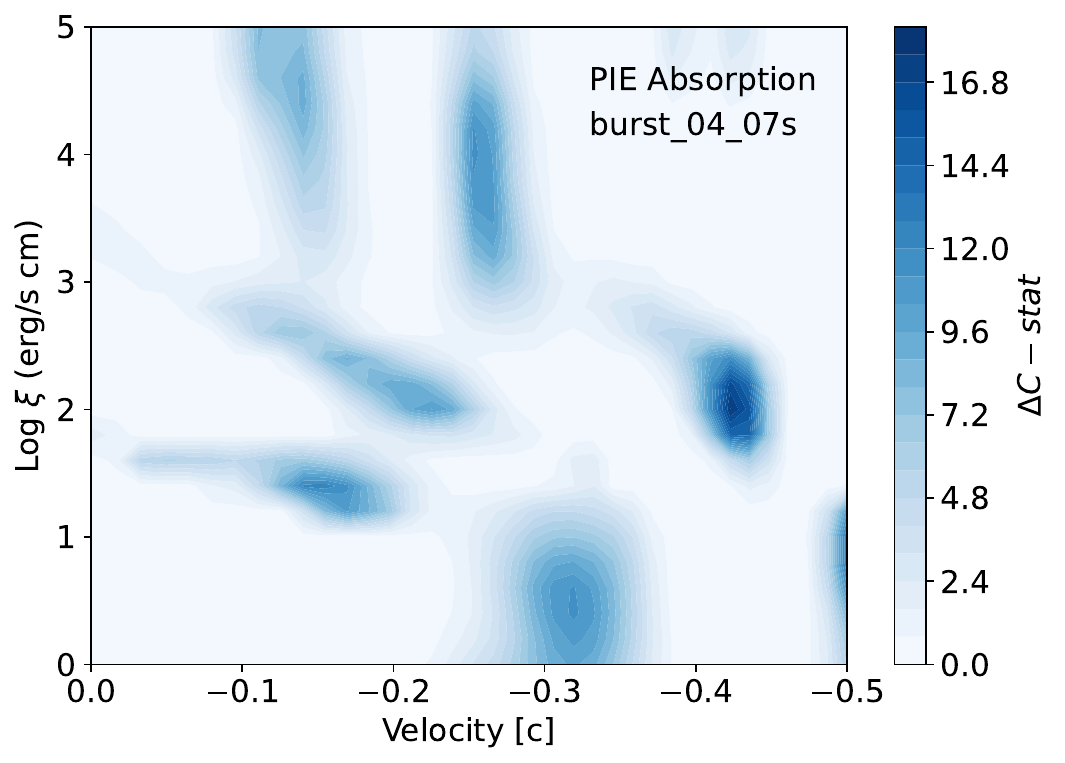}
    \includegraphics[width=0.30\textwidth]{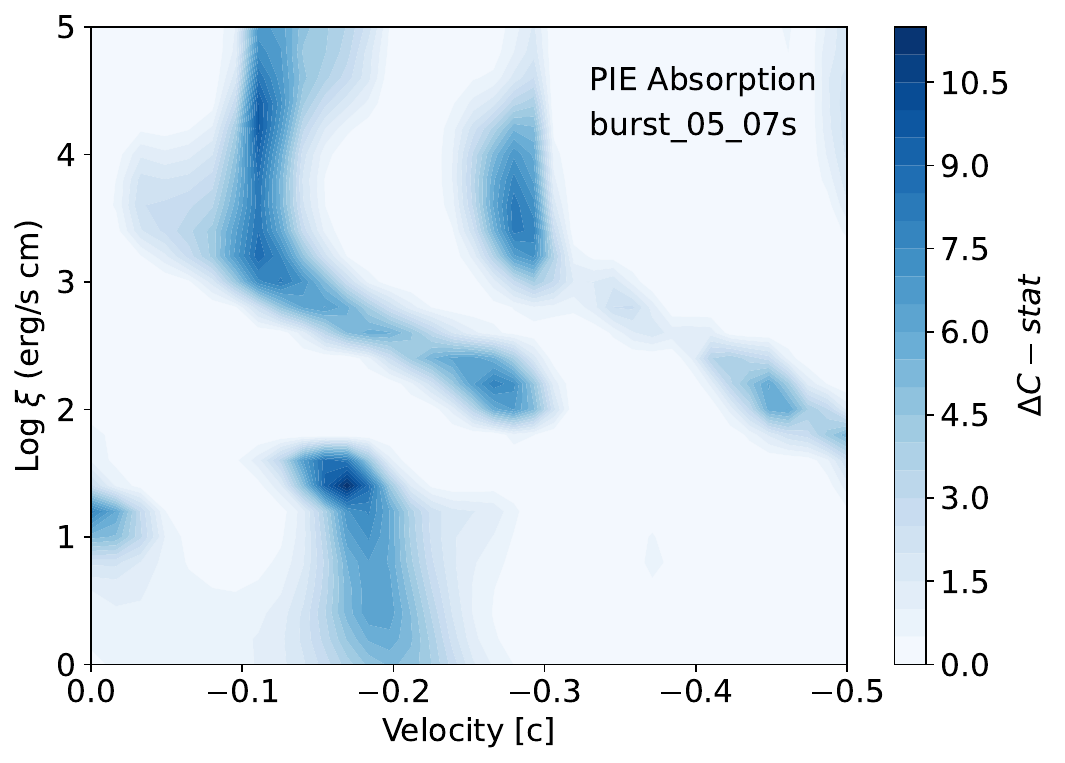}
    \includegraphics[width=0.30\textwidth]{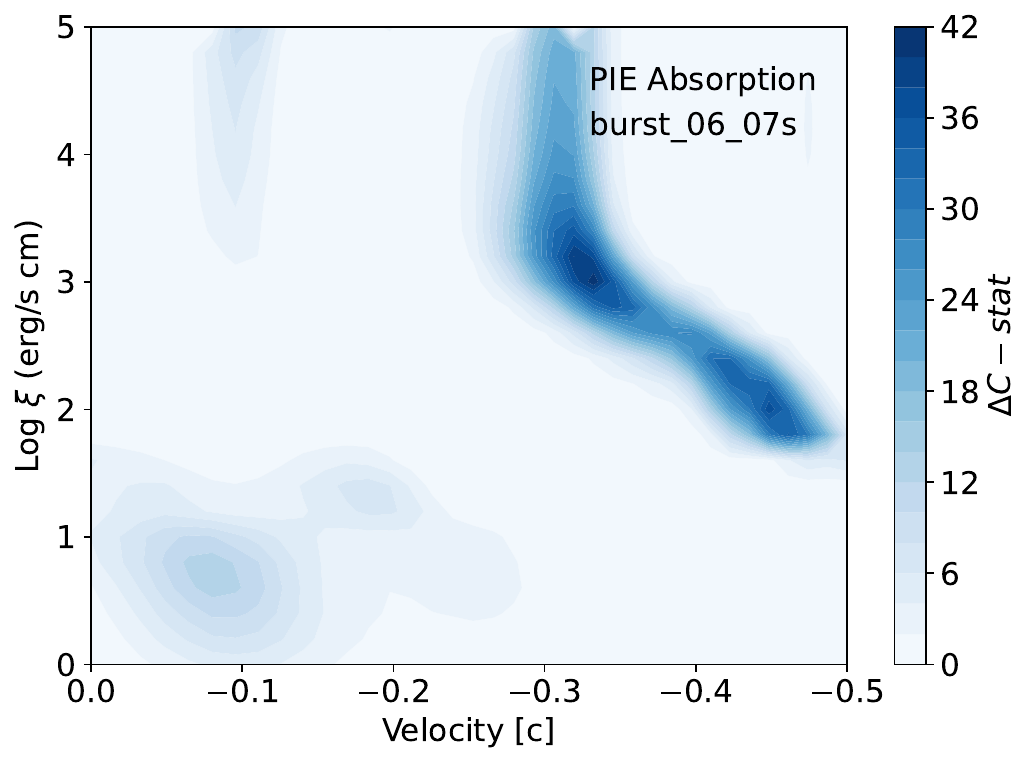}
    \includegraphics[width=0.30\textwidth]{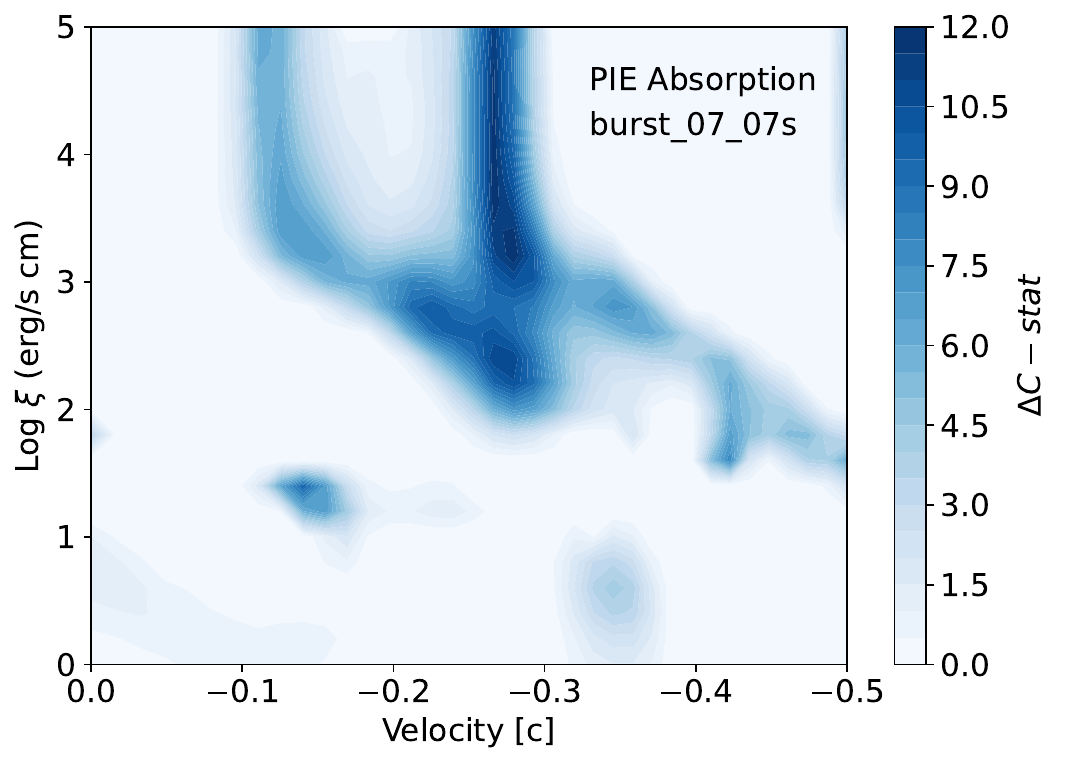}
    \includegraphics[width=0.30\textwidth]{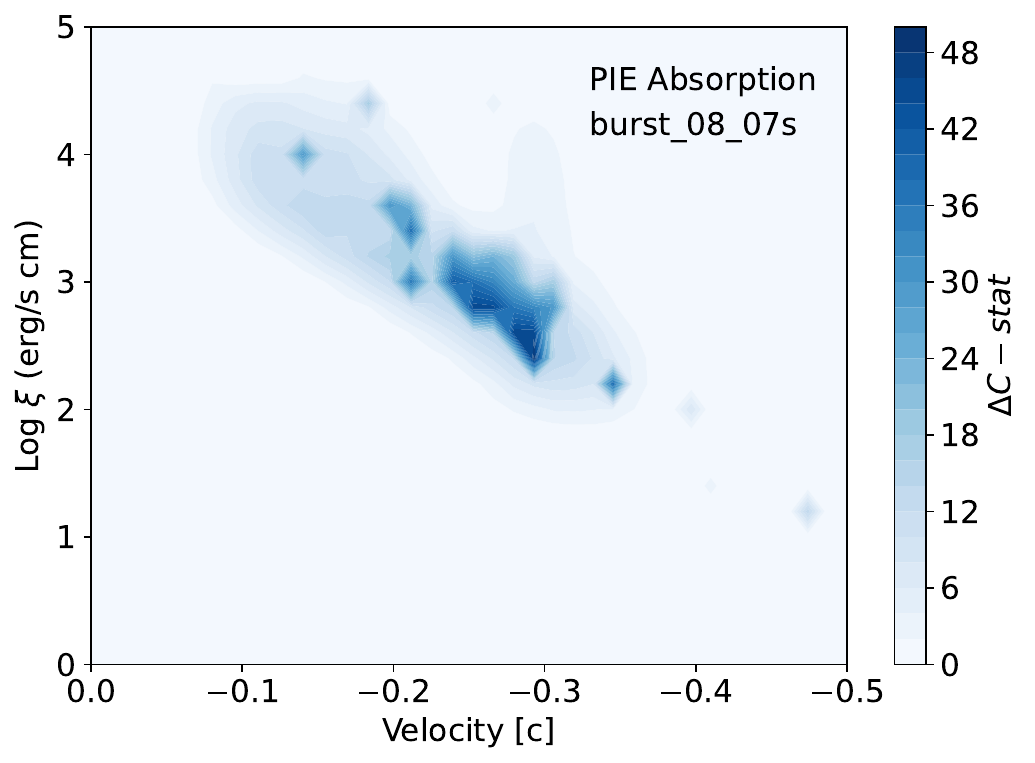}
    \includegraphics[width=0.30\textwidth]{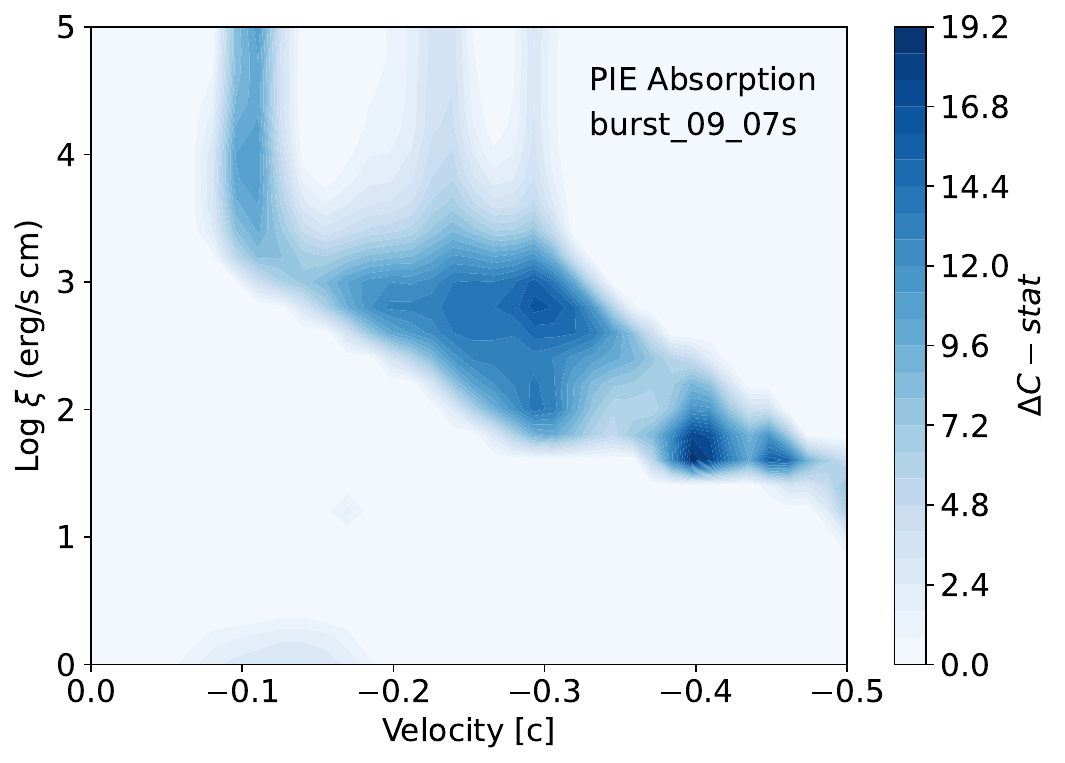}
    \includegraphics[width=0.30\textwidth]{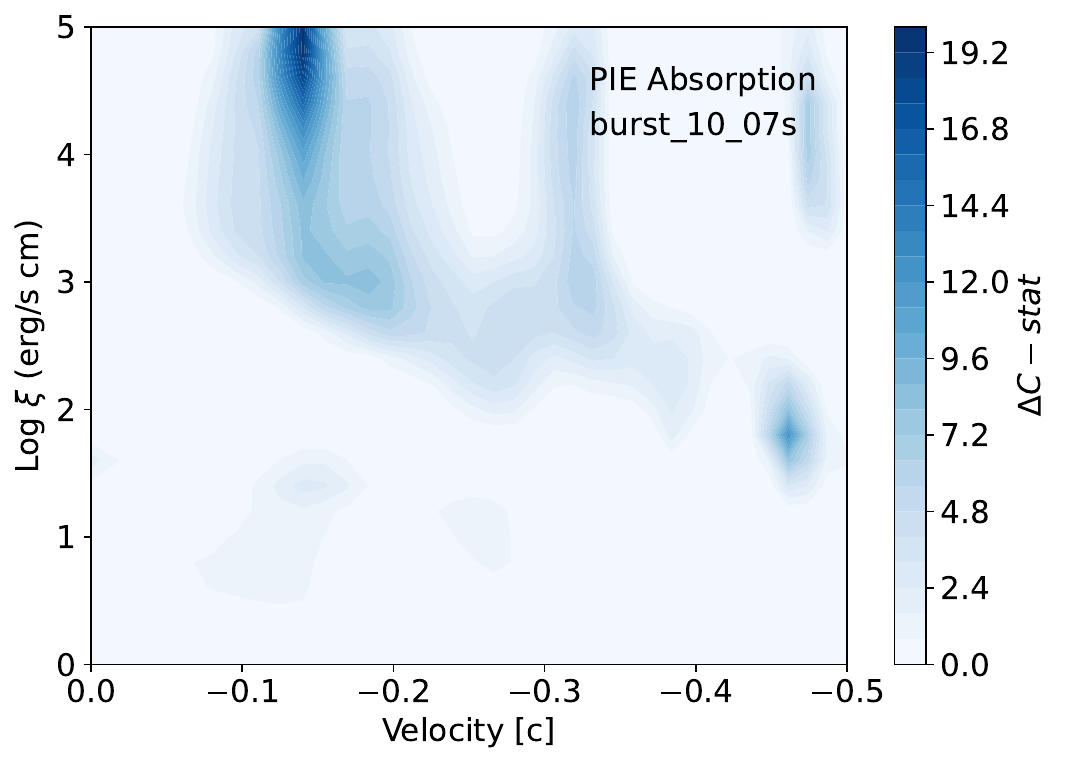}
    \includegraphics[width=0.30\textwidth]{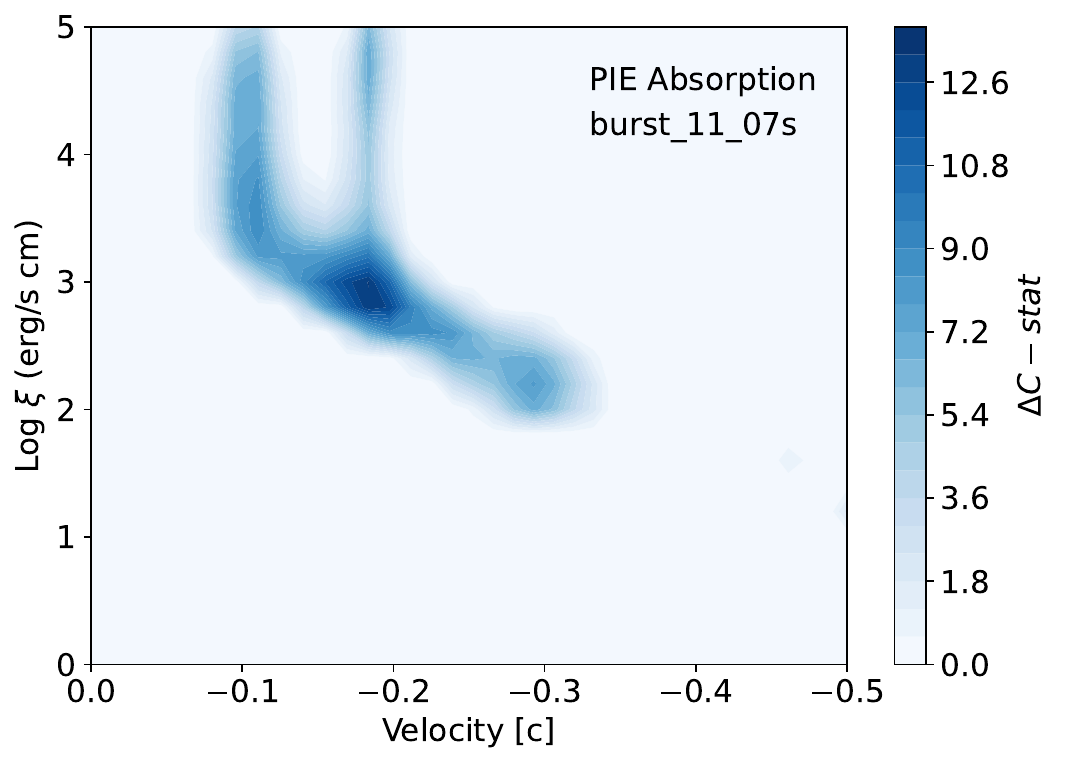}
    \includegraphics[width=0.30\textwidth]{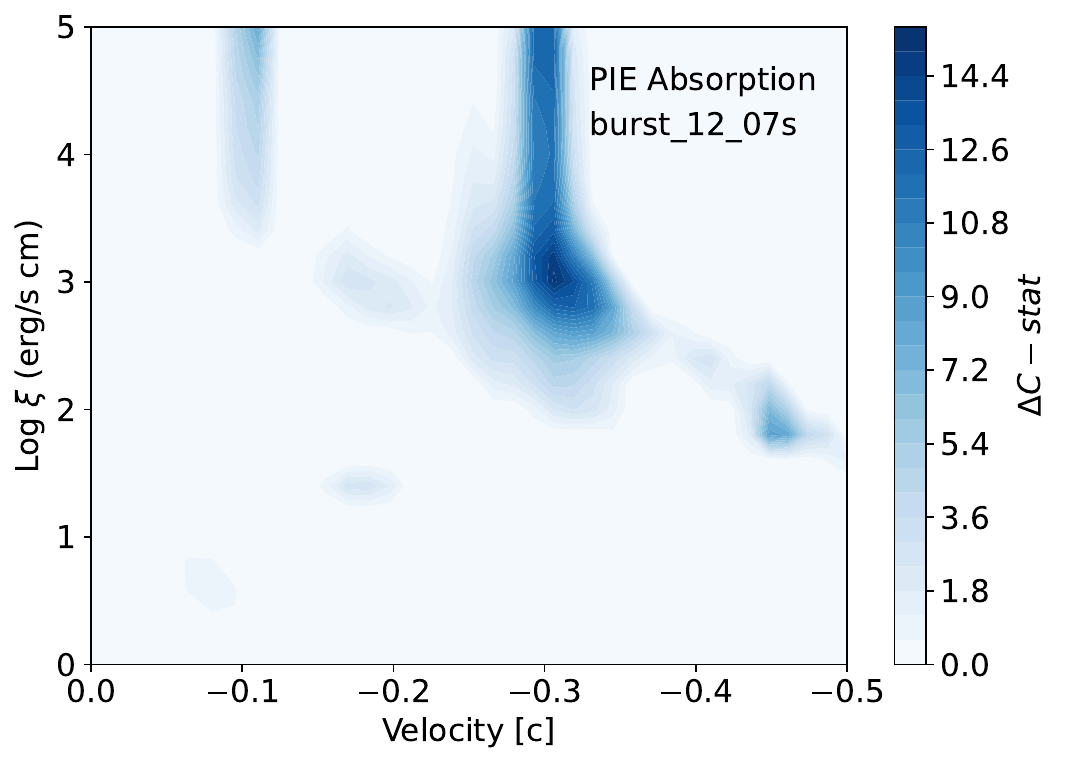}
       \caption{Multidimensional scan grids with the absorption photo-ionised plasma models (\textit{xabs}) for the 12 bursts. }
       \label{fig: XABS grids}
\end{figure*}

\newpage
\section{Multidimensional scan grids results}
\begin{figure*}[h!]
    \centering

    \includegraphics[width=0.53\textwidth]{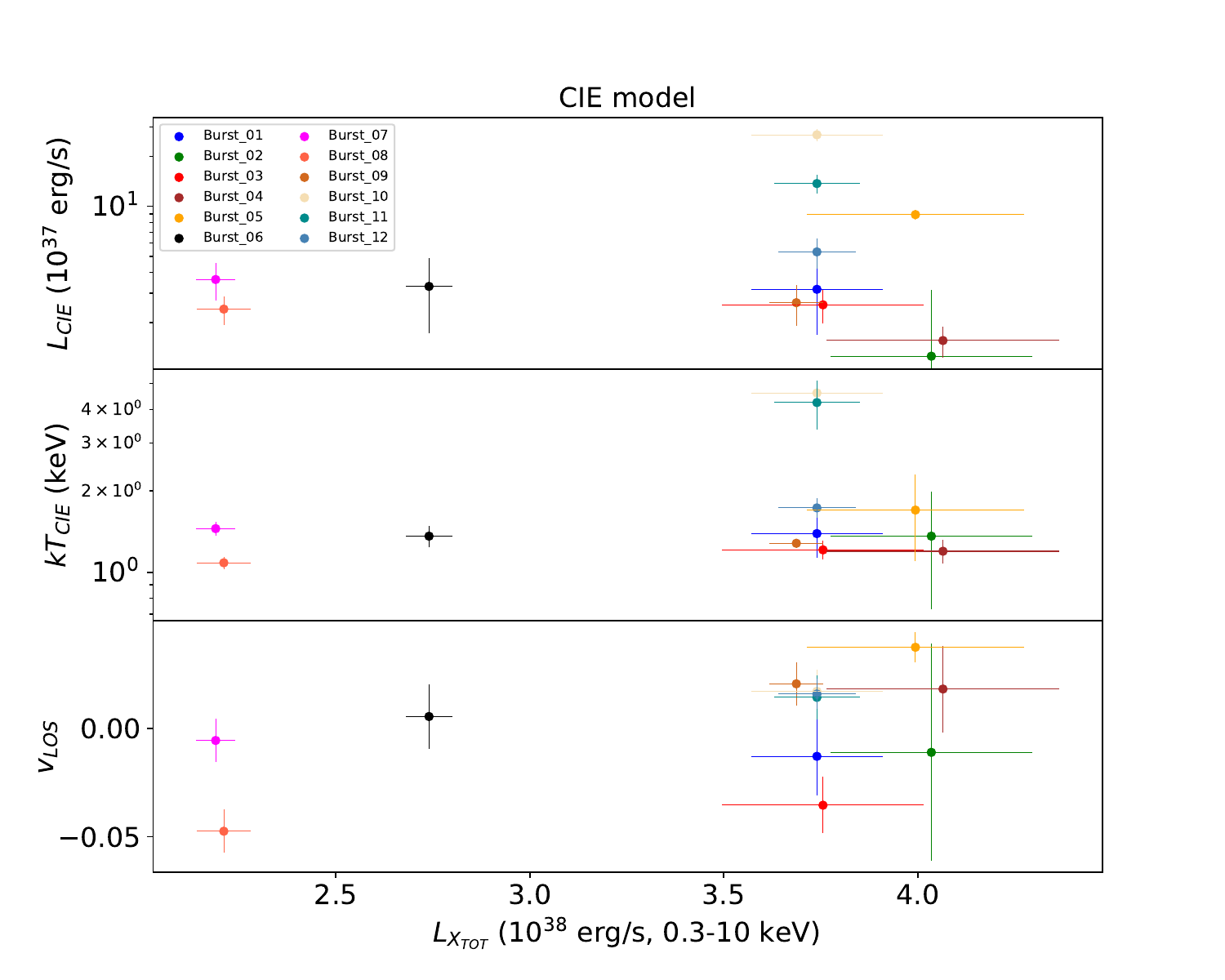}
    \includegraphics[width=0.53\textwidth]{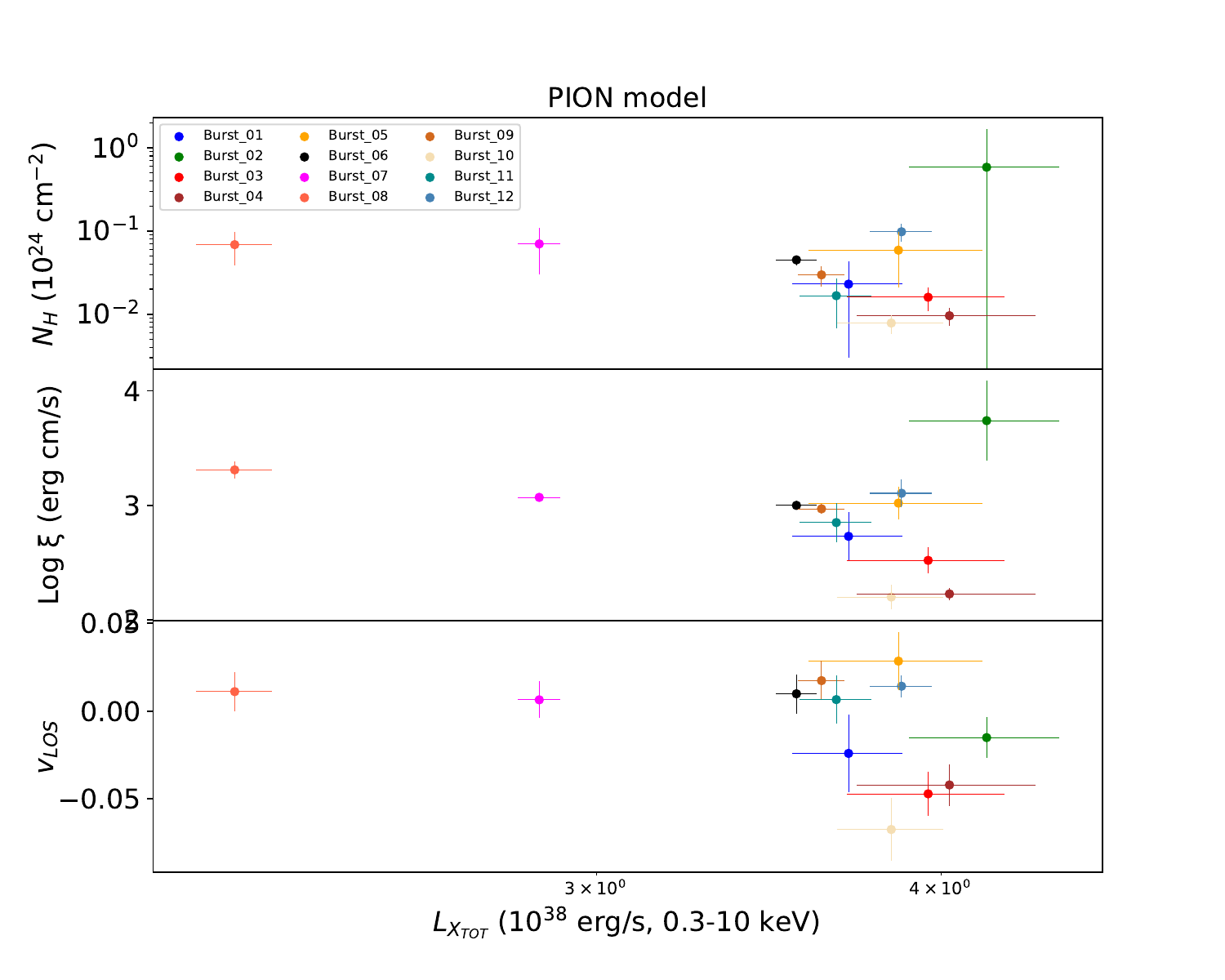}
    \includegraphics[width=0.53\textwidth]{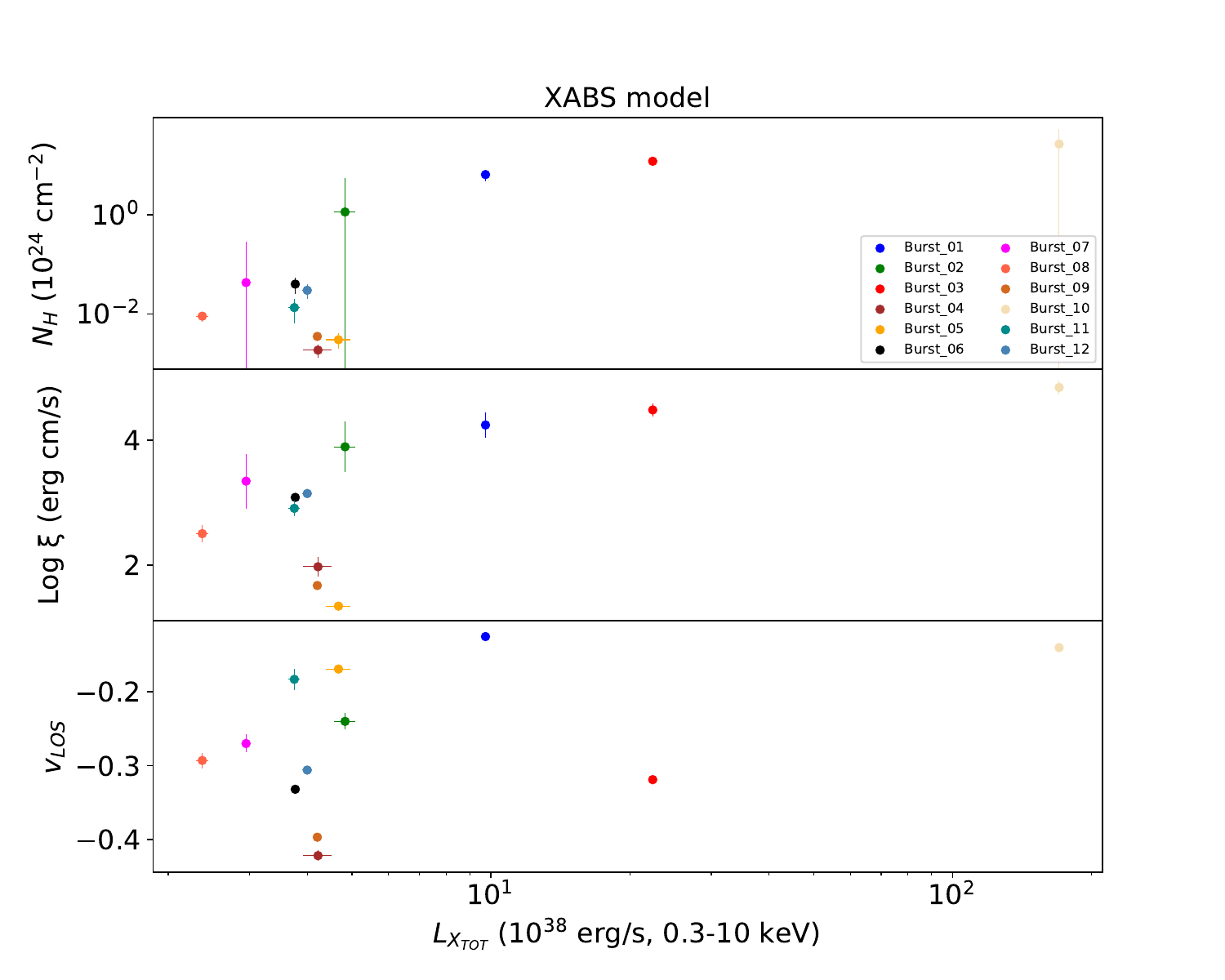}
    \caption{Results of the grids for the  \textit{cie} (top panel), \textit{pion} (middle panel) and \textit{xabs} (bottom panel) modelling. Top panel: $L_{\rm CIE}$, kT$_{\rm CIE}$ and $v_{\rm LOS}$ vs the total X-ray luminosity. Middle and bottom panel: $n_{\rm H}$, log $\xi$ and $v_{\rm LOS}$ vs the total X-ray luminosity. The luminosities are estimated in the 0.3-10 keV band. \\ }
        \label{fig:tables cie-pion-xabs}
\end{figure*}

\begin{center}
\begin{table*}[h!]
\caption{Results of the {\tt{PION - CIE - XABS}} modelling added to the continuum model for the 12 bursts.}  
 \renewcommand{\arraystretch}{1.}
 \small\addtolength{\tabcolsep}{-3pt}
 \vspace{0.1cm}
	\centering
	\scalebox{1.}{%
	\begin{tabular}{ccc @{\hspace{-27\tabcolsep}}c @{\hspace{-27\tabcolsep}}cccc}
    \hline\hline
    \toprule

Fit parameter                & Burst 1                           & Burst 2                                    & Burst 3                              & Burst 4                  & Burst 5                            & Burst 6                \\
\hline
                             &                                   &                                            &              PION   MODEL (EMISSION): \tt{hot * (pion + bb + comt)}                           &                          &                               &          \\
                             \hline

   $\rm n_{\rm H}$               & 0.02  $\pm$ 0.02                & 0.58 $\pm$ $_{0.46}^{1.74}$                    & 0.016 $\pm$ 0.005           & 0.010  $\pm$ 0.002       &   0.06  $\pm$ 0.04                  &  0.045 $\pm$ 0.006                   \\
      $\rm log \  \xi $            & 2.73  $\pm$ 0.21                   & 3.73  $\pm$ 0.35                           & 2.52   $\pm$ 0.11           & 2.23  $\pm$ 0.05        &   3.02  $\pm$ 0.14                  &  3.00 $\pm$ 0.04                   \\
   $\rm v_{\rm LOS}$                & -0.024  $\pm$ 0.021             & -0.015  $\pm$ 0.011                     & -0.05   $\pm$ 0.01         & -0.04  $\pm$ 0.01        &   0.03  $\pm$ 0.02               &  0.009  $\pm$ $_{0.008}^{0.014}$               \\
   $\rm L_{\rm PION}$                 & 0.15 $\pm$ 0.14             & 0.21  $\pm$ $_{0.13}^{0.61}$                    & 0.17  $\pm$ 0.05               & 0.21 $\pm$ 0.05      &   0.18 $\pm$ 0.12   &  0.16 $\pm$ 0.02         \\
    $\rm L_{ \rm Bol,tot}$    & 3.85  $\pm$ 0.20   &  4.31 $\pm$ 0.31 & 4.09 $\pm$ 0.35 &     4.15  $\pm$ 0.24 &   3.95 $\pm$ 0.33  & 3.70   $\pm$ 0.04 & \\
   $C_{\rm stat}/\rm d.o.f$          & 105/61                             & 56/62                                       & 88/63                       & 139/63                   &   82/62                             &  126/61                                     \\

   \hline 
                            &                                    &                                             &                CIE MODEL (EMISSION): \tt{hot * (cie + bb + comt)}                        &                          &                         &                     \\
                            \hline 
  $\rm Norm$                    & 164.2 $\pm$ 77.5                   & 50.42 $\pm$ $_{27.3}^{125.5}$              & 151.6 $\pm$ 35.0                     &   82.06 $\pm$  34.60                  &   272 $\pm$ $_{109}^{337}$             &  176.3 $\pm$ 84.5                                  \\
  $\rm T_{\rm CIE}$                 & 1.39  $\pm$ 0.03                   & 1.36  $\pm$ $_{0.24}^{1.02}$                  & 1.2  $\pm$ 0.1  & 1.19 $\pm$ 0.12      &   1.69 $\pm$ $_{0.27}^{0.92}$     &  1.37 $\pm$ $_{0.05}^{0.20}$             \\
  $\rm v_{\rm LOS}$             & -0.01 $\pm$ 0.02                   & -0.01 $\pm$ 0.05                          & -0.04  $\pm$ 0.01                     & 0.018 $\pm$ 0.019    &   0.037 $\pm$ 0.025  &  0.005 $\pm$ 0.024                \\
 $\rm L_{\rm CIE}$                 & 0.32 $\pm$ 0.15                  & 0.13  $\pm$  0.19                           & 0.26  $\pm$   0.06             & 0.16 $\pm$  0.03     &   0.89  $\pm$ $_{0.36}^{1.10}$   &  0.33 $\pm$ 0.16       \\
  $\rm L_{ \rm Bol,tot}$                 & 3.85 $\pm$ 0.15             & 4.15  $\pm$  0.19                             & 4.08  $\pm$   0.06             & 4.18 $\pm$  0.03     &   4.0 $\pm$ 0.7  &  3.55 $\pm$ 0.16            \\

  $C_{\rm stat}/\rm d.o.f$            & 107/61                             & 60/62                                      & 89/63                                        & 159/63                  &   83/62            &  128/61                                       \\

  \hline

                            &                                    &                                              &        XABS MODEL (ABSORPTION): \tt{hot * (xabs * (bb + comt))}                &                          &                         &                       \\ 
                            \hline

 $\rm n_{\rm H}$                  & 6.58 $\pm$ 0.07                    & 1.15 $\pm$ 0.76                              & 12.24  $\pm$ 0.74                    & 0.0018  $\pm$ 0.0006      &   0.0020 $\pm$ 0.0007 &  0.04   $\pm$ 0.01         \\
    $\rm log \  \xi $           & 4.24  $\pm$ 0.20                   & 3.89  $\pm$ 0.41                                & 4.48   $\pm$ 0.14                  & 1.97  $\pm$ 0.16        &   1.46  $\pm$ 0.09     & 3.08 $\pm$ 0.04                 \\
$\rm v_{\rm LOS}$                 & -0.125 $\pm$ 0.006                & -0.24  $\pm$ 0.01                           & -0.319  $\pm$ 0.006                  & -0.422 $\pm$ 0.007      &   -0.169 $\pm$ 0.012   &  -0.332 $\pm$ 0.006           \\

  $\rm L_{ \rm Bol,tot}$    &  1.27 $\pm$ 0.38 ($10^{39} \rm erg/s$)  &  5.86 $\pm$ $_{0.74}^{3.63}$ &  2.20 $\pm$ 0.20 ($10^{39} \rm erg/s$) & 4.33 $\pm$ 0.29 & 4.36 $\pm$ 0.29 &  0.945 $\pm$ 0.090      &  \\

  $C_{\rm stat}/\rm d.o.f$            & 113/61                             & 57/62                                         & 125/65                                   & 153/63                  &   93/62            &  107/61                                   \\
  \hline

    \end{tabular}}
\
\bigskip
\bigskip

 \renewcommand{\arraystretch}{1.}
 \vspace{0.1cm}
	\centering
	\scalebox{1.}{%
	\begin{tabular}{ccc @{\hspace{-31\tabcolsep}}c @{\hspace{-37\tabcolsep}}cccc}
    \hline\hline
    \toprule

Fit parameter                                   & Burst 7                       & Burst 8                  & Burst 9                           & Burst 10            & Burst 11                      & Burst 12               \\
\hline

                                               &                               &                          &      \hspace{1cm}                  PION   MODEL (EMISSION): \tt{hot * (pion + bb + comt)}                     &                     &                               &                       \\ 
    \hline   

   $\rm n_{\rm H}$                            & 0.07 $\pm$   0.04             & 0.07 $\pm$ 0.03           & 0.03 $\pm$ 0.01                   & 0.008 $\pm$ 0.002      & 0.016 $\pm$  0.010                     & 0.098 $\pm$ 0.023 \\
   $\rm log \  \xi $                       &  3.07  $\pm$ 0.12             & 3.31 $\pm$ 0.07         &  2.95  $\pm$ 0.05                 &  2.20 $\pm$  0.11   &  2.86 $\pm$ 0.17               &  3.10  $\pm$ 0.12         \\
   $\rm v_{\rm LOS}$                      &  0.006  $\pm$ 0.010             & 0.011  $\pm$ 0.011         &  0.029 $\pm$  $_{0.013}^{0.008}$                  &  -0.07  $\pm$ 0.02   &  0.006  $\pm$ 0.014           &  0.014  $\pm$ 0.006          \\
   $\rm L_{\rm PION}$                            &  0.19 $\pm$ 0.11      & 0.15 $\pm $ 0.06          &  0.15 $\pm$ 0.05                 &  0.16 $\pm$ 0.05    &   0.009 $\pm$ $_{0.004}^{0.008}$                                &  0.25  $\pm$  0.06             \\
    $\rm L_{ \rm Bol,tot}$    &  3.00 $\pm$ 0.10  &  2.44 $\pm$ 0.08  & 3.85 $\pm$ 0.08  &  4.04 $\pm$  0.28 & 3.59 $\pm$ 0.42 &   4.01 $\pm$ 0.09     &  \\
   
   $C_{\rm stat}/ \rm d.o.f$                     &  104/60                       & 109/54                   &  155/63                           &  118/62             &  101/61                       &  125/62                        \\
   \hline 
                                              &                                &                              &   \hspace{1cm}            CIE MODEL (EMISSION):  \tt{hot * (cie + bb + comt)}                         &                       &                                    &          \\
                                              \hline
  ${\rm Norm}$                            &  226.8 $\pm$ $_{35.5}^{79.9}$                       & 130.93 $\pm$ 25.25                  &  169.51 $\pm$ $_{14.74}^{77.84}$                            &  1497.8 $\pm$ 126.83               &  802.38 $\pm$ $_{44.59}^{159.46}$        &  381 $\pm$ $_{20}^{136}$                       \\
  $\rm T_{\rm{CIE}}$                          &  1.45 $\pm$ 0.08           & 1.08 $\pm $ 0.05       &  1.28 $\pm$ 0.05                 &  4.60 $\pm$ 0.54    &  4.24 $\pm$ $_{0.46}^{1.28}$            &  1.73 $\pm$ $_{0.28}^{0.10}$            \\
  $\rm v_{\rm LOS}$                      &  -0.006 $\pm$ 0.009           & -0.05 $\pm $ 0.01       &  0.02 $\pm$ 0.01                 &  0.02 $\pm$ 0.01    &  0.014 $\pm$ 0.011             &  0.016 $\pm$ 0.007            \\
  $\rm L_{\rm CIE}$                          &  0.36 $\pm$ $_{0.04}^{0.13}$           & 0.24 $\pm $ 0.05       &  0.26 $\pm$  $_{0.02}^{0.12}$               &  2.69 $\pm$ 0.22    &  1.37 $\pm$ $_{0.08}^{0.27}$            &  0.53 $\pm$ $_{0.03}^{0.19}$           \\
  $\rm L_{ \rm {Bol,tot}}$              &  3.04 $\pm$ 0.09           & 2.48 $\pm $ 0.05       &  3.90 $\pm$ 0.07                 &  4.80 $\pm$ 0.22    &  3.24 $\pm$ 0.27            &  4.12 $\pm$  0.11         \\

  $C_{\rm stat}/ \rm d.o.f$                &  108/60                       & 101/54                   &  131/60                           &  119/62             &  101/61                       &  129/62                         \\
\hline

                                   &                                &                              &     \hspace{1cm}             XABS MODEL (ABSORPTION): \tt{hot * (xabs * (bb + comt))}                    &                       &                                    &          \\
                                   \hline

 $\rm n_{\rm H}$                &  0.04  $\pm$ $_{0.01}^{0.49}$    & 0.009 $\pm$ 0.002 &  0.003 $\pm$ 0.001           &  27.25 $\pm$ $_{15.00}^{29.80}$   &  0.013  $\pm$ 0.007 &  0.03   $\pm$  0.01         \\
    $\rm log \  \xi $           &  3.34  $\pm$ $_{0.07}^{0.81}$              & 2.50 $\pm$ 0.13         &  1.67 $\pm$ 0.02                 &  4.8 $\pm$  0.1   &  2.90 $\pm$ 0.13             &  3.14  $\pm$ 0.05          \\
$\rm v_{\rm LOS}$             &  -0.27 $\pm$ 0.01            & -0.293 $\pm $ $_{0.001}^{0.019}$       &  -0.397 $\pm$ 0.006                 &  -0.140 $\pm$ 0.004    &  -0.183 $\pm$ 0.014             &  -0.306 $\pm$ $_{0}^{0.003}$            \\

$\rm L_{ \rm Bol,tot}$    & 3.08 $\pm$ 0.09  & 2.61 $\pm$ 0.41  & 4.38 $\pm$ 0.26 & 1.22 $\pm$  0.854 $(10^{40}\rm erg/s)$   & 3.86  $\pm$ $_{0.13}^{0.58}$ &  4.15  $\pm$ 0.99       &  \\

  $C_{\rm stat}/ \rm d.o.f$              &  127/60                       & 107/54                   &  126/60                           &  126/62             &  103/61                       &  147/62                         \\
        
\hline
    \end{tabular}}

    \label{table: Results of the lines modeling for the 12 bursts.}
\tablefoot{The luminosities $\rm L$ for the {\tt{pion}} and {\tt{cie}} model and $\rm L_{Bol,tot}$ are calculated in the 0.3-10 keV  and 0.1-20 keV band, respectively, and expressed in $10^{38}$ erg/s unit. The line of sight velocities $v_{\rm LOS}$ are expressed are expressed in c unit. The ionisation parameter $\xi$ is expressed in erg/s cm unit. The $n_{\rm H}$ column density parameter is expressed in $10^{24}$/$\rm cm^{2}$ unit.} 

     \vspace{-0.3cm}
\end{table*}
\end{center}
\end{appendix}
\clearpage

\end{document}